\documentclass[aps,superscriptaddress,eqsecnum,nofootinbib,preprintnumbers]{revtex4}
\usepackage{graphicx,epsfig}
\usepackage{amsmath}
\usepackage{amssymb}
\usepackage{subfigure}

\usepackage{relsize}

\usepackage{textcomp}

\newcommand{\p}{\partial}

\begin{document}

\title{
A solution of the dark energy and its coincidence problem based on local antigravity sources without
fine-tuning or new scales
}

\author{Georgios Kofinas}
\email{gkofinas@aegean.gr}
\affiliation{Research Group of Geometry, Dynamical Systems and Cosmology,\\
Department of Information and Communication Systems Engineering,\\
University of the Aegean, Karlovassi 83200, Samos, Greece}

\author{Vasilios Zarikas}
\email{vzarikas@teilam.gr}
\affiliation{Central Greece University of Applied Sciences,\\
Department of Electrical Engineering, 35100 Lamia, Greece}
\affiliation{Nazarbayev University, School of Engineering, Astana, Republic of Kazakhstan, 010000}

\begin{abstract}

A novel idea is proposed for a natural solution of the dark energy and its cosmic
coincidence problem. The existence of local antigravity sources, associated with astrophysical
matter configurations distributed throughout the universe, can lead to a recent cosmic
acceleration effect. Various physical theories can be compatible with this idea, but here,
in order to test our proposal, we focus on quantum originated spherically symmetric metrics matched
with the cosmological evolution through the simplest Swiss cheese model.
In the context of asymptotically safe gravity, we have explained the observed amount of dark
energy using Newton's constant, the galaxy or cluster length scales, and dimensionless order
one parameters predicted by the theory, without fine-tuning or extra unproven energy scales.
The interior modified Schwarzschild-de Sitter metric allows us to approximately interpret this
result as that the standard cosmological constant is a composite quantity made of the above
parameters, instead of a fundamental one.

\end{abstract}

\maketitle

\section{Introduction}
\label{Introduction}

The so called cosmological constant problem is nothing more than the simple observation,
due to Zeldovich, that the quantum vacuum energy density should unavoidably contribute to the
energy-momentum of the Einstein equations in the form of a cosmological constant. The recent
discovery of a Higgs-like particle at the CERN Large Hadron Collider provides the experimental
verification of the existence of the electroweak vacuum energy, and thus of the reality of the
cosmological constant problem. The obvious absence of such a huge vacuum energy in the universe
has led to theoretical and phenomenological attempts to cancel out this vacuum energy, which are
all so far subject to fine-tuning problems
\cite{Weinberg:1988cp, Sahni:1999gb, Peebles:2002gy, Sola:2013gha}.

On the other hand, the observational evidence of the accelerated expansion of the universe
\cite{Riess:1998cb, Perlmutter:1998np, Knop:2003iy, Spergel:2006hy, Komatsu:2010fb, Ade:2013zuv,
Huterer:1998qv} has introduced the notion of dark energy. One option is that the dark energy is due
to a cosmological constant $\Lambda$ in the $\Lambda$CDM model, where in this case the explanation
of the huge discrepancy between this observed $\Lambda$ and the expected quantum vacuum energy is
rather more pertinent. Even if the dark energy has nothing to do with the quantum vacuum energy,
but is due to quite different phenomena, the cosmological constant problem remains as a hard problem
in physics. In any case, the existence of dark energy is associated with another cosmological puzzle,
the so called cosmic coincidence problem. The latter refers to the need for an
explanation of the recent passage from a deceleration era to present acceleration cosmic phase.

The aim of the present work is twofold: First, to propose a novel idea for a natural solution to
the dark energy issue and its associated cosmic coincidence problem of recent acceleration. Second,
to implement this idea through an interesting and concrete scenario, among others, which explains the
correct amount of dark energy without the introduction of new and arbitrary scales or fine-tuning.

{\emph{The proposed solution is based on the simple idea that the acceleration/dark energy can be due to
infrared modifications of gravity at intermediate astrophysical scales which effectively generate
local antigravity effects. The cosmological consequence of all these homogeneously distributed local
antigravity sources is an overall cosmic acceleration through the matching between the local and the
cosmic patches. Before the appearance of astrophysical structures (galaxies, clusters of galaxies),
such antigravity effects do not exist, and therefore, the recent emergence of dark energy
is not a coincidence but an outcome of the recent formation of structure.}} Before the appearance
of structure and the emergence of sufficient repulsive effects, the conventional deceleration
scenario is expected.

Various physical theories (alternative gravities, extra-dimensional gravities, quantum gravities,
e.tc.) can be implemented and be compatible with the previous general idea, providing intermediate
distance infrared modifications which act as local antigravity sources. It is worth to notice that
when the same physical theories are applied directly at the far infrared cosmic scales do not
necessarily give comparable or significant cosmological effects. So, a dark energy of local origin
in the universe is not an equivalent or alternative description, but can be a necessity in order to
reveal the relevant phenomena at intermediate scales. Quantum theories of gravity, in particular,
provide types of models where local repulsive effects are naturally expected (for example, a quantum
gravity origin of negative pressure can be formed in the interior of astrophysical black holes).
Asymptotically safe (AS) gravity \cite{ASreviews} is one of the promising quantum gravity frameworks
that we will elaborate more thoroughly in the following sections in relation to the previous ideas.
{\emph{We will show in our most successful scenario that the observed dark energy can be explained
from the Newton's constant, the galaxy or cluster length scales, and dimensionless order one
parameters predicted by AS theory, without fine-tuning or introduction of new scales. This can
approximately be interpreted as that the observed cosmological constant $\Lambda$ is not a fundamental
parameter, but it is composite and naturally arises from other fundamental quantities.}}

In order to study the effect of all local sources of antigravity in the cosmic evolution, we adopt
in the present work a simple Swiss cheese model by matching a homogeneously and isotropic spacetime
with the appropriate local spherically symmetric metrics \cite{Einstein:1946ev}, and this formulation
is presented in Section II. As an introductory step to set up the Swiss cheese evolution equations,
we work out in Section III the classical Schwarzschild metric. In Section IV we provide some general
thoughts on the relation between a possible locally originated dark energy and the coincidence
problem. In Section V, the Schwarzschild-de Sitter black hole is
discussed with respect to the above scenario. In Section VI quantum improved Schwarzschild-de Sitter
metrics are considered; quantum gravity effects may indeed introduce an explicit or effective
cosmological constant which arises from ultraviolet or infrared modifications of gravity
\cite{Dymnikova:2001fb, Hayward:2005gi, Kiritsis:2009rx, Babichev:2013cya, Rodrigues:2015hba}.
Finally, Section VII is the largest and most important one, where the AS theory is applied in the
context of our ideas. The first subsection VIIA discusses the running of the cosmological constant
close to the Gaussian fixed point of the AS evolution and the resulting cosmology is practically
indistinguishable from the $\Lambda$CDM scenario with the same fine-tuning problems. The last
subsection VIIB discusses in detail, for the running of the cosmological constant close to the
infrared (IR) fixed point of the AS evolution, the quite interesting emergence of the dark energy
out of known physical scales and parameters predicted by the theory, and provides a natural
explanation to the recent cosmic acceleration without obvious observational conflicts with internal
dynamics of galaxies or clusters. We finish with the conclusions in Section VIII.

It is worth mentioning that attempts to explain acceleration without a dark energy component, or
also to produce dark energy, all due to structure formation, have already appeared in the literature
(e.g. \cite{inho}, \cite{inhoswiss}, \cite{structure} and references therein). The existence of
structure formation in the universe implies a non-linear local evolution, while the distribution
of the non-linear regions is homogeneous and isotropic above a today homogeneity scale of the
order $100\text{Mpc}$. The apparent recent cosmic acceleration could be the effect of inhomogeneities
and/or anisotropies on the average expansion rate, broadly referred as back-reaction. This
approach can potentially solve the coincidence problem too. However, it is fair to say that our
viewpoint in the present work is different. In the averaging procedure, the matter is treated
as a usual pressureless ideal fluid in the context of General Relativity, gravity has the
standard attractive behaviour inside the structure, and cosmic acceleration arises due to the
non-trivial complexity of the considered solution; there are no explicit antigravity forces and
repulsive effects come only through averaging. Here, on the contrary, the acceleration
and the dark energy come from the existence of antigravity sources related to the astrophysical
structures in a as simple as possible spacetime and no averaging is performed; in the present work,
these repulsive forces are basically of quantum origin as AS suggests, although in general, they
can be of some other geometric nature generated by some modified gravity theory with IR gravity
modifications. In our approach, this simple spacetime is described as a first step by the
homogeneous Swiss cheese model with its known Schucking matching surface, although a better
approximation would be to use inhomogeneous Swiss cheese models (e.g. some analogues of
Lemaitre-Tolman-Bondi or Szekeres); averaging processes in this case are expected to enhance the
cosmic acceleration found here. A different scenario, where structure is responsible for acceleration,
was presented in \cite{Kofinas:2011pq}; in a five-dimensional setup, a brane-bulk energy exchange
in the interior of galactic core black holes produces a sufficient negative dark pressure to
play the role of dark energy.

\section{Swiss cheese models}
\label{bker}

The Swiss cheese cosmological model, first introduced by Einstein and Strauss \cite{Einstein:1946ev},
is solution of General Relativity that globally respects homogeneity and isotropy, while locally
describes a spherically symmetric solution. Other more general Swiss cheese models refer to
inhomogeneous solutions. A Swiss cheese model with spherical symmetry overcomes the difficulty of
how to glue a static solution of the theory at hand within a larger time-dependent homogeneous and
isotropic spacetime. The idea is to assume a very large number of local objects homogeneously and
isotropically distributed in the universe. The matching of a spatially homogeneous metric as the
exterior spacetime to a local interior solution has to be realized across a spherical boundary
that stays at a fixed coordinate radius in the cosmological frame while evolves in the interior frame.

Let us consider a four-dimensional manifold $M$ with metric $g_{\mu\nu}$ and a timelike
hypersurface $\Sigma$ which splits the spacetime $M$ into two parts.
The spacetime coordinates are denoted
by $x^{\mu}$ ($\mu,\nu,...$ are four-dimensional coordinate indices) and can be different between
the two regions. The coordinates on $\Sigma$ are denoted by $\chi^{i}$ ($i,j,...$ are
three-dimensional coordinate
indices on $\Sigma$). The embedding of $\Sigma$ in $M$ is given by some functions $x^{\mu}(\chi^{i})$.
The unit normal vector $n^{\mu}$ to $\Sigma$ points inwards the two regions.
The first relevant quantity characterizing $\Sigma$ is the induced metric $h_{\mu\nu}=g_{\mu\nu}
-n_{\mu}n_{\nu}$ coming from the spacetime in which it is embedded.
The second quantity is the extrinsic curvature $K_{\mu\nu}=h_{\mu}^{\kappa}
h_{\nu}^{\lambda}n_{\kappa;\lambda}$, where a $;$ denotes covariant differentiation with respect
to $g_{\mu\nu}$. In the adapted frame where $x^{\mu}$, say $\bar{x}^{\mu}$,
contains $\bar{x}^{i}$ with $\bar{x}^{i}|_{\Sigma}=\chi^{i}$ and some extra transverse coordinate, it
is $h_{ij}=g_{ij}$. However, this quantity can be expressed in terms of arbitrary spacetime
coordinates $x^{\mu}$ as
\begin{equation}
h_{ij}=g_{\mu\nu}\frac{\p x^\mu}{\p \chi^i}\frac{\p x^\nu}{\p \chi^j}\,.
\label{kee3}
\end{equation}
Similarly, for the extrinsic curvature it is $K_{ij}=n_{i;j}$, and can be expressed as
\begin{eqnarray}
K_{ij}&=&\Big(\frac{\p n_{\mu}}{\p\chi^{j}}
-\Gamma^\lambda_{\,\,\,\mu\nu}n_\lambda\frac{\p x^\nu}{\p \chi^j}\Big)
\frac{\p x^{\mu}}{\p\chi^{i}}
\label{eq:SFF}\\
&=&-n_{\lambda}\Big(\frac{\p^{2}x^{\lambda}}{\p\chi^{i}\p\chi^{j}}+\Gamma^{\lambda}_{\,\,\,\mu\nu}
\frac{\p x^{\mu}}{\p\chi^{i}}\frac{\p x^{\nu}}{\p\chi^{j}}\Big)\,,
\label{juer}
\end{eqnarray}
where $\Gamma^\kappa_{\,\,\,\mu\nu}$ are the Christoffel symbols of $g_{\mu\nu}$.

Continuity of the spacetime across the hypersurface $\Sigma$ implies that $h_{ij}$ is continuous
on $\Sigma$, which means that $h_{ij}$ is the same when computed on either side of $\Sigma$.
If we consider Einstein gravity with a regular spacetime matter content and vanishing distributional
energy-momentum tensor on $\Sigma$, then the Israel-Darmois matching conditions \cite{Israel:1966rt}
imply that the sum of the two extrinsic curvatures computed on the two sides of $\Sigma$ is zero.

The model of Einstein-Strauss refers to the embedding of a Schwarzschild mass into FRW cosmology.
Here, we shall assume a general static spherically symmetric metric which matches smoothly to a
homogeneous and isotropic cosmological metric. In spherical coordinates the cosmological metric takes
the form
\begin{equation}
ds^2=-dt^2+a^2(t)\left[\frac{dr^2}{1-\kappa r^2}+r^2\left(d\theta^2+\sin^2\!\theta \,d\varphi^2\right)
\right]\, ,
\label{eq:FRW}
\end{equation}
where $a(t)$ is the scale factor and $\kappa=0, \pm 1$ characterizes the spatial curvature.
In these coordinates, a ``spherical'' boundary is defined to have a fixed coordinate radius
$r=r_{\Sigma}$, with $r_{\Sigma}$ constant. Of course, this boundary is seen by a cosmological
observer to expand, following the universal expansion. If $\bar{x}^{\mu}=(t,r,\theta,\varphi)$ are the
coordinates of the metric (\ref{eq:FRW}), then the hypersurface $\Sigma$ is determined by
the function $\bar{f}(\bar{x}^{\mu})=r-r_{\Sigma}=0$ and the cosmological metric occurs for
$r\geq r_{\Sigma}$. From the coordinates $\bar{x}^{\mu}$ one can parametrize $\Sigma$ by
the coordinates $\chi^{i}=\bar{x}^{i}|_{\Sigma}=(t,\theta,\varphi)$, and therefore on $\Sigma$
it is $\bar{x}^{\mu}(\chi^{i})=(t,r_{\Sigma},\theta,\varphi)$.
The unit normal vector can be calculated from
\begin{equation}
\bar{n}_\mu=\frac{\bar{f}_{,\mu}}{\sqrt{|\bar{g}^{\kappa\lambda}\bar{f}_{,\kappa}
\bar{f}_{,\lambda}|}}\,,
\label{eq:unit norm eq}
\end{equation}
where a comma means differentiation with respect to $\bar{x}^{\mu}$. Obviously, the plus sign in
(\ref{eq:unit norm eq}) makes certain that $\bar{n}^{\mu}$ is inward the cosmological region
(to the direction of increasing $r$). Thus,
\begin{equation}
\bar{n}_{\mu}=\Big(0\,,\,\frac{a}{\sqrt{1\!-\!\kappa r_{\Sigma}^{2}}}\,,\,0\,,\,0\Big)\,.
\label{nfi}
\end{equation}
Note that $\bar{n}^{\mu}$ is spacelike, $\bar{n}^{\mu}\bar{n}_{\mu}=1$, as expected.

The interior region $r\leq r_{\Sigma}$ is replaced by another metric which has the following form
\begin{equation}
ds^2=-J(R)F(R)dT^2+\frac{dR^2}{F(R)}+R^2\left(d\theta^2+\sin^2\!\theta\, d\varphi^2\right)\,,
\label{generalBH}
\end{equation}
where $J,F>0$.
This metric represents a static spherically symmetric spacetime in Schwarzschild-like coordinates.
The functions $F(R)$ and $J(R)$ are given by the specific metric in use. Since the two-dimensional
sphere $(\theta,\varphi)$ is the common fiber for both metrics (\ref{eq:FRW}), (\ref{generalBH}),
the position of $\Sigma$ in the spacetime described by (\ref{generalBH}) does not depend on
$\theta,\varphi$ and is given by the functions $T=T_{S}(t), R=R_{S}(t)$. The subscript $S$
refers to Schucking, $R_{S}$ is called Schucking radius and it is time-dependent.
Therefore, the spherical boundary does not remain in constant radial coordinate
distance in the Schwarzschild-like patch as the universe expands.
The coordinates $\hat{x}^{\mu}=(T,R,\theta,\varphi)$ of the metric (\ref{generalBH})
take on $\Sigma$ the form $\hat{x}^{\mu}(\chi^{i})=(T_{S}(t),R_{S}(t),\theta,\varphi)$.
The unit normal vector
$\hat{n}^{\mu}$ cannot be calculated now directly from a formula as (\ref{eq:unit norm eq}),
since the function $\hat{f}(\hat{x}^{\mu})$ of the matching surface is now unknown. However,
due to the symmetry it is expected that $\hat{n}_{\theta}=\hat{n}_{\varphi}=0$, therefore the
orthonormality of $\hat{n}^{\mu}$ will provide two conditions for $\hat{n}_{T},\hat{n}_{R}$.
Indeed, since the three vectors $\frac{\p \hat{x}^{\mu}}{\p \chi^{i}}$ are tangent to $\Sigma$,
the condition $\hat{n}_{\mu}\frac{\p \hat{x}^{\mu}}{\p \chi^{i}}=0$ implies
$\hat{n}_{\theta}=\hat{n}_{\varphi}=0$ and
\begin{equation}
\frac{dT_{S}}{dt}\hat{n}_{T}+\frac{dR_{S}}{dt}\hat{n}_{R}=0\,.
\label{S2cond}
\end{equation}
Furthermore, from $\hat{n}^{\mu}\hat{n}_{\mu}=1$ one obtains
\begin{equation}
\frac{1}{JF}\hat{n}_{T}^{2}-F\hat{n}_{R}^{2}=-1\,,
\label{eq:normal S1 cond}
\end{equation}
where $J,F$ are located at $R_{S}$.

So far, we have established the geometrical setting on the two sides of the boundary hypersurface.
The junction of the two regions on $\Sigma$ demands $\bar{h}_{ij}=\hat{h}_{ij}$, which provides
through (\ref{kee3}) the conditions
\begin{equation}
JF\Big(\frac{dT_{S}}{dt}\Big)^{2}-\frac{1}{F}\Big(\frac{dR_{S}}{dt}\Big)^{2}=1
\label{eq:First FF1}
\end{equation}
and
\begin{equation}
R_{S}=a r_{\Sigma}\,.
\label{eq:First FF2}
\end{equation}
The two equations (\ref{eq:First FF1}), (\ref{eq:First FF2}) can also arise easier from the
two expressions for the induced metric on $\Sigma$ coming from (\ref{eq:FRW}), (\ref{generalBH})
\begin{eqnarray}
ds_{\Sigma}^{2}&=&-dt^{2}+a^{2}r_{\Sigma}^{2}(d\theta^{2}+\sin^{2}\!\theta\,d\varphi^{2})\\
&=&-\Big[JF\Big(\frac{dT_{S}}{dt}\Big)^{2}-\frac{1}{F}\Big(\frac{dR_{S}}{dt}\Big)^{2}\Big]dt^{2}
+R_{S}^{2}(d\theta^{2}+\sin^{2}\!\theta\,d\varphi^{2})\,.
\label{oker}
\end{eqnarray}
Solving (\ref{S2cond}), (\ref{eq:normal S1 cond}) for $\hat{n}_{T},\hat{n}_{R}$ and using
(\ref{eq:First FF1}) we find the normal vector $\hat{n}^{\mu}$ from
\begin{equation}
\hat{n}_{\mu}=\Big(\epsilon\sqrt{J}\,\frac{dR_{S}}{dt}\,,\,-\epsilon\sqrt{J}\,\frac{dT_{S}}{dt}
\,,\,0\,,\,0\Big)\,,
\label{eq:normal K}
\end{equation}
where $\epsilon=\pm 1$.
The demand that $\hat{n}^{\mu}$ is inward the central void (to the direction of decreasing
$R$) implies $\hat{n}_{R}<0$. Additionally, the forms of $\bar{n}_{\mu},\hat{n}_{\mu}$ show that
the directions defined by the coordinate axes $T,R$ are different than those of $t,r$,
however, the centers of the two coordinate systems coincide.

What remains is the matching of the two extrinsic curvatures on $\Sigma$, i.e. the demand
$\bar{K}_{ij}+\hat{K}_{ij}=0$. Due to the simple form of $\bar{n}_{\mu}$, the extrinsic curvature
$\bar{K}_{ij}$ in the cosmological region can be easily computed from either equation
(\ref{eq:SFF}) or (\ref{juer}) as
\begin{equation}
\bar{K}_{ij}=-\bar{n}_{r}\bar{\Gamma}^{r}_{\,\,\,ij}=\frac{1}{2}\bar{n}_{r}\bar{g}^{rr}
\bar{g}_{ij,r}
\label{jert}
\end{equation}
and finally
\begin{equation}
(\bar{K}_{tt},\bar{K}_{\theta\theta},\bar{K}_{\varphi\varphi})=
\sqrt{1\!-\!\kappa r_{\Sigma}^{2}}\,\,ar_{\Sigma}\,(0,\,1,\,\sin^{2}\!\theta)\,.
\label{kert}
\end{equation}
In the interior region the computation is more involved and it is slightly more convenient to
use the expression (\ref{juer}) to compute $\hat{K}_{ij}$. The corresponding non-vanishing
Christoffel symbols
are $\hat{\Gamma}^{R}_{\,\,\,\,TT}=JF^{2}\,\hat{\Gamma}^{T}_{\,\,\,\,TR}=\frac{F}{2}(JF'+FJ')$,
$\hat{\Gamma}^{R}_{\,\,\,\,RR}=-\frac{F'}{2F}$, $\hat{\Gamma}^{R}_{\,\,\,\,\varphi\varphi}
=\sin^{2}\!\theta\,\hat{\Gamma}^{R}_{\,\,\,\,\theta\theta}=-RF\sin^{2}\!\theta$,
$\hat{\Gamma}^{\theta}_{\,\,\,\,R\theta}=\hat{\Gamma}^{\varphi}_{\,\,\,\,R\varphi}=\frac{1}{R}$,
$\hat{\Gamma}^{\theta}_{\,\,\,\,\varphi\varphi}=-\sin{\theta} \cos{\theta}$,
$\hat{\Gamma}^{\varphi}_{\,\,\,\,\theta\varphi}=\cot{\theta}$, where a prime denotes differentiation
with respect to $R$. Then, it arises that all
$\hat{K}_{ij}=0$ for $i\neq j$, while
$\hat{K}_{\varphi\varphi}=\sin^{2}\!\theta\hat{K}_{\theta\theta}$,
\begin{equation}
\hat{K}_{\theta\theta}=-\hat{n}_{R}\hat{\Gamma}^{R}_{\,\,\,\,\theta\theta}=R_{S}F\hat{n}_{R}
\label{iwre}
\end{equation}
\begin{equation}
\hat{K}_{tt}=-\hat{n}_{T}\frac{d^{2}T_{S}}{dt^{2}}-\hat{n}_{R}\frac{d^{2}R_{S}}{dt^{2}}
-\hat{n}_{R}\hat{\Gamma}^{R}_{\,\,\,\,TT}\Big(\frac{dT_{S}}{dt}\Big)^{2}
-\hat{n}_{R}\hat{\Gamma}^{R}_{\,\,\,\,RR}\Big(\frac{dR_{S}}{dt}\Big)^{2}
-2\hat{n}_{T}\hat{\Gamma}^{T}_{\,\,\,\,TR}\frac{dT_{S}}{dt}\frac{dR_{S}}{dt}\,.
\label{wike}
\end{equation}
Finally, the condition $\bar{K}_{\theta\theta}+\hat{K}_{\theta\theta}=0$ (or equivalently for
$\varphi$) gives the consistency equation
\begin{equation}
\frac{dT_{S}}{dt}=\frac{\epsilon\sqrt{1\!-\!\kappa r_{\Sigma}^{2}}\,\,ar_{\Sigma}}{R_{S}F\sqrt{J}}\,,
\label{jerf}
\end{equation}
which, with the use of (\ref{eq:First FF2}), takes the form
\begin{equation}
\frac{dT_{S}}{dt}=\frac{\epsilon\sqrt{1\!-\!\kappa r_{\Sigma}^{2}}}{F\sqrt{J}}\,.
\label{jwei}
\end{equation}
It then follows from (\ref{eq:First FF1}) that
\begin{equation}
\Big(\frac{dR_{S}}{dt}\Big)^{2}=1\!-\!\kappa r_{\Sigma}^{2}-F(R_{S})\,.
\label{gyyu}
\end{equation}
Therefore, equations (\ref{jwei}), (\ref{gyyu}) determine the position of $\Sigma$ in the space
$(T,R)$. From (\ref{jwei}) it is obvious that indeed it is $\hat{n}_{R}<0$.

The final task is the examination of the matching condition $\bar{K}_{tt}+\hat{K}_{tt}=0$, i.e.
$\hat{K}_{tt}=0$. This equation contains the second time derivatives of $T_{S},R_{S}$ that we need to
calculate. From equations (\ref{jwei}), (\ref{gyyu}) it arises
\begin{eqnarray}
&&\frac{d^{2}T_{S}}{dt^{2}}=-\epsilon\sqrt{1\!-\!\kappa r_{\Sigma}^{2}}\,\frac{(F\sqrt{J})'}{F^{2}J}
\,\frac{dR_{S}}{dt}\\
\label{kejr}
&&\frac{d^{2}R_{S}}{dt^{2}}=-\frac{F'}{2}\,.
\label{kevf}
\end{eqnarray}
Using all the previous expressions in (\ref{wike}), it turns out that $J'=0$, which means
$J'(R_{S})=0$. This relation, due to (\ref{eq:First FF2}), implies in general an algebraic equation
for $a(t)$, which will be inconsistent with equation (\ref{gyyu}). There are however
various functions $J(R)$ which satisfy this equation. For example, a consistent choice is that $J(R)$
is constant throughout (as happens in Schwarzschild metric), and in this case without loss of
generality we can rescale $T$ so that this constant is one. Another consistent case would be $J(R)$
to be a power series of the form $J(R)=J(R_{S})+c_{1}(R\!-\!R_{S})^{2}+...$ . The successful matching
has proved that the choice of the matching surface $\Sigma$ was the appropriate one.

\section{General Relativity black holes and the ensuing FRW cosmology}

If the black hole is described by the classical Schwarzschild solution, i.e.
\begin{equation}
F(R)=1-\frac{2G_{\!N} M}{R}\,\,\,\,\,,\,\,\,\,\,J(R)=1\,,
\label{karv}
\end{equation}
equation (\ref{gyyu}) provides through (\ref{eq:First FF2}) the cosmic evolution of the scale
factor $a$. Namely, we take
\begin{equation}
H^2=\frac{\dot{a}^2}{a^2}=\frac{2G_{\!N} M}{r_{\Sigma}^{3} a^3}-\frac{\kappa}{a^2}\,,
\label{fr1c}
\end{equation}
where a dot denotes differentiation with respect to cosmic time $t$.
This equation is qualitatively similar to the standard FRW evolution with dust (zero pressure)
as its cosmic fluid and a possible curvature term.
Of course, in order for this solution to be physically realistic and represent a spatially
homogeneous universe, not just a single sphere of comoving radius $r_{\Sigma}$ should be present,
but a number of such spheres are uniformly distributed throughout the space. Otherwise, there
would exist a preferred position in the universe. Each such sphere can be physically realized by an
astrophysical object, such as a galaxy (with its extended spherical halo) or a cluster of galaxies,
which we assume that it has a typical mean mass $M$. It will be seen that the Schuching
radius lies outside the real border of the astrophysical object, therefore, $F(R_{S})$ in
(\ref{gyyu}) is provided by the value of the expression (\ref{karv}) (otherwise we would meet the
inconvenient situation to consider an interior Oppenheimer-Volkoff type of solution or some other
more realistic matter profile). This means that for the value $F(R_{S})$, which is our only interest
in order to make the matching and derive the cosmological metric, it is like if all the mass $M$ is
gathered at the center of the spherical symmetry. The same is true for other spherically symmetric
metrics, modifications of Schwarzschild solution, to be discussed later. Furthermore, equation
(\ref{kevf}) gives
\begin{equation}
\frac{\ddot{a}}{a}=-\frac{G_{\!N}M}{r_{\Sigma}^{3}a^{3}}\,,
\label{fr2c}
\end{equation}
which indicates a decelerated expansion.

In order for equations (\ref{fr1c}), (\ref{fr2c}) to describe precisely a standard matter
dominated universe, the matter dilution term in (\ref{fr1c}) should be $\frac{8\pi G_{\!N}}{3}\rho$,
where $\rho$ is the cosmic matter energy density, and the term in (\ref{fr2c}) should be
$-\frac{4\pi G_{\!N}}{3}\rho$. Therefore, we make the standard assumption of Swiss cheese models
that the matching radius $r_{\Sigma}$ is such that when its interior region is filled with energy
density equal to the cosmic matter density $\rho$, the interior energy equals $M$.
Namely, we set
\begin{equation}
\rho=\frac{M}{\frac{4\pi}{3}R_{S}^{3}}=\frac{3M}{4\pi r_{\Sigma}^{3}a^{3}}\,.
\label{lper}
\end{equation}
This condition can also equivalently be interpreted that the mass $M$ of the object is uniformly
stretched up to the radius $r_{\Sigma}$. Since in any case the mass $M$ can be considered that is
located at the center of spherical symmetry, the above definition of $r_{\Sigma}$ offers a simple way
to determine the spheres where the matching with the cosmological metric occurs. Although this
definition is certainly ad-hoc and uses the mass $M$ of the object and the cosmic density $\rho$, it
has the merit that it avoids to use other details of the structure, such as the size of the object
and the distance between similar structures. However, still
in the Swiss cheese model the cosmic evolution remains exactly the same as in the cosmological picture.

It is now clear that equations (\ref{fr1c}), (\ref{fr2c}) become
\begin{eqnarray}
H^{2}&=&\frac{8\pi G_{\!N}}{3}\rho-\frac{\kappa}{a^{2}}\label{wird}\\
\frac{\ddot{a}}{a}&=&-\frac{4\pi G_{\!N}}{3}\rho\,.\label{erdd}
\end{eqnarray}
If we define the matter density parameter in the conventional way
\begin{equation}
\Omega_{m}=\frac{8\pi G_{\!N}\rho}{3H^{2}}\,,
\label{jdrv}
\end{equation}
it is found
\begin{equation}
r_{\Sigma}=\Big(\frac{2G_{\!N}M}{\Omega_{m0}a_{0}^{3}H_{0}^{2}}\Big)^{\!\frac{1}{3}}\,,
\label{wkef}
\end{equation}
where a subscript 0 denotes the today value.

Let us suppose we want to model a universe consisting of two types of dust with different densities,
$\rho_1$ and $\rho_2$ (e.g. dark matter and stars or black holes).
In order to avoid unnecessary technical complexity arising from
inhomogeneous placement of dusts, it would be a fair approximation to describe the cosmic evolution
assuming that the universe is filled with a homogeneous distribution of spherical configurations that
consist of two spherical objects that have different masses $M_{1}$ and $M_{2}$ within the Schucking
radius $R_{S}$. The quantities $\rho_{1},M_{1}$ satisfy equation (\ref{lper}) and similarly
for the other ingredient. Since the total cosmic energy density $\rho$ is the sum of the two energy
densities $\rho_{1},\rho_{2}$, it is implied that the matching is performed as before with the
difference that now the mass $M$ of the central object is the sum of the two masses, i.e.
$M=M_{1}+M_{2}$. In this case, the same equations (\ref{fr1c}), (\ref{fr2c}) apply, with $M$ this
total mass.

Let us finish with a few numerics. The present value of the Hubble parameter will be taken as
$H_{0}=0.72\times 10^{-10}\text{yr}^{-1}$. In the present work, we are not going to perform fittings
to real data, where $H_{0}$ could also be considered as a fitted parameter. We will use as a
typical mass for a galaxy $M=10^{11}M_{\odot}$ and for a cluster of galaxies $M=10^{15}M_{\odot}$,
(with $M_{\odot}$ the solar mass) in order to give some estimates.
If we ignore the term of spatial curvature in (\ref{wird}), then $\Omega_{m}=1$,
and equation (\ref{wkef}) gives for the galaxy $r_{\Sigma}=0.56\text{Mpc}$ and for the cluster
$r_{\Sigma}=12\text{Mpc}$ (for a realistic $\Omega_{m0}$ these distances become larger).
Since the typical radius of a spiral galaxy (including its dark matter
halo) is $R_{b}\approx 0.15\text{Mpc}$, it is obvious that $r_{\Sigma}$ is a few times larger
than the galactic radius. Moreover, the mean distance between galaxies is a few $\text{Mpc}$,
thus, the Schucking radii of two neighboring galaxies do not overlap.
As for clusters, they have radii from $R_{b}\approx 0.5\text{Mpc}$
to $R_{b}\approx 5\text{Mpc}$, and therefore, $r_{\Sigma}$ is again outside the cluster.
If a mean distance between the borders of two adjacent clusters is something like
$20\text{Mpc}$, the two Schucking radii still do not intersect.

\section{A perspective for the coincidence problem}
\label{eudk}

In the $\Lambda$CDM model, $\Lambda$ has been found observationally to be of the order $H_{0}^{2}$
(more precisely, $\Lambda=3\Omega_{\Lambda 0}H_{0}^{2}$, $\Omega_{\Lambda 0}\approx 0.7$).
This means that the energy scale defined by $\sqrt{\Lambda}$ is extremely small compared to
the Planck mass scale $M_{\rm{Pl}}$ (which is a typical scale of gravity), and also its energy
density $\rho_{\Lambda}=\frac{\Lambda}{8\pi G_{\!N}}\approx 2.8\times 10^{-11}\rm{eV}^{4}
\sim (10^{-3}\rm{eV})^{4}$ is many orders of magnitude smaller than the theoretical vacuum energy
value $\rho_{\rm{vac}}$ estimated by quantum corrections of Quantum Field Theory with any sensible
cut-off. This discrepancy is called cosmological constant problem, which is the most severe
hierarchy problem in modern physics. It is also called fine-tuning problem since adding a bare
cosmological constant of opposite sign in the action to cancel $\rho_{\rm{vac}}$, this should be
tuned to extreme accuracy in order to give the effective value $10^{-3}\rm{eV}$ above (if
supersymmetry is restored in the high energy, extreme and unnatural fine-tuning is still needed).
Even if the present dark energy in the universe has nothing to do with a cosmological constant,
the question of understanding why the estimated quantum vacuum energy cancels out and does not
contribute, still remains and may need quantum gravity or other physics to be discovered.

Beyond the previous problem, why $\Lambda$ is so extraordinarily small, there is an extra
question named coincidence problem, related to the specific value of $\Lambda\sim H_{0}^{2}$.
Since the energy density $\rho$ falls like $\rho\sim a^{-3}$ starting from a huge (if not infinite)
value, why does it happen today to be $8\pi G_{\!N}\rho_{0}\sim \Lambda$ (actually
$8\pi G_{\!N}\rho_{0}\approx 0.4\Lambda$), and not a very big or a very small proportionality
factor to be present? Why are dark matter and dark energy of the same order today,
$\rho_{0}\sim\rho_{\Lambda}$? Moreover, since $\sqrt{\Lambda}\sim H_{0}$, the time scale
$t_{\Lambda}\sim 1/\sqrt{\Lambda}$ is of the same order as the age of the universe $H_{0}^{-1}$,
something that did not need necessarily to be the case. There are three unrelated quantities
$\rho_{0},\Lambda,G_{\!N}$ and there is no obvious reason why they should be related like that.
To be more precise, the same relation $8\pi G_{\!N}\rho\sim\Lambda$ holds recently, for
$0\leqslant z\lesssim \mathcal{O}(1)$, which means for a few billion years (taking into account
the time, it may be thought that the problem is not so sharp). On the contrary, such a relation
between $G_{\!N}\rho$ and $\Lambda$ could have happened in the very past, at even larger redshifts
(which is probably precluded by anthropic arguments), and this implies that today we would have a
universe full of cosmological constant and negligible matter contribution. Or, finally, such a
relation could occur in the very future and today we would observe matter domination with negligible
$\Lambda$. It is the same to say that although the Hubble parameter started in the past from huge
values (if not infinite), recently it is $H^{2}\sim \Lambda\sim 8\pi G_{\!N}\rho$, and not
$H^{2}\approx\Lambda/3$ or $H^{2}\approx 8\pi G_{\!N}\rho/3$.
To realize better the clear sensitivity of the recent coincidence on the value of $\Lambda$, let
us assume that the cosmological constant was just one hundred times larger than the observed one.
Then its coincidence with the matter would have occurred at a redshift almost 5 and today the dark
matter would be less than just one percent of the dark energy. Or at the other end, if $\Lambda$
was one hundred times smaller than the observed value, its coincidence with the matter would occur
at a redshift almost $-0.7$ and today the dark energy would be almost two percent of the dark
matter. Therefore, the coincidence problem appears because $\Lambda$ takes a value inside a very
narrow range of the $\rho$ values. In terms of the flatness parameters the coincidence problem is
stated by a relation of the form $\Omega_{m0}\sim\Omega_{\Lambda 0}$, and not
$\Omega_{m0}\ll\Omega_{\Lambda 0}$ or $\Omega_{m0}\gg\Omega_{\Lambda 0}$. Ignoring $\kappa$, it
holds $\Omega_{m}\approx 1$, $\Omega_{\Lambda}\approx 0$ for a broad range of redshifts in the past
until recently, where it is $\Omega_{m}\sim\Omega_{\Lambda}$. In the far future it will be
$\Omega_{m}\approx 0$, $\Omega_{\Lambda}\approx 1$. Acceleration exists as long as
$2\Omega_{\Lambda}>\Omega_{m}$.

In general dark energy models the coincidence problem is formulated through the observational
today acceleration along with the relation $\Omega_{m0}\sim\Omega_{DE,0}$, while
in the past it is strongly believed, due to structure formation reasons, that it was
$\Omega_{m}\approx 1$, $\Omega_{DE}\approx 0$. The dark energy density $\rho_{DE}$ defined
by $8\pi G_{\!N}\rho_{DE}=3\Omega_{DE}H^{2}$ obeys $\rho_{0}\sim\rho_{DE,0}$.
Depending on the particular dark energy model, the quantity $\rho_{DE,0}$ contains integration
constants reflecting initial conditions of possible fields involved (e.g. $\phi_{0},\dot{\phi}_{0}$
for a scalar field, or $\rho_{0}$ itself in a geometrical modification of gravity), dimensionfull
or dimensionless couplings/parameters of the theory, and probably other quantities (e.g. of
astrophysical nature). The previous coincidence relation between the energy densities provides an
equation between all these quantities which have to be appropriately adjusted.
Usually, in dark energy models the scale defined by $\Lambda$ is exchanged by another scale
of the same order describing some new physics. It looks like the coincidence problem is a question
of naturalness between integration constants and other parameters, and analyzing naturalness is
not an issue easily quantified.

A proposal as a solution of the coincidence problem will consist of one more dark energy model,
which may be more or less natural, may introduce new physics or not, may introduce new scales or
not, but its verification will come not out of concept but out of experimental evidence of the
particular model. And finally, when the origin of dark energy has been apprehended, it will become
obvious what the independent scales and initial conditions created by Nature are. Even a model which
contains new scales, that at present are unrelated from the rest of physics, may be very close
to reality. This is why analyzing a dark energy model, the values of $\Omega_{m0},\Omega_{DE,0}$,
as well as the rest of the parameters/initial conditions, are in general extracted after fittings to
the observational data and there is no special concern about a deeper understanding which would mean
to express these values in terms of other more fundamental ones. Of course, if such an explanation
for the recent emergence of dark energy through the coincidence relation $\rho\sim\rho_{DE}$ can be
provided in terms of quantities that already play a role in Nature or/and other quantities
theoretically predicted in the context of a theory, this would possess extra naturalness and might
render the particular dark energy model promising (this is the case for one of the models to be
presented in the present work). Another idea that alleviates the coincidence problem comes
through the realization of the current state of the universe close to a global fixed point
(saddle or preferably attractor) of the cosmic evolution, since then, the coincidence problem
becomes an issue of only the parameters of the model and not also of the initial conditions
\cite{Wetterich:1994bg}. In order for this to be possible with the present acceleration and a
scaling behaviour between $\Omega_{m},\Omega_{DE}$, the violation of the standard energy-momentum
conservation of the matter is necessary \cite{Kofinas:2005hc}.

The proposal introduced in the present paper is that the dark energy observed
recently in the universe may be the result of local gravity effects occurring in the interior of
astrophysical objects, such as massive structures (galaxies, clusters) or even black holes,
and these effects will directly determine the cosmic evolution.
These local effects can arise from an arbitrary gravitational theory (alternative/modified gravity,
extra-dimensional gravity, quantum gravity, e.tc.). The main point is that the specific
gravitational theory is not applied directly to cosmology in the conventional way, with the matter
described as a usual perfect fluid, in order to obtain time dependent differential equations for
the geometry (e.g. scale factor) and the other ingredients; the reason for this is that it is not
clear how the cosmic effective energy-momentum tensor can be quantified taking into account the
extra contributions of local origin. So, even if the theory is managed to be applied directly to
cosmology, the result will in general be different that the one arising from the process described
here because, depending on the scales of the theory, the dark energy can be suppressed in one of
the two derived cosmologies and be considerable in the other. Thus, if a gravity
effect becomes substantial only at an intermediate infrared scale (astrophysical one), it cannot
be revealed at the far infrared cosmological scale.
In addition, integration constants emanating from possible integrations (due to extra
fields or geometrical effects) in the local metric will be of quite different nature that
the cosmological ones; these constants might be specified or at least estimated from quantities
characterizing the astrophysical object itself or from regularity arguments at the center of the
structure, contrary to the specification of the integration constants of a cosmological quantity
which needs a quite different treatment.
Applying the gravitational theory first inside the structure means to find the gravitational and the
other possible fields in the interior of the object. Because the astrophysical structures
are not point-like but they are extended (the galaxies have a luminous profile which is
surrounded by the dark matter halo and the clusters contain a distribution of galaxies and dark matter)
or may be described by a collapsing phase, this task can be complicated.
However, depending on the ansatz for the local and the cosmological metrics and how these are
interrelated, it may be enough to find the static spherically symmetric solution of the theory
(as happens in the Swiss cheese model described previously), or other more complete solutions may
be needed to describe an inhomogeneous universe with more realistic structures. In the present work
the Swiss cheese model will be adopted as the simplest but not necessarily the most realistic
construction, and therefore, the ensuing cosmologies will arise through the matching of the
interior metric with the exterior FRW metric on the Schucking surface. It is obvious that such a
local gravity effect should not contradict with observations at the relevant astrophysical scale.

In the context where the dark energy owes its origin to the presence of structure, either
due to the reasons elaborated here or due to averaging process (as e.g. in \cite{inho},
\cite{inhoswiss}, \cite{structure}), since the various structures are formed during the cosmic
evolution recently at small redshifts, dark energy also appears recently not as a coincidence but
as an emerging effect of the structure. With this in mind, that the coincidence problem can be
a guiding line for studying cosmology, it becomes tempting to see what are the
new scales and integration constants introduced by some gravitational theory in the
above context, so that the corresponding cosmology confronts (if possible)
with the acceleration and other data, and especially with the relation $\rho\sim\rho_{DE}$.
In the following sections, we will implement the previous ideas to derive a few dark energy models,
where in the last subsection our most promising model will appear.

\section{Black holes with cosmological constant}

When a constant cosmological term is added to the Schwarzschild metric, the well-known
Schwarzschild-de Sitter metric arises
\begin{equation}
ds^2=-\Big(1-\frac{2G_{\!N} M}{R}-\frac{1}{3}\Lambda R^2\Big)dT^2
+\frac{dR^2}{1-\frac{2G_{\!N} M}{R}-\frac{1}{3}\Lambda R^2}
+R^2\left(d\theta^2+\sin^2\theta d\varphi^2\right)\, .
\label{ASFM}
\end{equation}
It is apparent that it is $J(R)=1$. Equation (\ref{gyyu}) provides through (\ref{eq:First FF2}) the
cosmic evolution of the scale factor
\begin{equation}
\frac{\dot{a}^2}{a^2}+\frac{\kappa}{a^2}=\frac{2G_{\!N} M}{r_{\Sigma}^{3} a^3}
+\frac{\Lambda}{3}\,.
\label{frke}
\end{equation}
Furthermore, equation (\ref{kevf}) gives
\begin{equation}
\frac{\ddot{a}}{a}=-\frac{G_{\!N}M}{r_{\Sigma}^{3}a^{3}}+\frac{\Lambda}{3}\,,
\label{frjw}
\end{equation}
which indicates the well-known late-times accelerated expansion when the two terms on the r.h.s.
of (\ref{frke}) become comparable.
Using the Swiss cheese condition (\ref{lper}), equations (\ref{frke}), (\ref{frjw}) are also written as
\begin{eqnarray}
H^{2}+\frac{\kappa}{a^{2}}&=&\frac{8\pi G_{\!N}}{3}\rho+\frac{\Lambda}{3}\label{wiiw}\\
\frac{\ddot{a}}{a}&=&-\frac{4\pi G_{\!N}}{3}\rho+\frac{\Lambda}{3}\,.\label{erjq}
\end{eqnarray}
Since $\Omega_{\rm{m0}}$ is close to $0.30$ according to the most recent constraints
\cite{Ade:2015xua}, equation (\ref{wkef}) gives for a galaxy with $M=10^{11}M_{\odot}$ that
$r_{\Sigma}=0.83\text{Mpc}$ and for a cluster with $M=10^{15}M_{\odot}$ that $r_{\Sigma}=18\text{Mpc}$.

Equations (\ref{wiiw}), (\ref{erjq}) are the standard cosmological equations of $\rm{\Lambda}CDM$ model.
In this model, the cosmological constant $\Lambda$ is considered as a universal constant related
to the vacuum energy. However, in the context of the present work, $\Lambda$ arises from the
interior black hole solution (\ref{ASFM}) and has a quite different origin and meaning.
This $\Lambda$ is of astrophysical origin and is the total cosmological constant coming from the
sum of all antigravity sources inside the Schucking radius of the galaxy or cluster.
It is expected that, through some concrete quantum gravity theory, matter is related to the
generation of an explicit or effective cosmological constant.
For example, in the centres of astrophysical black holes, the avoidance of singularity could
be achieved due to the presence of a repulsive pressure of quantum origin balancing the
attraction of gravity. Another important difference with this $\Lambda$ is that since, according to our
proposal, the antigravity sources are connected to either massive structures (galaxies, clusters)
or astrophysical black holes, therefore, before the appearance of all these objects the total
$\Lambda$ is zero. As a result, this $\Lambda$ becomes a function of cosmic time, suppressed
at larger redshifts where the antigravity effect is weaker. A constant $\Lambda$ is expected to
be only an approximation at late times.

The metric (\ref{ASFM}) contains the Newtonian term $\frac{2G_{\!N}M}{R}$ and the cosmological
constant term $\frac{1}{3}\Lambda R^{2}$. For distances $R$ close to the border with coordinate
distance $R_{b}$, the matter can be considered as being gathered at the origin, as mentioned above.
We will give an estimate of the corresponding values of the potential and the force due to the
cosmological constant. In the weak field limit the force corresponding to
the Newtonian term is $-\frac{G_{\!N}M}{R^{2}}$, while the cosmological constant force is
$\frac{1}{3}\Lambda R$ and is repulsive.
The ratio of the magnitudes of the cosmological constant force to the Newtonian force is
$\frac{2\Omega_{\Lambda 0}}{\Omega_{m0}}(\frac{R}{r_{\Sigma}})^{3}$, therefore the significance of
the cosmological constant increases with distance.
At the border of a typical galaxy with mass $10^{11}M_{\odot}$ and radius $0.15\text{Mpc}$ the
Newtonian term has a value approximately $6\times 10^{-8}$, while the
cosmological constant term is almost $9\times 10^{-10}$, which
is therefore two orders of magnitude smaller than the former term. Of course, the two
potentials at the Schucking radius are of the same order since dark matter and dark energy
today are of the same order.
Similarly, the repulsive force at the border is also almost two orders of magnitude
smaller that the Newtonian force. Therefore, the $\Lambda$ term is ignorable
at the galaxy level and the galaxy dynamics is not disturbed by this antigravity effect.

For the clusters there is a larger variability of the range of their radii
and the corresponding masses. At the Schucking radius still the two potentials are of the same order.
For a mass $10^{15}M_{\odot}$ and radius $0.5\text{Mpc}$ the Newtonian term is $2\times 10^{-4}$ at
the border, while the cosmological constant term is $10^{-8}$. As a result, the
repulsive force is four orders of magnitude smaller that the attractive force. For a radius of
$5\text{Mpc}$, the $\Lambda$ force is still smaller than the Newtonian force, but just one
order of magnitude. For a mass $10^{14}M_{\odot}$ and radius $5\text{Mpc}$, the two forces
become equal in magnitude. If the cosmological constant is indeed generated at the cluster scales,
then this constant should be present in all clusters. Therefore, more investigation is needed
at particular clusters that could show off some abnormal dynamics and if this can be explained
through a constant $\Lambda$.

The previous discussion shows that $\Lambda$ could be generated inside astrophysical objects,
either without affecting their dynamics or signaling some observable deviations in this dynamics,
and at the same time to create the standard $\Lambda$CDM cosmology. Of course, this constant
$\Lambda$ does not offer any alleviation to the coincidence puzzle.

\section{Black holes with varying cosmological constant of quantum origin}

The Schwarzschild-de Sitter metric can be progressed to a quantum improved Schwarzschild-de Sitter
metric describing the astrophysical object. This metric has the form
\begin{equation}
ds^2=-\Big(1-\frac{2G_k M}{R}-\frac{1}{3}\Lambda_k R^2\Big)dT^2
+\frac{dR^2}{1-\frac{2G_k M}{R}-\frac{1}{3}\Lambda_k R^2}
+R^2\left(d\theta^2+\sin^2\theta d\varphi^2\right)\,,
\label{ASBH}
\end{equation}
where the quantities $G_{k},\Lambda_{k}$ are functions of a characteristic energy scale $k$
and $F(R)=1-\frac{2G_{k}M}{R}-\frac{1}{3}\Lambda_{k}R^{2}$.
The functional behaviour of $G_{k},\Lambda_{k}$ is determined by the underlying quantum theory of
gravity. This energy scale $k$ is related to the distance from the center of the object and the
exact dependence arises from the particular quantum corrections. Therefore, $G_{k},\Lambda_{k}$ are
also related to the distance. In the Swiss cheese analysis, however, only the front value $R_{S}$
of the distance at the matching surface influences the cosmic evolution, thus only the corresponding
energy value $k_{S}$ will be relevant. As mentioned before, in a real galaxy or cluster the total
mass consists of either stars, dark matter, or black holes (classical or quantum modified) that
we collectively denote $M$. Although the various objects are distributed throughout, in our approach
it is sufficient to consider that these materials are gathered together at the center of spherical
symmetry.

As it is known, the Israel matching conditions are only applicable in Einstein gravity with some
regular energy-momentum tensor. In an alternative/modified gravity, either containing extra fields
or not, the corresponding matching conditions are in general modified. One might wonder if the
Israel conditions are still applicable in our case, with a metric of the form (\ref{ASBH}). The
answer is positive and we will explain this in the following. A quantum originated spherically
symmetric metric, as the one described above, does not in general arise as a solution of some
classical field equations for the metric, but is obtained by considering some quantum corrections
beyond the classical Einstein term. For example, in AS gravity, the solution of the Renormalization
Group (RG) flow equations gives $G_{k},\Lambda_{k}$. Therefore, a metric, such as (\ref{ASBH}),
is quite reasonable and necessary for our Swiss cheese approach to be interpreted as a solution
of a coupled gravity-matter system satisfying Einstein equations
$G_{\mu\nu}=8\pi G_{\!N}T_{\mu\nu}^{(\text{tot})}$.
This $T_{\mu\nu}^{(\text{tot})}=T_{\mu\nu}+T_{\mu\nu}^{(\text{eff})}$ contains, apart from a possible
real matter energy-momentum tensor $T_{\mu\nu}$ (which for us is zero since the mass is just an
integration constant), an effective energy-momentum tensor $T_{\mu\nu}^{(\text{eff})}$ of
gravitational origin which takes into account the quantum corrections (for an interpretation
of such a $T_{\mu\nu}^{(\text{eff})}$ in terms of fluid variables see \cite{Torres:2014gta}).
Since (\ref{ASBH}) expresses the quantum corrections of the classical Schwarzschild metric, the
tensor $T_{\mu\nu}^{(\text{eff})}$ appears as the correction beyond the Einstein equations of
motion and not beyond some other modified classical equations of motion. To find this
$T_{\mu\nu}^{(\text{eff})}$, we need to compute the Einstein tensor $G_{\mu\nu}$ of the metric
(\ref{ASBH}). This could lead to a non-trivial situation, where the Israel matching conditions are
satisfied or not, depending on the form of this effective energy-momentum tensor. However, for the
whole analysis of the present paper, just the Israel conditions arise. Indeed, the Einstein tensor
$G^{\mu}_{\,\,\,\nu}$ which is constructed from the metric (\ref{ASBH}) has the following
non-vanishing components
\begin{equation}
G^{T}_{\,\,T}=G^{R}_{\,\,R}=\frac{1}{R^{2}}(RF'+F-1)\,\,\,\,\,\,\,\,,\,\,\,\,\,\,\,\,
G^{\theta}_{\,\,\theta}=G^{\varphi}_{\,\,\varphi}=\frac{1}{2R}(RF''+2F')\,.
\label{kilo}
\end{equation}
Since the Schwarzschild metric with $F_{\text{Sch}}=1-\frac{2G_{\!N}M}{R}$ satisfies the vacuum
Einstein equations, we get
\begin{equation}
G^{T}_{\,\,T}=G^{R}_{\,\,R}=\frac{1}{R^{2}}(R\mathcal{Q}'+\mathcal{Q})\,\,\,\,\,\,\,\,,\,\,\,\,\,\,\,\,
G^{\theta}_{\,\,\theta}=G^{\varphi}_{\,\,\varphi}=\frac{1}{2R}(R\mathcal{Q}''+2\mathcal{Q}')\,,
\label{repo}
\end{equation}
where the quantity
\begin{equation}
\mathcal{Q}=\frac{2(G_{\!N}\!-\!G_{k})M}{R}-\frac{1}{3}\Lambda_{k}R^{2}
\label{muti}
\end{equation}
is defined by $\mathcal{Q}\equiv F-\big(1-\frac{2G_{\!N}M}{R}\big)$, i.e. it is the deviation of
the metric component $F$ from $1-\frac{2G_{\!N}M}{R}$. The quantity $\mathcal{Q}$ will be seen
that is well-defined, depending on the assumptions of the quantum theory. Therefore, in the interior
regime with the metric (\ref{ASBH}), the Einstein equations
$G^{\mu}_{\,\,\,\nu}=8\pi G_{\!N}T^{\mu(\text{eff})}_{\,\,\,\nu}$ acquire a well-defined
$T^{\mu(\text{eff})}_{\,\,\,\nu}$, which is given by the right hand sides of (\ref{repo})
and parametrized by the quantity $\mathcal{Q}$,
\begin{equation}
T^{T(\text{eff})}_{\,\,\,T}=T^{R(\text{eff})}_{\,\,\,R}
=\frac{1}{8\pi G_{\!N}R^{2}}(R\mathcal{Q}'+\mathcal{Q})\,\,\,\,\,\,\,\,,\,\,\,\,\,\,\,\,
T^{\theta(\text{eff})}_{\,\,\,\theta}=T^{\varphi(\text{eff})}_{\,\,\,\varphi}=
\frac{1}{16\pi G_{\!N}R}(R\mathcal{Q}''+2\mathcal{Q}')\,.
\label{koin}
\end{equation}
Moreover, equating the right hand sides of the expressions (\ref{kilo}), (\ref{repo}), we find
the ordinary differential equations of $F$,
\begin{equation}
RF'+F-1=R\mathcal{Q}'+\mathcal{Q}\,\,\,\,\,\,\,\,,\,\,\,\,\,\,\,\,
RF''+2F'=R\mathcal{Q}''+2\mathcal{Q}'\,.
\label{muja}
\end{equation}
Differentiating the first equation of (\ref{muja}), we get the second equation. We are not
particularly interested in the precise function $F(R)$ since for cosmology the most important is
only the value of $F$ at the Schucking radius. We discern now the following cases.
\newline
---\,\,If $G_{k},\Lambda_{k}$ are functions of $R$ in an explicit algebraic form (as happens in the
subsection VII-B-1), then the metric component $F$, the quantity $\mathcal{Q}$ of (\ref{muti}) and
$T^{\mu(\text{eff})}_{\,\,\,\nu}$ of (\ref{koin}) become also such functions of $R$. The second
(as well as the first) derivatives of the metric component $F$ in the Einstein equations
(\ref{muja}) are contained only on their left hand sides, which come from $G^{\mu}_{\,\,\,\nu}$,
while the $\mathcal{Q}$ terms on the right hand sides, which come from
$T^{\mu(\text{eff})}_{\,\,\,\nu}$, act as a sort of given potentials that modify $F(R)$ from the
Schwarzschild metric. Although, this picture is rather trivial due to that the evaluation is
performed on-shell, on the explicit metric component $F(R)$, however, it is still meaningful since
the function $F(R)$ still satisfies the differential equations (\ref{muja}). As it is well known,
the matching conditions are extracted from the Einstein equations
$G^{\mu}_{\,\,\,\nu}=8\pi G_{\!N}T^{\mu(\text{eff})}_{\,\,\,\nu}$
focusing only to the second derivatives of the metric components. Since these derivatives are only
inside the Einstein tensor $G^{\mu}_{\,\,\,\nu}$ and not inside
$T^{\mu(\text{eff})}_{\,\,\,\nu}$, the Israel matching conditions naturally arise in this case.
\newline
---\,\,In the subsections VII-A, VII-B-2, we will discuss another situation, which is non-trivial
in the sense that $F$ is not a known function of $R$. At the same time, this situation
is actually more promising in its cosmological results and also more favorable theoretically.
Here, $\Lambda_{k}$ is not an explicit function of $R$, but it contains an integral of $F(R)$.
Since $F(R)$ is not known, but is the metric component to be found by solving (\ref{muja}), this
integral cannot be performed. This means that the quantity $\mathcal{Q}$ contains an independent
geometrical field $D(R)$ with its own equation of motion. The important point is that this equation
of motion is only of first order, i.e. it contains $D'$ and not $D''$, and therefore, there is no
discontinuity of $D'$ at the matching surface. As a result,
$T^{\mu(\text{eff})}_{\,\,\,\nu}$ in (\ref{koin}) becomes a function of $R,D,F,F'$, while no
$F''$ is present in $T^{\mu(\text{eff})}_{\,\,\,\nu}$. Therefore, again $F''$ is only
contained in the tensor $G^{\mu}_{\,\,\,\nu}$ of the Einstein equations
$G^{\mu}_{\,\,\,\nu}=8\pi G_{\!N}T^{\mu(\text{eff})}_{\,\,\,\nu}$ and the Israel conditions remain
intact. We will provide more precise explanations about this issue at the appropriate point later.
The metric component $F(R)$ can be found by solving for $F,D$ the coupled system of the
first equation in (\ref{muja}), i.e. equation $RF'+F-1=R\mathcal{Q}'(R,D,F)+\mathcal{Q}(R,D)$,
together with the first order differential equation for $D$.
\newline
---\,\,Finally, if we assume, merely as a mathematical extension of the above, the more complicated
situation where $G_{k},\Lambda_{k}$ contain double integrations of $F,F'$, this implies that the
equation of motion of the $D$ field is of second order. Then, a discontinuity of $D'$ could be
present at the matching surface and the Israel matching conditions might be modified. We only
mention such a possibility, without regarding it in our analysis, to show how the Israel conditions
could be violated in extreme and artificial situations, where no clear physical motivation
justifies such constructions.

To summarize, we have proved that the Israel matching conditions are still applicable for the
metric (\ref{ASBH}) in the context of our analysis. Therefore, in the Swiss cheese approach,
equations (\ref{gyyu}), (\ref{kevf}) can be used to derive the cosmological evolution.

If $G_{k}$ has the constant observed value $G_{\!N}$ and the underlying theory provides only
ultraviolet (UV) corrections to $\Lambda$, it is not possible to get cosmic acceleration since
$\Lambda$ is suppressed at large distances and $\Lambda(R_{S})$ almost vanishes. Therefore,
infrared corrections of $\Lambda$ are necessary.
For example, a well known phenomenological description of a quantum corrected non-singular black hole
is provided by the Hayward metric \cite{Hayward:2005gi} where
\begin{equation}
F(R)=1-\frac{2G_{\!N}MR^{2}}{R^{3}\!+\!2G_{\!N}ML^{2}}=1-\frac{2G_{\!N}M}{R}
-2G_{\!N}M\Big(\frac{1}{R^{3}\!+\!2G_{\!N}ML^{2}}\!-\!\frac{1}{R^{3}}\Big)R^{2}\,.
\label{kede}
\end{equation}
The length scale $L$ controls the ultraviolet correction close to the origin
and smoothes out the singularity. In this metric the effective
cosmological constant becomes a function of $R$, namely $\Lambda(R)=
6G_{\!N}M\Big(\frac{1}{R^{3}+2G_{\!N}ML^{2}}\!-\!\frac{1}{R^{3}}\Big)$.
At distances where the Newtonian potential is very weak and $R\gtrsim L$, it arises that
the potential of this cosmological ``constant'' is negligible compared to the Newtonian
potential and its corresponding weak-field force (which is repulsive) is well suppressed
compared to the Newtonian force. Equation (\ref{kevf}) for the acceleration gives
\begin{equation}
\frac{\ddot{a}}{a}=G_{\!N}M\frac{4G_{\!N}ML^{2}\!-\!r_{\Sigma}^{3}a^{3}}
{(2G_{\!N}ML^{2}\!+\!r_{\Sigma}^{3}a^{3})^{2}}\,.
\label{mher}
\end{equation}
This metric does not provide a recent cosmic acceleration since at late times deceleration
emerges. Similarly, the solution presented in \cite{Dymnikova:2001fb} fails for the same reason.

\section{The case of Asymptotically Safe Gravity}

A concrete realization of the functions $G_{k},\Lambda_{k}$ is provided by the asymptotically
safe scenario of quantum gravity. In Appendix A, we present a few basic elements of the
theory and of its application in cosmology. According to the AS program, both Newton's constant
$G_k$ and cosmological constant $\Lambda_k$ are energy dependent,
$G_k=G(k)=g_{k}k^{-2}\,,\,\Lambda_k=\Lambda(k)=\lambda_{k}k^{2}$, where $k$ is an energy measure
of the system and $g_{k},\lambda_{k}$ are the dimensionless running couplings governed by
some RG flow equations.
The exact RG flow of the couplings from the Planck regime down to the
present epoch is not yet known. So, it is not clear what is the real trajectory in the space
$g_{k},\lambda_{k}$ followed by the universe and how the classical General Relativity regime
with a constant $G_{\!N}$ and negligible $\Lambda$ can be obtained.
However, even if it was possible to predict the flow of the decrease of $\Lambda_{k}$ all the way
down to the current cold cosmic energy scale and a very small value of $\Lambda$ could be realized,
the coincidence problem, having to do with the precise order of magnitude of $\Lambda$,
would still be manifest and profound. Furthermore, in the picture with a constant
$G$ and a $\Lambda_{k}$ monotonically decreasing with time, a recent passage from a
deceleration to an acceleration phase would not be possible. The far infrared limit of
$g_{k},\lambda_{k}$ certainly covers the late-times cosmological scales. However, even if a
late-times behaviour with an increasing $G$ does exists, it will describe the future and not
the present universe which possesses rather an antiscreening instead of a screening behaviour.
Additionally, there is the possibility that the far infrared corrections of the cosmological constant
at cosmic scales are too small to affect the present universe evolution and drive into acceleration.

Anyway, the cosmic scale corrections of $G,\Lambda$ are not of interest in our approach. Our
interest is focused on the intermediate infrared corrections occurring at the astrophysical
structures scales. The hope is that these intermediate scale quantum corrections will be
significant enough to have direct influence on the current cosmology and on the observed dark
energy component, but at the same time they will not conflict in an obvious way with the
local dynamics. Indeed, we will show until the end of the paper that the recent cosmic acceleration
can be the result of such quantum corrections of the cosmological constant at the galactic or cluster
of galaxies scale.

As before, the universe will be described by the Swiss cheese model and the matching will take
place on the surface between a cosmological metric and a quantum modified spherically symmetric metric.
Several interesting approaches of RG improved black hole metrics appear in literature. In our
analysis we will work with the quantum improved Schwarzschild-de Sitter metric (\ref{ASBH}),
as e.g. presented in \cite{Koch:2014cqa}. Of course, the precise form of the metric will arise from
the forms adopted for $G_{k},\Lambda_{k}$, as these are predicted or motivated by AS. A different
treatment would be to substitute the functions $G_{k},\Lambda_{k}$ inside some consistent
quantum corrected gravitational equations of motion or inside some consistent action, and derive
first the quantum corrected spherically symmetric solution. As mentioned, here we will follow for
simplicity the method of obtaining a quantum corrected metric by starting from a classical solution
and promoting $G_{\!N},\Lambda$ to energy dependent quantities according to the AS program.

In order to proceed further, the energy measure $k$ has to be connected with a length scale $L$,
i.e. $k=\xi/L$, where $\xi$ is a dimensionless parameter which is expected to be of order one.
As we approach the center of spherical
symmetry, the mean energy increases and $k$ is a measure of the energy scale that is encoded in
Renormalisation Group approaches to Quantum Gravity. A simple option is to set as $L$ the
coordinate distance $R$ of the spherically symmetric metric. Then, the value of the cosmological
constant $\Lambda_{k}$, which provides a value for the local vacuum energy density, depends
explicitly on the distance from the center, $\Lambda_{k}=\Lambda(R)$. The same is true for $G_{k}=G(R)$,
and so, the metric component $F$ of the metric (\ref{ASBH}) becomes an explicit function of $R$,
$F=F(R)$. Let us remind that the function $F(R)$ is precise when the mass $M$
is located at the origin. In our approach where we want to describe real astrophysical objects
with a mass profile, $F(R)$ in the interior should be treated with caution.

A more natural option is to set as $L$ the proper distance $D>0$
\cite{Bonanno:2001xi}. This case is more involved physically and technically. Following a radial
curve defined by $dT=d\theta=d\varphi=0$ to reach a point with coordinate $R$, we have
\begin{equation}
D(R)=\int_{R_{1}}^{R}\frac{d\mathcal{R}}{\sqrt{F(\mathcal{R})}}\,.
\label{joet}
\end{equation}
This is a formal expression until one realizes what is its meaning and the meaning of $R_{1}$.
Now, the function $F$ is not an explicit function of $R$ since $D$ is also contained in $F$ by
construction, so $F(\mathcal{R})$ in (\ref{joet}) basically means $F(\mathcal{R},D(\mathcal{R}))$.
So, (\ref{joet}) is not a simple integral but it is an integral equation which can be converted to
the more useful differential equation
\begin{equation}
D'(R)=\frac{1}{\sqrt{F(R)}}
\label{hteg}\,.
\end{equation}
This is a well-defined, but complicated, differential equation for $D$ since $F$ contains again
$D(R)$ (or in different words, $F$ is given by (\ref{muja})).
Integration of equation (\ref{hteg}) will provide an integration constant $\sigma$, so the solution
of (\ref{hteg}) is $D(R;\sigma)$. Plugging this $D$ in $F$ will provide $F(R;\sigma)$. For a specific
$\sigma$, a value $R_{1}(\sigma)$ should exist that makes the equation (\ref{joet}) meaningful for
$R>R_{1}$. The positiveness of $D$ may also provide some restrictions on $R$. Even if a minimum horizon
distance $R_{H}$ exists where $F(R_{H};\sigma)=0$, $R_{1}$ does not necessarily coincide with $R_{H}$,
since after $D(R;\sigma)$ has been found, it is possible that the arising function
$1/\sqrt{F(R;\sigma)}$ is non-integrable around $R_{H}$. So, the variable $D(R)$ is a proper distance,
but not in the conventional sense of an integration in a prefixed background. It can rather be
considered as a new dynamical field of geometrical nature with its own equation of motion (\ref{hteg}),
where the spacetime metric is determined through $D(R;\sigma)$. The role of the function $D(R)$ will
be crucial in our analysis. The integration constant $\sigma$ could be determined by some assumption,
for example if $R_{H}$ exists, it could be set $D_{H}=0$ or $D_{H}=R_{H}$. In our case of Swiss cheese
models, $\sigma$ will be determined from the demand of having the correct amount of dark energy today.
However, the interesting thing, especially in relation to the coincidence problem,
is that the corresponding $D(R)$ will have throughout natural values of the order of the length of
the astrophysical object, while at the same time it will provide the correct estimate of dark
energy.

Let us now complete the discussion started in section VI about the validity of the Israel
matching conditions when the choice (\ref{joet}) is made. The cosmological constant becomes
a function of $D$, i.e. it is $\Lambda_{k}=\Lambda_{k}(D)$. If $G_{k}=G_{\!N}$ (as we will assume
in our analysis), the quantity $\mathcal{Q}$ in (\ref{muti}) becomes
$\mathcal{Q}=-\frac{1}{3}\Lambda_{k}R^{2}$, so it is a function of $R,D$, i.e.
$\mathcal{Q}=\mathcal{Q}(R,D)$. Making use of (\ref{hteg}), we can easily compute
$\mathcal{Q}'=-\frac{2}{3}\Lambda_{k}R+\frac{\xi R^{2}}{3D^{2}\sqrt{F}}
\frac{\partial\Lambda_{k}}{\partial k}=\mathcal{Q'}(R,D,F)$,
$\mathcal{Q}''=-\frac{2}{3}\Lambda_{k}+\frac{\xi R}{6D^{3}F^{3/2}}
(8DF\!-\!4R\sqrt{F}\!-\!RDF')\frac{\partial\Lambda_{k}}{\partial k}-
\frac{\xi^{2}R^{2}}{3D^{4}F}\frac{\partial^{2}\Lambda_{k}}{\partial k^{2}}=\mathcal{Q''}(R,D,F,F')$.
It is now clear, as also mentioned above, that the right hand sides of the Einstein equations
(\ref{muja}), i.e. the components of $T^{\mu(\text{eff})}_{\,\,\,\nu}$, are only functions of
$R,D,F,F'$ and not of $F''$. Since $F''$ is only contained in $G^{\mu}_{\,\,\,\nu}$ on the
left hand sides of (\ref{muja}), and the evolution of $D$ is governed by the first order
differential equation (\ref{hteg}), the Israel matching conditions arise.

In our Swiss cheese approach of cosmology, the matching between the interior and the exterior metric
occurs at the Schucking radius which only enters the cosmological evolution. The front
value $k_{S}$ at the Schucking radius is inversely proportional to a characteristic length of the
metric. For the choice $L=R$ we get
\begin{equation}
k_{S}=\frac{\xi}{R_{S}}\,.
\label{meyc}
\end{equation}
For $L=D$ it is
\begin{equation}
k_{S}=\frac{\xi}{D_{S}}\,,
\label{jaet}
\end{equation}
where
\begin{equation}
D_{S}=\int_{R_{1}}^{R_{S}}\frac{dR}{\sqrt{F(R)}}
\label{ekry}
\end{equation}
is the proper distance of the Schucking radius. Therefore, the front values $\Lambda(R_{S}),G(R_{S})$
at the Schucking radius are the ones where the matching occurs and determine the cosmological
evolution from (\ref{gyyu}), (\ref{eq:First FF2}) as
\begin{equation}
\frac{\dot{a}^{2}}{a^{2}}+\frac{\kappa}{a^{2}}=\frac{2G(R_{S})M}{r_{\Sigma}^{3}a^{3}}
+\frac{1}{3}\Lambda(R_{S})\,.
\label{yehg}
\end{equation}

We finish with a comment which doesn't have a special significance.
The dependence of $\Lambda$, $G$ on the distance inside the object does not seem to affect our
cosmology. However, there is an indirect influence which affects the parameters.
Indeed, the total cosmic energy of the cosmological portion that is excised from the Swiss cheese
should be equal to the energy provided by the various masses inside the astrophysical object plus
the vacuum energy due to the cosmological constant. Since the vacuum energy of a cosmological
constant is $\frac{\Lambda}{8\pi G}$, we have approximately the equation
\begin{equation}
\frac{4\pi}{3}R_{S}^{3}\rho_{\text{tot}}=M+\int_{0}^{R_{S}}\!\frac{\Lambda(R)}{8\pi G(R)}
\,4\pi R^{2}dR\,,
\label{kwrt}
\end{equation}
where $\rho_{\text{tot}}=\rho+\rho_{DE}$ is the total cosmic energy density
(dark matter plus dark energy) \cite{Bhattacharya:2013tq}. Equation (\ref{kwrt}) evaluated
at the today values, according to (\ref{eq:First FF2}), sets a restriction between the various
parameters. Unfortunately however, we cannot make use of this equation since we do not know the
precise functions $\Lambda(k),G(k)$ from the UV with $k=\infty$ up to $k_{S}\sim R_{S}^{-1}$.
The difficulty is rather basically due to $\Lambda(k)$ since $G(k)$ rapidly evolves to its constant
value $G_{\!N}$. So, we will have one more parameter left free in our analysis.

\subsection{First RG flow behaviour: close to the Gaussian fixed point}
\label{kiee}

There is a fixed point of the RG flow equations, the Gaussian fixed point (GFP) \cite{Reuter:2001ag},
which is saddle and is located at $g=\lambda=0$. A appropriate class of trajectories in the
Einstein-Hilbert truncation of the RG flow can be linearized about the GFP, where the dimensionless
couplings are pretty small. These trajectories possess interesting qualitative properties such as a
long classical regime (long $\ln{k}$ ``time'' due to the vanishing of beta functions) and a small
positive cosmological constant in  the infrared, features that seem relevant to the description of
gravitational phenomena in the real universe. The analysis is fairly clear and in the vicinity of
the GFP it arises that $\Lambda$ has a running $k^{4}$ and $G$ has an approximately constant value
which is interpreted as $G_{\!N}$. Therefore,
\begin{equation}
G_{k}=G_{\!N}\,\,\,\,\,\,\,\,,\,\,\,\,\,\,\,\,\Lambda_{k}=\alpha+\beta k^{4}\,,
\label{kerf}
\end{equation}
where $\alpha,\beta$ are positive constants. Moreover it is $\beta=\nu G_{\!N}$, where
$\nu=\mathcal{O}(1)$. In terms of the dimensionless couplings it is
$\lambda_{k}=\alpha k^{-2}+\beta k^{2}$, $g_{k}=G_{\!N}k^{2}$.
These equations are valid if $\lambda_{k}\ll 1$, $g_{k}\ll 1$. While the above segment which lies
inside the linear regime of the GFP can be continued with the flow equation into the UV fixed point,
this approximation breaks down in the IR where $\lambda_{k}$ approaches the value $1/2$.
Therefore, as our first choice, we will assume that within a certain range of $k$-values
encountered in an astrophysical object, the RG trajectory is approximated by (\ref{kerf}).

For the choice (\ref{jaet}), equation (\ref{yehg}) provides
the cosmic evolution of the scale factor $a$ as
\begin{equation}
\frac{\dot{a}^{2}}{a^{2}}+\frac{\kappa}{a^{2}}=\frac{2G_{\!N}M}{r_{\Sigma}^{3}a^{3}}
+\frac{\alpha}{3}+\frac{\beta\xi^{4}}{3D_{S}^{4}}\,.
\label{iweg}
\end{equation}
The choice (\ref{meyc}) is not interesting, since the last term in equation (\ref{iweg}) would be a
radiation term $a^{-4}$, and (\ref{iweg}) would lead to $\Lambda$CDM model with a radiation term.
From (\ref{ekry}) it can be easily found that the time evolution of $D_{S}$ is given by the equation
\begin{equation}
\dot{D}_{S}=\frac{r_{\Sigma}aH}{\sqrt{1-\frac{2G_{\!N}M}{r_{\Sigma}a}-\frac{\alpha}{3}r_{\Sigma}^{2}
a^{2}-\frac{\beta\xi^{4}r_{\Sigma}^{2}a^{2}}{3D_{S}^{4}}}}\,.
\label{kebt}
\end{equation}
Equations (\ref{iweg}), (\ref{kebt}) form a system of two coupled differential equations for $a,D_{S}$.
We can bring this system in a more standard form defining
\begin{equation}
\chi=\frac{\alpha}{3}+\frac{\beta\xi^{4}}{3D_{S}^{4}}\,,
\label{kegr}
\end{equation}
and then
\begin{eqnarray}
\frac{\dot{a}^{2}}{a^{2}}+\frac{\kappa}{a^{2}}&=&\frac{2G_{\!N}M}{r_{\Sigma}^{3}a^{3}}+\chi
\label{jekw}\\
\dot{\chi}&=&-\frac{4\cdot 3^{\frac{1}{4}}r_{\Sigma}aH(\chi-\frac{\alpha}{3})^{\frac{5}{4}}}
{\xi\beta^{\frac{1}{4}}\sqrt{1-\frac{2G_{\!N}M}{r_{\Sigma}a}-r_{\Sigma}^{2}a^{2}\chi}}\,.
\label{kewra}
\end{eqnarray}
Note from (\ref{kegr}) that $\chi-\frac{\alpha}{3}>0$. From (\ref{jekw}) it is seen that the
quantity $\chi$ plays the role of dark energy. Namely, it is
\begin{equation}
H^{2}+\frac{\kappa}{a^{2}}=\frac{8\pi G_{\!N}}{3}(\rho+\rho_{DE})\,,
\label{ekaf}
\end{equation}
where $\rho$ is given by (\ref{lper}) and $\rho_{DE}=\frac{3}{8\pi G_{\!N}}\chi$.

Note in passing that equation (\ref{iweg}), combined with equation (\ref{ekaf}), gives
\begin{equation}
\frac{4\pi}{3}R_{S}^{3}\rho_{\text{tot}}=M+\frac{\Lambda(R_{S})R_{S}^{3}}{6G_{\!N}}\,.
\label{kwow}
\end{equation}
Comparing this equation with (\ref{kwrt}), it arises that the total energy due to the
cosmological constant is given by two expressions, first by the integral in (\ref{kwrt})
and second by the last term in (\ref{kwow}). There is no contradiction with that, since
the equality of these two expressions at the today values simply provides the additional constraint
on the parameters mentioned above.

The density parameters are defined in the standard way
\begin{equation}
\Omega_{m}=\frac{8\pi G_{\!N}\rho}{3H^{2}}\,\,\,\,\,\,\,,\,\,\,\,\,\,\,
\Omega_{DE}=\frac{8\pi G_{\!N}\rho_{DE}}{3H^{2}}\,.
\label{kebh}
\end{equation}
From the first of equations (\ref{kebh}), using the today values of the variables,
a relation between the parameters $M$ and $r_{\Sigma}$ can be found
\begin{equation}
r_{\Sigma}=\Big(\frac{2G_{\!N}M}{\Omega_{m0}a_{0}^{3}H_{0}^{2}}\Big)^{\!\frac{1}{3}}\,.
\label{lews}
\end{equation}
For the typical masses we use, it was found above that $r_{\Sigma}=0.83\text{Mpc}$ for a galaxy,
and $r_{\Sigma}=18\text{Mpc}$ for a cluster of galaxies.

It is more convenient to work with the redshift $z=\frac{a_{0}}{a}-1$, where the today
value $a_{0}$ of the scale factor can be set to unity. From (\ref{kewra}), (\ref{lews})
the evolution of $\chi(z)$ is found from
\begin{equation}
\frac{d\chi}{dz}=\frac{4\cdot
3^{\frac{1}{4}}(\chi-\frac{\alpha}{3})^{\frac{5}{4}}}
{\xi\beta^{\frac{1}{4}}(1\!+\!z)^{2}\sqrt{\frac{1}{r_{\Sigma}^{2}a_{0}^{2}}-
\Omega_{m0}H_{0}^{2}(1\!+\!z)-\frac{\chi}{(1+z)^{2}}}}\,.
\label{sery}
\end{equation}
After having solved (\ref{sery}), the evolution of the Hubble parameter as a function of $z$ is
given by the expression
\begin{equation}
H^{2}=\Omega_{m0}H_{0}^{2}(1\!+\!z)^{3}+\chi-\frac{\kappa}{a_{0}^{2}}(1\!+\!z)^{2}\,.
\label{jwfr}
\end{equation}
Using (\ref{ekaf}), (\ref{lews}), the quantity $\Omega_{m}$ can also be found as a function of the
redshift
\begin{equation}
\Omega_{m}=\Big[1+\frac{1}{\Omega_{m0}H_{0}^{2}}\frac{\chi}{(1\!+\!z)^{3}}-
\frac{\kappa}{a_{0}^{2}\Omega_{m0}H_{0}^{2}}\frac{1}{1\!+\!z}\Big]^{-1}\,.
\label{heth}
\end{equation}

For the numerical investigation of the system we will need the today value $\chi_{0}$ of $\chi$.
From (\ref{jekw}) or (\ref{jwfr}) we find
\begin{equation}
\chi_{0}=\Omega_{DE,0}H_{0}^{2}\,.
\label{bdet}
\end{equation}
The differential equation (\ref{sery}) contains the parameters
$\xi,\beta=\nu G_{\!N},r_{\Sigma},\alpha$. The parameters $\xi$ and $\nu$ are of order unity.
The parameter $r_{\Sigma}$ was found from (\ref{lews}). Finally, the parameter $\alpha$ is
free in order to try to achieve the correct phenomenology. With these parameters and $\chi_{0}$ given
in (\ref{bdet}), we can solve numerically (\ref{sery}) and find $\chi(z)$. Then we can plot
$\Omega_{m}(z)$ from (\ref{heth}).

From (\ref{jekw}), (\ref{kewra}) one can obtain a Raychaudhuri type of equation.
Differentiating (\ref{jekw}), and using (\ref{kewra}) and (\ref{jekw}) itself, we get
\begin{equation}
\frac{\ddot{a}}{a}=-\frac{1}{2}\Omega_{m0}H_{0}^{2}(1\!+\!z)^{3}+\chi-\frac{2\cdot
3^{\frac{1}{4}}(\chi-\frac{\alpha}{3})^{\frac{5}{4}}}
{\xi\beta^{\frac{1}{4}}(1\!+\!z)\sqrt{\frac{1}{r_{\Sigma}^{2}a_{0}^{2}}-
\Omega_{m0}H_{0}^{2}(1\!+\!z)-\frac{\chi}{(1+z)^{2}}}}\,.
\label{ower}
\end{equation}
The same equation also arises from (\ref{kevf}). The deceleration parameter is given by
$q=-H^{-2}\frac{\ddot{a}}{a}$.
From (\ref{ower}), for $\kappa=0$, and using (\ref{bdet}), the condition
for acceleration today $\ddot{a}|_{0}>0$ is written as
\begin{equation}
\frac{1}{r_{\Sigma}^{2}a_{0}^{2}H_{0}^{2}}>\frac{4\cdot\sqrt{3}}{\xi^{2}\sqrt{\nu}}\,
\frac{(1-\Omega_{m0}-\frac{\alpha}{3H_{0}^{2}})^{\frac{5}{2}}}{(1-\frac{3}{2}
\Omega_{m0})^{2}}\,\frac{1}{H_{0}\sqrt{G_{\!N}}}+1\,,
\label{jeir}
\end{equation}
which is approximated by
\begin{equation}
0<1-\Omega_{m0}-\frac{\alpha}{3H_{0}^{2}}< \Big(\frac{\xi^{2}\sqrt{\nu}}{4\sqrt{3}}\Big)^{\!\frac{2}{5}}
\Big(1\!-\!\frac{3}{2}\Omega_{m0}\Big)^{\!\frac{4}{5}}\Big(\frac{H_{0}\sqrt{G_{\!N}}}
{r_{\Sigma}^{2}a_{0}^{2}H_{0}^{2}}\Big)^{\!\frac{2}{5}}\,.
\label{poew}
\end{equation}
It is found from (\ref{poew}) that $1-\Omega_{m0}-\frac{\alpha}{3H_{0}^{2}}\lesssim 10^{-22}$, from
where it is obvious that $\alpha>0$. For such values of $\alpha$, the conditions
$\lambda_{k}\ll 1$, $g_{k}\ll 1$ are seen from (\ref{kegr}), (\ref{poew}) that are easily satisfied.
Of course, such values form an extreme fine tuning for $\alpha$ since $\alpha$
has to be extremely close to the value $3(1-\Omega_{m0})H_{0}^{2}$, which is also the value of the
cosmological constant.
Additionally, for such values of $\alpha$ it can be seen that $\dot{\Omega}_{m}|_{0}<0$,
which assures a local increase of $\Omega_{m}$ in the past. There is also the option that
$1-\Omega_{m0}-\frac{\alpha}{3H_{0}^{2}}$ is of order unity, and then from (\ref{poew}) it
turns out that $\nu>10^{106}$, however such values are not predicted by the quantum theory.

Finally, since the pressure of the matter component vanishes, $p=0$, equation (\ref{ower}) can be
brought into the form
\begin{equation}
2\dot{H}+3H^{2}+\frac{\kappa}{a^{2}}=-8\pi G_{\!N}(p+p_{DE})\,,
\label{lerf}
\end{equation}
where the dark energy pressure is given by
\begin{equation}
-8\pi G_{\!N}p_{DE}=3\chi-\frac{4\cdot
3^{\frac{1}{4}}(\chi-\frac{\alpha}{3})^{\frac{5}{4}}}
{\xi\beta^{\frac{1}{4}}(1\!+\!z)\sqrt{\frac{1}{r_{\Sigma}^{2}a_{0}^{2}}-
\Omega_{m0}H_{0}^{2}(1\!+\!z)-\frac{\chi}{(1+z)^{2}}}}\,.
\label{iwie}
\end{equation}
A significant parameter for the study of the late-times cosmology is the
equation-of-state parameter for the effective dark energy sector $w_{DE}=\frac{p_{DE}}{\rho_{DE}}$,
which is found from (\ref{iwie}) to be
\begin{equation}
w_{DE}=-1+\frac{4\cdot
3^{-\frac{3}{4}}(\chi-\frac{\alpha}{3})^{\frac{5}{4}}}
{\xi\beta^{\frac{1}{4}}(1\!+\!z)\chi\sqrt{\frac{1}{r_{\Sigma}^{2}a_{0}^{2}}-
\Omega_{m0}H_{0}^{2}(1\!+\!z)-\frac{\chi}{(1+z)^{2}}}}\,.
\label{kjet}
\end{equation}
Therefore, $w_{DE}$ cannot take phantom values smaller than -1. Its today value becomes for $\kappa=0$
\begin{equation}
w_{DE,0}\approx -1+\frac{4\cdot 3^{-\frac{3}{4}}}{\xi\nu^{\frac{1}{4}}(1\!-\!\Omega_{m0})}
\Big(1\!-\!\Omega_{m0}\!-\!\frac{\alpha}{3H_{0}^{2}}\Big)^{\frac{5}{4}}
\Big(\frac{r_{\Sigma}^{2}a_{0}^{2}H_{0}^{2}}{H_{0}\sqrt{G_{\!N}}}\Big)^{\frac{1}{2}}
\label{srfr}
\end{equation}
and, according to (\ref{poew}), it can take any value $-1<w_{DE,0}<-1+(\frac{2}{3}-\Omega_{m0})
(1-\Omega_{m0})^{-1}\approx -0.48$.

Because of the above extreme fine-tuning in $\alpha$, numerical study of the equations is not
possible and we need to perform an analytical study in order to understand the behaviour of the
system. This analysis is shown in the Appendix B for $\kappa=0$. We summarize here
the results of this analysis. Concerning the evolution of the Hubble parameter and the parameter
$\Omega_{m}$, the $\Lambda$CDM behaviour arises practically in all the regime of applicability of
the model all the way down to the present epoch, with an exception in the very first steps
of the AS evolution. As for the deceleration parameter and the dark energy equation-of-state parameter,
there are intervals where these parameters have a behaviour different from that of $\Lambda$CDM.
This discrepancy between the behaviour of the parameters $H(z),\Omega_{m}(z)$ and the behaviour
of the parameters $\ddot{a}(z),w_{DE}(z)$ is peculiar and is due to the
presence of the higher time derivatives contained in the acceleration, which
can lead to significant contribution from terms which are negligible in $H$, $\Omega_{m}$.
However, for the numerical values of the various constants of astrophysical and cosmological origin
describing our model, the final result is that in all physically interesting cases the acceleration
properties of the model cannot be practically discerned from the $\Lambda$CDM scenario,
therefore our model is indistinguishable from $\Lambda$CDM. There is a possibility, as always in
cosmology, that the perturbations of the model evolve in a distinct way from $\Lambda$CDM, however
the previous analysis of the background, with the particular values encountered, leaves little hope
for observational evidence of the deviations from $\Lambda$CDM.

\subsection{Second RG flow behaviour: close to the infrared}
\label{jurt}

There are encouraging indications that for $k\rightarrow 0$ the cosmological
constant runs proportional to $k^{2}$, so $\Lambda_{k}=\lambda_{\ast}^{\text{IR}}k^{2}$,
where $\lambda_{\ast}^{\text{IR}}>0$ is the infrared fixed point of the $\lambda$-evolution.
At the same time, it seems that $G_{k}$ increases a lot, for example $g_{k}$ could converge to
an IR value $g_{\ast}^{\text{IR}}>0$ or even diverge. This postulated fixed point \cite{Bonanno:2001hi}
can be considered as the IR counterpart of the UV Non-Gaussian fixed point (NGFP) \cite{Reuter:2009kq}.
This non-trivial IR running can be assumed that is due to quantum fluctuations with very small
momenta, corresponding to distances larger than the largest localized structures in the universe.
Since we are interested in generating the dark energy at astrophysical scales, the exact realization
of the above IR fixed point at cosmological scales is not our purpose. There are works,
not concerning cosmology, which consider the possibility that such a fixed point can act already at
astrophysical scales \cite{Reuter:2004nx}, \cite{Esposito:2007xz}. Our approach will be different.
We will consider, as it is pretty reasonable, that the above IR fixed point has not yet been
reached at the intermediate astrophysical scales, and therefore, some deviations from the above
functions $G_{k},\Lambda_{k}$ should be present inside the objects of our interest.

Concerning the gravitational constant $G_{k}$, since at moderate scales, well beyond
the NGFP, $G_{k}$ is approximately constant, we will assume that it has the constant value
$G_{\!N}$ over a broad range of scales ranging from the submillimeter up to the galaxy or cluster
scale. Actually, we will not make use of the submillimeter lower bound, so we can just restrict
ourselves above some observable macroscopic distances. There is the possibility that at large
astrophysical scales, $G$ acquires an IR correction beyond $G_{\!N}$, however, we will not be
concerned about this in our introductory treatment here, since we do not want to assume arbitrary
functional forms. Already the relatively successful explanation of the galactic or cluster dynamics
using the standard Newton's law makes our adoption $G_{k}=G_{\!N}$ legal enough.

We will be restricted
on the single effect of the variability of the cosmological constant $\Lambda_{k}$. Since the
functional form of the deviation of $\Lambda_{k}$ from its IR form $k^{2}$ is not known, we will
assume that the running of $\Lambda_{k}$ is described by a power law dependence on the energy,
$\Lambda_{k}\sim k^{b}$. Of course, this is a simple ad-hoc parametrization and the analysis will
show how far one can go with the single parameter $b$. However, if $\Lambda_{k}$ differs slightly
from its IR form $k^{2}$, what is actually not unexpected since astrophysical structures are
already ``large'' enough, then $b$ will be close to the value 2 and the power-law dependence $k^{b}$
will be a very good approximation of the running behaviour. Moreover, moving with the deformed law
$k^{b}$ from the IR cosmological law $k^{2}$ down to the astrophysical scales, it seems more probable
for $b$ to be slightly larger than 2, instead of smaller than 2. This is due to that since $k$
increases at smaller length scales, the parameter $b$ should also increase in order to have
a significant astrophysical $\Lambda_{k}$, otherwise $k^{b}$ would be neutralized and $\Lambda_{k}$
would be suppressed.

We summarize by writing our ansatz
\begin{equation}
G_{k}=G_{\!N}\,\,\,\,\,\,\,\,,\,\,\,\,\,\,\,\,\Lambda_{k}=\gamma k^{b}\,,
\label{helg}
\end{equation}
where $\gamma>0, b$ are constants. The parameter $\gamma$ has dimensions mass to the power $2-b$
and will be parametrized as $\gamma=\tilde{\gamma} G_{\!N}^{\frac{b}{2}-1}$ with
$\tilde{\gamma}$ dimensionless. If an order $\mathcal{O}(1)$ value for $\tilde{\gamma}$ arises,
this will mean that no new mass scale is needed (no new physics is needed for the explanation of
dark energy other than AS gravity and the knowledge of structure) and the coincidence problem might
be resolved from already controllable physics without fine-tuning or new scales.
Indeed, it will be shown that in the most faithful case, discussed in the last
subsection VII-B-2, it turns out that $b$ is a little higher than 2 and $\tilde{\gamma}=\mathcal{O}(1)$,
which means that our model with the dark energy originated from IR quantum corrections of the
cosmological constant at the astrophysical level can be pretty successful.
With the assumption (\ref{helg}), the metric (\ref{ASBH}) contains the Newtonian term
$\frac{2G_{\!N}M}{R}$ and the non-trivial cosmological constant term $\frac{1}{3}\Lambda_{k} R^{2}$
which becomes $\frac{\xi^{b}\tilde{\gamma}}{3}\big(\frac{R}{L}\big)^{\!b}
\big(\frac{\sqrt{G_{\!N}}}{R}\big)^{\!b-2}$.

For $L=R$ we will see that current acceleration requires that $b<1.57$ (actually $b$ should be smaller
than 1 in order to have a reasonable $w_{DE}$ today), therefore we cannot be in the most interesting
IR range of $b$ a little larger than 2 and the law $k^{b}$ loses its theoretical significance. Then, it
will arise that $\tilde{\gamma}$ has to be various decades of orders of magnitude smaller than unity
in order to arrange the amount of dark energy, which means that new scales are introduced through
$\tilde{\gamma}$. All this situation, which will be examined in the next subsection VII-B-1, although
not an obvious fail, certainly it cannot be considered as a big success in relation to the coincidence
problem. If such values are acceptable, then the internal dynamics presents some measurable deviations
from the standard Newtonian dynamics which need more investigation to check their consistency with
observations. As for the force corresponding to the new cosmological constant term, this is
$\frac{\tilde{\gamma}\xi^{b}(2-b)}{6\sqrt{G_{\!N}}}\big(\frac{\sqrt{G_{\!N}}}{R}\big)^{\!b-1}$, and
therefore it is repulsive. The ratio of the new cosmological constant term to the Newtonian term turns
out to be $\frac{\Omega_{DE,0}}{\Omega_{m0}}\big(\frac{R}{r_{\Sigma}}\big)^{\!3-b}$, while the
corresponding ratio of the forces gets an extra factor $b\!-\!2$. Thus, for large clusters these ratios
at the border are larger than 0.1 for any $b$ and the ratios can reach up to 0.35 for the largest
$b$. For small clusters the ratios are suppressed to less than one per cent. Finally, for galaxies
the ratios are almost 0.2 for the largest $b$ and decrease considerable for smaller $b$. Certainly,
inside the structures, the matter profiles have to be taken into account to obtain accurate results.

What it will turn out to be the most exciting case is $L=D$, which will be examined in the last
subsection VII-B-2. In this case, the dark energy, along with its acceleration, will be explained
with $b$ a little larger than 2 and $\tilde{\gamma}$ some number of order unity. More precisely, the
quantity $D$ at the Schucking radius today, $D_{S,0}$, which can be considered as the integration
constant of the differential equation for $D(R)$, has to be arranged to provide the correct amount
of dark energy. This implies that $D_{S,0}$ is of order of $r_{\Sigma}$, and more generally it will be
shown numerically that the whole function $D(R)$ turns out to be of order of $r_{\Sigma}$. This is an
extra naturalness for our model, since $D$, which has some meaning of proper distance, is of the length
of the astrophysical object, and $D_{S,0}$ does not introduce a new scale. Therefore, the
new cosmological constant term becomes
$\frac{1}{3}\Lambda_{k}R^{2}=\frac{\xi^{b}\tilde{\gamma}}{3}\big[\frac{1}{G_{\!N}}
\big(\frac{\sqrt{G_{\!N}}}{r_{\Sigma}}\big)^{\!b}\big]\big(\frac{r_{\Sigma}}{D}\big)^{\!b}R^{2}$, where
we can set $D\sim r_{\Sigma}$ in order to make an estimate. Of course, the deviations of the new
cosmological constant term from the $R^{2}$ law and the precise functional form of $D(R)$ are very
important and will provide a cosmology distinctive from $\Lambda$CDM.
However, we already observe that the quantity
$\frac{1}{G_{\!N}}\big(\frac{\sqrt{G_{\!N}}}{r_{\Sigma}}\big)^{\!b}$ has dimensions of cosmological
constant, and for $b$ close to $2.1$ it is very close to the order of magnitude of the standard
cosmological constant $\Lambda=4.7\times10^{-84}\text{GeV}^{2}$ of $\Lambda$CDM. The factor
$\big(\frac{r_{\Sigma}}{D}\big)^{\!b}$ offers a small distance-dependent deformation of the constant,
which however, as mentioned, is important for the derived cosmology. Therefore, our model will
provide a natural deformation of the $\Lambda$CDM model, without introducing arbitrary scales or
fine-tuning. Similarly, the constant prefactor $\xi^{b}\tilde{\gamma}$ can contribute to one order
of magnitude. {\it{As a result, beyond the general idea, motivated by the coincidence problem, of
generating the dark energy from local antigravity sources, the most important thing to emerge from
the present work is the introduction of the quantity
\begin{equation}
\frac{1}{G_{\!N}}\Big(\frac{\sqrt{G_{\!N}}}{r_{\Sigma}}\Big)^{\!b}\,,
\label{jurd}
\end{equation}
which arises in the context of infrared AS gravity and plays a role similar to the standard
cosmological constant $\Lambda$. This quantity, with $b$ a little larger (for galaxies) or a little
smaller (for clusters) than 2.1, has the quite interesting property that it has the same order of
magnitude as the standard $\Lambda$.}} For example, for galaxies with $b=2.13$ the quantity
(\ref{jurd}) is $2.1\times10^{-84}\text{GeV}^{2}$, while for clusters with $b=2.08$ it becomes
$2.6\times10^{-84}\text{GeV}^{2}$. It remains to AS gravity to confirm or not if such values of $b$
are predicted at the astrophysical scales. Therefore, in some loose rephrasing, it can be said that
the standard cosmological constant is not any longer an arbitrary independent quantity, as in
$\Lambda$CDM, but it is constructed out of $G_{\!N},r_{\Sigma},b$.

As a direct consequence of the above discussion for $L=D$, the new cosmological constant term itself,
$\frac{1}{3}\Lambda_{k}R^{2}$, becomes at the matching surface $R=r_{\Sigma}$ of the order of
$\big(\frac{\sqrt{G_{\!N}}}{r_{\Sigma}}\big)^{\!b-2}$. For galaxies with $b=2.13$ and typical
mass $M=10^{11}M_{\odot}$, this term is $4\times 10^{-8}$, while the Newtonian term is $10^{-8}$.
For clusters with $b=2.08$ and mass $M=10^{15}M_{\odot}$, this term is $2\times 10^{-5}$, while
the Newtonian term is $5\times 10^{-6}$. Therefore, the two gravitational potentials are comparabe
at the Schucking radius. We can see that approximately a change in $b$ less than
five per cent from the above values, gives a change in this cosmological constant term less than
one order of magnitude. This shows clearly a sensitivity on the value of $b$, which however cannot
be considered as fine-tuning or at least extreme fine-tuning. Additionally, the fact that $b$ acquires
smaller values at cluster scales relatively to galaxy scales is consistent with the IR value $b=2$ at
cosmological scales or beyond. At an extreme limit of the solar scale, a even larger value of $b$ is
expected. For a solar distance $R=1$A.U. the quantity $\sqrt{G_{\!N}}/R$ takes the value $10^{-46}$,
while the Newtonian potential of the sun at this distance is $10^{-8}$. Already for $b=2.25$ the
quantity $\big(\frac{\sqrt{G_{\!N}}}{R}\big)^{\!b-2}$, which gives an estimation of the new term,
becomes 3 orders of magnitude smaller than the Newtonian potential and this discrepancy becomes
huge for larger $b$. Therefore, it is reasonable to assume that the stringent solar-system, as well as
the laboratory tests of gravity, remain unaffected by the new term. Finally, as for the force
of the new term, this is $\frac{\xi^{b}\tilde{\gamma}}{6}\big(2-b\frac{R}{D\sqrt{F}}\big)
\big[\frac{1}{G_{\!N}}\big(\frac{\sqrt{G_{\!N}}}{r_{\Sigma}}\big)^{\!b}\big]
\big(\frac{r_{\Sigma}}{D}\big)^{\!b}R$, where $F(R)=1-\frac{2G_{\!N}M}{R}-\frac{1}{3}\Lambda_{k}R^{2}$.
The study of the new potential and its force at the border and inside the object is more complicated
due to the presence of $D(R)$, so in order to get relatively reliable results, we will need a more
detailed treatment. A preliminary analysis is given in the last subsection, where the result is that
there is no obvious inconsistency with the internal dynamics.

We continue with the implication in cosmology of the previous new cosmological constant term for
$L=D$. While the Newtonian term is certainly responsible for the dust term $\Omega_{m0}H_{0}^{2}/a^{3}$
on the right hand side of (\ref{yehg}), the above running cosmological constant term is responsible
for the dark energy term
$\frac{\xi^{b}\tilde{\gamma}}{3}\frac{1}{G_{\!N}}\big(\frac{\sqrt{G_{\!N}}}{D_{S}}\big)^{\!b}$.
Therefore, $\Omega_{DE,0}$ is equal to
$\frac{\xi^{b}\tilde{\gamma}}{3}\big(\frac{r_{\Sigma}}{D_{S,0}}\big)^{\!b}\frac{1}{G_{\!N}H_{0}^{2}}
\big(\frac{\sqrt{G_{\!N}}}{r_{\Sigma}}\big)^{\!b}$. Since $D_{S,0}\sim r_{\Sigma}$ and
$\tilde{\gamma}$ is taken to be of order one, it is implied from the discussion on the value of
the quantity (\ref{jurd}) that $\Omega_{DE,0}\sim 1$. This can be considered as a natural
explanation of the coincidence problem. The hard coincidence of the value of the standard cosmological
constant $\Lambda\sim H_{0}^{2}$ is exchanged to a mild coincidence of the index $b$ close to 2.1, which
happens to satisfy
\begin{equation}
\frac{1}{G_{\!N}}\Big(\frac{\sqrt{G_{\!N}}}{r_{\Sigma}}\Big)^{\!b}\sim H_{0}^{2}.
\label{mutn}
\end{equation}
Strictly speaking it is the today value $\Lambda(R_{S,0})$ of the time-dependent function
$\Lambda(R_{S})$ that has a value of order of $H_{0}^{2}$.

As discussed in Sec.~\ref{eudk}, the form (\ref{helg}) for the running cosmological constant can be
applied directly at conventional cosmologies without the Swiss cheese description to define a different
cosmological evolution. In this cosmology, the dark energy term is
$\frac{\xi^{b}\tilde{\gamma}}{3}\frac{1}{G_{\!N}}\big(\frac{\sqrt{G_{\!N}}}{L}\big)^{\!b}$.
The quantity $L$ is a cosmological distance scale instead of an astrophysical one used in the
previous paragraph and this creates a huge difference. It is a common and sensible
approach to assume $L=H^{-1}$ in this case, and the today dark energy term becomes
$\frac{\tilde{\gamma}\xi^{b}}{3}(\sqrt{G_{\!N}}H_{0})^{b-2}H_{0}^{2}$. This quantity
acquires the correct order of magnitude, $H_{0}^{2}$, if $|b-2|<0.01$, which means that $b$ should be
very close to the IR value 2. It is not that the two approaches, the one defined by
(\ref{mutn}) and the one discussed here, differ only quantitatively. They can also differ
qualitatively, in the sense that one approach can be relevant for the explanation of dark energy and
the other not. For example, if AS predicts for the index $b$ at the current cosmological scales the
value 2.04, this automatically means that the cosmological interpretation of (\ref{helg}) is irrelevant
today and the IR value of $b$ approximately 2 will be useful for the future universe evolution. In our
work we consider the values of $b$ which imply the satisfaction of equation (\ref{mutn}).

\subsubsection{Energy-coordinate distance scale relation}
\label{hggwj}

We first choose the simpler case where the energy scale $k$ is inversely proportional to the
coordinate radius $R$, therefore for the front value of $k$ at the Schucking radius we have
the relation (\ref{meyc}). From equation (\ref{yehg}) the Hubble equation becomes
\begin{equation}
\frac{\dot{a}^{2}}{a^{2}}+\frac{\kappa}{a^{2}}=\frac{2G_{\!N}M}{r_{\Sigma}^{3}a^{3}}+
\frac{\gamma\xi^{b}}{3r_{\Sigma}^{b}a^{b}}\,.
\label{kepe}
\end{equation}
This is a simple equation for the evolution of the scale factor, which is also written as
\begin{equation}
H^{2}+\frac{\kappa}{a^{2}}=\frac{8\pi G_{\!N}}{3}(\rho+\rho_{DE})\,,
\label{ekafs}
\end{equation}
where $\rho$ is given by (\ref{lper}) and $\rho_{DE}=\frac{\gamma \xi^{b}}{8\pi G_{\!N}
r_{\Sigma}^{b}a^{b}}$\,.

The density parameters are defined as in (\ref{kebh}) and the value
of $r_{\Sigma}$ for a galaxy or a cluster is found from (\ref{lews}) as before.
The evolution of the Hubble parameter as a function of $z$ is
\begin{equation}
H^{2}=\Omega_{m0}H_{0}^{2}(1\!+\!z)^{3}+\frac{\gamma\xi^{b}}{3r_{\Sigma}^{b}a_{0}^{b}}
(1\!+\!z)^{b}-\frac{\kappa}{a_{0}^{2}}(1\!+\!z)^{2}\,,
\label{jrfd}
\end{equation}
where $a_{0}$ is set to unity. Similarly, the matter density abundance $\Omega_{m}$ becomes
\begin{equation}
\Omega_{m}=\Big[1+\frac{\gamma\xi^{b}}{3r_{\Sigma}^{b}a_{0}^{b}\Omega_{m0}H_{0}^{2}}
\frac{1}{(1\!+\!z)^{3-b}}-\frac{\kappa}{a_{0}^{2}\Omega_{m0}H_{0}^{2}}\frac{1}{1\!+\!z}\Big]^{-1}\,.
\label{ujer}
\end{equation}
These equations contain the parameters $\xi$ (which is of order one), $r_{\Sigma}$
(which is known from (\ref{lews}) for the typical masses we use), and $\gamma$, $b$.
For $\kappa=0$, in order to have $\Omega_{m}$ an increasing function of $z$ in
agreement with observations, it should be $b<3$.

Now, from (\ref{ujer}) we take the condition for the correct amount of today dark energy
\begin{equation}
\xi^{b}\tilde{\gamma}=3\Omega_{DE,0}\,G_{\!N}H_{0}^{2}\Big(\frac{r_{\Sigma}a_{0}}{\sqrt{G_{\!N}}}
\Big)^{b}\,.
\label{keiw}
\end{equation}
Equation (\ref{keiw}) is analogous to the relation $\Lambda=3\Omega_{\Lambda 0}H_{0}^{2}$.
As was also explained above, here it is the varying cosmological constant that creates the dark energy
based on the parameters $\tilde{\gamma},b$ characterizing $\Lambda_{k}$, and the astrophysical
scale $r_{\Sigma}$. For arbitrary values of $b$, the parameter $\tilde{\gamma}$ will be very large
or very small, introducing in this way a new massive scale $\gamma=\tilde{\gamma}G_{\!N}^{\frac{b}{2}-1}$.
The cosmological constant $\Lambda_{k}=\gamma k^{b}$ should be generated inside
the structure due to quantum corrections, so the parameters $\tilde{\gamma},b$ should be given by
the AS theory at the astrophysical scales. Therefore, equation (\ref{keiw}), as
an equation between orders of magnitude, forms a coincidence similar to that of $\Lambda$
(actually it can be considered as increased coincidence) among these two parameters $\tilde{\gamma},b$,
the astrophysical value $r_{\Sigma}$ and the cosmological parameter $H_{0}$. What we find particularly
interesting in relation to the coincidence problem is the situation with $\tilde{\gamma}\sim 1$,
because then, no new scale is introduced for the explanation of dark energy, other than the
astrophysical scale. In this favorite case of ours, the relation (\ref{mutn}) is valid with the
index $b$ having values close to 2.1 as explained before, and the hard coincidence of (\ref{keiw})
reduces and is just rendered to the mild coincidence of the value of $b$. Unfortunately, we will
see below that such $b$ are not allowed here due to acceleration reasons. We should note additionally
that the situation with equation (\ref{keiw}) is different than the picture where the dark energy is
due to some extra field having its own equation of motion. In that case, the corresponding of
equation (\ref{keiw}) would form a coincidence relation of dark matter-dark energy where some
integration constants of the extra field would be involved, while the parameters of the theory would
remain free to accommodate some other observation. This will actually be the case of the next subsection.

The acceleration is found to be
\begin{equation}
\frac{\ddot{a}}{a}=-\frac{1}{2}\Omega_{m0}H_{0}^{2}(1\!+\!z)^{3}
+\frac{(2\!-\!b)\gamma \xi^{b}}{6r_{\Sigma}^{b}a_{0}^{b}}(1\!+\!z)^{b}\,.
\label{wehr}
\end{equation}
The condition for today acceleration $\ddot{a}|_{0}>0$ becomes
\begin{equation}
\frac{(2\!-\!b)\gamma\xi^{b}}{3r_{\Sigma}^{b}a_{0}^{b}\Omega_{m0}H_{0}^{2}}>1\,,
\label{ierf}
\end{equation}
which implies that a necessary condition is $b<2$. Equation (\ref{wehr}) implies that we have the
correct behaviour with a past deceleration and a today acceleration.
Due to (\ref{keiw}), the inequality (\ref{ierf}) takes the form
\begin{equation}
b<2-\frac{\Omega_{m0}}{\Omega_{DE,0}}\,,
\label{wtjg}
\end{equation}
which means that $b<1.57$. Then, it arises from (\ref{keiw}) that $\tilde{\gamma}\lesssim 10^{-30}$
(and even smaller for galaxies), so $\tilde{\gamma}$ is various decades of orders of magnitude
smaller than unity, and finally, no alleviation is offered to the coincidence problem. We continue
in the following with a little more investigation of the present case, but it has already become
obvious that the new massive scale $\gamma=\tilde{\gamma}G_{\!N}^{\frac{b}{2}-1}$, very different
than the one provided by $G_{\!N}^{\frac{b}{2}-1}$, is necessarily introduced. And this is due to
the values of $b$ implied by the subtle issue of acceleration. Note that for the standard
$\Lambda$ it is $\Lambda=3.1\times 10^{-122}G_{\!N}^{-1}$.

\begin{figure}[ht]
\begin{center}
\includegraphics[width=0.80\textwidth]{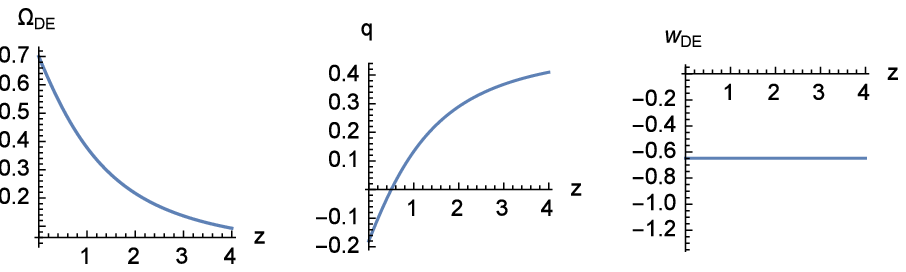}
\caption{{\it{The late-times cosmological evolution for a spatially flat
universe, for the parameter choice $b=1.06$, $\xi=1$, $\tilde{\gamma}=2.8\times10^{-60}$,
$M=10^{15}M_{\odot}$ and $\Omega_{m0}=0.3$.
In the left graph we depict the evolution of the
dark energy density parameter as a function of the redshift $z$. In the middle graph we
present the evolution of the deceleration parameter.
Finally, the dark energy equation-of-state
parameter is depicted in the right graph.}}} \label{basicplot2}
\end{center}
\end{figure}

We have also to assure that the transition point from deceleration to acceleration is recent.
Combining equations (\ref{ujer}), (\ref{wehr}), we find that this transition occurs at redshift
$z_{t}$ which satisfies the equation
\begin{equation}
\frac{(1+z_{t})^{3-b}}{2\!-\!b}=
\frac{1}{\Omega_{m0}}-1+\frac{\kappa}{a_{0}^{2}\Omega_{m0}H_{0}^{2}}\,.
\label{erfw}
\end{equation}
So, $z_{t}$ is basically determined from the parameter $b$ (for example, for the value
$z_{t}\approx 0.5$ it is $b\approx 1$). Ignoring $\kappa$, it is found from (\ref{erfw}) that
$z_{t}\lesssim 0.67$. The condition (\ref{wtjg}) is equivalent to $z_{t}>0$. The deceleration
parameter becomes today
\begin{equation}
q_{0}=\frac{1}{2}\Omega_{m0}-\frac{2-b}{2}\Omega_{DE,0}\,.
\label{gher}
\end{equation}
So, depending on $b$, it is $-0.55<q_{0}<0$. The lowest value $q_{0}=-0.55$ characterizes the
$\Lambda$CDM model and here is attached to $b\rightarrow 0$ with $z_{t}=0.67$. The upper value
$q_{0}=0$ is associated to the maximum value $b=1.57$ with $z_{t}\rightarrow 0$ (for the
intermediate example with $b\approx 1$ it is $q_{0}\approx -0.2$). If we are to insist to be close
to the familiar value $q_{0}=-0.55$, then $b$ should be close to zero and the model is just a
slight variation of the $\Lambda$CDM model, since there are no extra parameters to create some
degeneracy. For such $b$ close to zero, the dark energy term has a slow evolution instead of the
constancy of the standard $\Lambda$ and it is something like $\tilde{\gamma}\lesssim 10^{-100}$.
We finish with writing the pressure of dark energy for a general $b$ as
\begin{equation}
8\pi G_{\!N}p_{DE}=-\frac{(3\!-\!b)\gamma \xi^{b}}{3r_{\Sigma}^{b}a_{0}^{b}}(1\!+\!z)^{b}\,,
\label{dhes}
\end{equation}
from where the equation of state of dark energy is
\begin{equation}
w_{DE}=-1+\frac{b}{3}\,,
\label{kjdl}
\end{equation}
which is constant and is not compatible with a seemingly varying $w_{DE}$.

We present in Fig. \ref{basicplot2} the cosmological evolution as a function of $z$
for a spatially flat universe, with the quantum cosmological constant originated at the cluster level
with $M=10^{15}M_{\odot}$, $\Omega_{m0}=0.3$, and for the parameter choice $b=1.06$, $\xi=1$,
$\tilde{\gamma}=2.8\times 10^{-60}$, which provide $z_{t}=0.5$.
We depict in the left graph the evolution of the dark energy density parameter $\Omega_{DE}$ from
equation (\ref{ujer}), where it appears a typical decreasing behaviour for larger redshifts. In
the middle graph the evolution of the deceleration parameter $q$ is shown, where a passage from
deceleration to acceleration at late times can be seen. The dark energy equation-of-state parameter
$w_{DE}$ remains constant in the right graph. It is obvious that in order to have a $w_{DE,0}$
close to -1, according to observations, the value of the parameter $b$ has to be close to zero,
and thus, the model is finally very close to the $\Lambda$CDM model.

Although the present model is not taken seriously as it is not theoretically favorable in relation
to the coincidence problem, and also it should be very close to the $\Lambda$CDM model (by choosing
$b$ close to zero) in order to have some compatibility with simple observational tests, we find
interesting to leave it a little space to breath by noticing two points. First, the parameter $b$
is not exactly constant, but more accurately, it is $z$-dependent since at different cosmic epochs
the running couplings move at different points of the AS renormalization group phase portrait.
Therefore, in general, $w_{DE}$ becomes time-dependent. This point is also related to the specific
model of structure growth one has to assume for a typical galaxy or cluster in order to make a more
realistic implementation of the model. Since, during the collapsing phase of a structure (larger
length scales), the index $b$ is expected to be smaller than the today value of the formed object,
the mean value of $b$ is smaller, and thus the real $w_{DE}$ is reduced. Second, in view of the
inhomogeneous/anisotropic models discussed in the Introduction (\cite{inho}, \cite{inhoswiss},
\cite{structure}), the description of a inhomogeneous universe with more realistic structures
would provide through averaging processes enhancement of cosmic acceleration, and therefore, $w_{DE}$
would become even smaller. In any case, the scenario discussed here will be improved in various
aspects in the next subsection where the energy scale will be determined from the length scale $L=D$.

\subsubsection{Energy-proper distance scale relation}
\label{kjwknj}

Here we use the proper distance as measure of the energy scale, which seems to be more realistic,
so we assume the law (\ref{jaet}). From equation (\ref{yehg}) the Hubble evolution is given by
\begin{equation}
\frac{\dot{a}^{2}}{a^{2}}+\frac{\kappa}{a^{2}}=\frac{2G_{\!N}M}{r_{\Sigma}^{3}a^{3}}
+\frac{\gamma\xi^{b}}{3D_{S}^{b}}\,,
\label{ksrg}
\end{equation}
where
\begin{equation}
\dot{D}_{S}=\frac{r_{\Sigma}aH}{\sqrt{1-\frac{2G_{\!N}M}{r_{\Sigma}a}
-\frac{\gamma\xi^{b}r_{\Sigma}^{2}a^{2}}{3D_{S}^{b}}}}\,.
\label{kewr}
\end{equation}
The quantity $D_{S}$, which expresses the proper distance of the matching surface, acts as a new
cosmological field of geometrical nature with its own equation of motion (\ref{kewr}).
Equations (\ref{ksrg}), (\ref{kewr}) form a system of two coupled differential equations for
$a,D_{S}$. Defining
\begin{equation}
\psi=\frac{\gamma\xi^{b}}{3D_{S}^{b}}\,,
\label{kiet}
\end{equation}
which should be positive, we bring the system to the more standard form
\begin{eqnarray}
&&\frac{\dot{a}^{2}}{a^{2}}+\frac{\kappa}{a^{2}}=\frac{2G_{\!N}M}{r_{\Sigma}^{3}a^{3}}
+\psi\label{krds}\\
&&\dot{\psi}=-\frac{3^{\frac{1}{b}}br_{\Sigma}aH\psi^{1+\frac{1}{b}}}
{\xi\gamma^{\frac{1}{b}}\sqrt{1-\frac{2G_{\!N}M}{r_{\Sigma}a}
-r_{\Sigma}^{2}a^{2}\psi}}\label{sjdg}\,,
\end{eqnarray}
where $\psi$ plays the role of dark energy. We also have
\begin{equation}
H^{2}+\frac{\kappa}{a^{2}}=\frac{8\pi G_{\!N}}{3}(\rho+\rho_{DE})\,,
\label{ketr}
\end{equation}
where $\rho$ is given by (\ref{lper}) and $\rho_{DE}=\frac{3}{8\pi G_{\!N}}\psi$.

The density parameters are defined as in (\ref{kebh}) and the value
of $r_{\Sigma}$ for a galaxy or a cluster is found from (\ref{lews}) as before.
It is more convenient to work with the redshift $z$ and the evolution of $\psi(z)$ is given by
\begin{equation}
\frac{d\psi}{dz}=\frac{3^{\frac{1}{b}}b\psi^{1+\frac{1}{b}}}
{\xi\gamma^{\frac{1}{b}}(1\!+\!z)^{2}\sqrt{\frac{1}{r_{\Sigma}^{2}a_{0}^{2}}
-\Omega_{m0}H_{0}^{2}(1\!+\!z)-\frac{\psi}{(1+z)^{2}}}}\,.
\label{wskt}
\end{equation}
After having solved (\ref{wskt}), the evolution of the Hubble parameter as a function of $z$ is
\begin{equation}
H^{2}=\Omega_{m0}H_{0}^{2}(1\!+\!z)^{3}+\psi-\frac{\kappa}{a_{0}^{2}}(1\!+\!z)^{2}\,,
\label{rewd}
\end{equation}
while, the matter density abundance $\Omega_{m}$ becomes
\begin{equation}
\Omega_{m}=\Big[1+\frac{1}{\Omega_{m0}H_{0}^{2}}\frac{\psi}{(1\!+\!z)^{3}}-
\frac{\kappa}{a_{0}^{2}\Omega_{m0}H_{0}^{2}}\frac{1}{1\!+\!z}\Big]^{-1}\,.
\label{ierr}
\end{equation}

For the numerical investigation of the system we will need the today value $\psi_{0}$ of $\psi$.
In terms of the other parameters it is
\begin{equation}
\psi_{0}=\Omega_{DE,0}H_{0}^{2}\,.
\label{pewr}
\end{equation}
The differential equation (\ref{wskt}) contains the parameters $\xi$ (which is of order one),
$r_{\Sigma}$ (which is known from (\ref{lews}) for the typical masses we use), and $\gamma$, $b$.
With these parameters and $\psi_{0}$ given in (\ref{pewr}), we can solve numerically (\ref{wskt})
and find $\psi(z)$. Then we can plot $\Omega_{m}(z)$ from (\ref{ierr}).

It is illuminating to define the variable $\tilde{\psi}=\psi/H_{0}^{2}$, and then, equations
(\ref{wskt}), (\ref{rewd}), (\ref{ierr}) become respectively
\begin{eqnarray}
&&\frac{d\tilde{\psi}}{dz}=\frac{3^{\frac{1}{b}}b(G_{\!N}H_{0}^{2})^{\frac{1}{b}-\frac{1}{2}}
\tilde{\psi}^{1+\frac{1}{b}}}
{\xi\tilde{\gamma}^{\frac{1}{b}}(1\!+\!z)^{2}\sqrt{\frac{1}{r_{\Sigma}^{2}a_{0}^{2}H_{0}^{2}}
-\Omega_{m0}(1\!+\!z)-\frac{\tilde{\psi}}{(1+z)^{2}}}}\label{jeth}\\
&&\frac{H^{2}}{H_{0}^{2}}=\Omega_{m0}(1\!+\!z)^{3}+\tilde{\psi}+\Omega_{\kappa 0}
(1\!+\!z)^{2}\label{uhne}\\
&&\Omega_{m}=\Big[1+\frac{1}{\Omega_{m0}}\frac{\tilde{\psi}}{(1\!+\!z)^{3}}+
\frac{\Omega_{\kappa 0}}{\Omega_{m0}}\frac{1}{1\!+\!z}\Big]^{-1}\,,\label{munf}
\end{eqnarray}
where $\tilde{\psi}_{0}=\Omega_{DE,0}$ and $\Omega_{\kappa 0}=-\kappa/(a_{0}^{2}H_{0}^{2})$.
Since $\Omega_{\kappa 0}\ll 1$, $\Omega_{m0}\approx 0.3$ and $\Omega_{m}$ should basically increase
in the past, it arises from (\ref{munf}) that for recent redshifts, where our scenario makes sense, it
should be $\tilde{\psi}/(1+z)^{3}\lesssim 1$, otherwise $\Omega_{m}$ would drop to unacceptably small
values in the past. As a result, $\tilde{\psi}/(1+z)^{2}\lesssim 10$, which
is actually even smaller. On the other hand, the quantity $1/(r_{\Sigma}^{2}H_{0}^{2})$ in the square
root of (\ref{jeth}) is approximately $2\times 10^{7}$ for a galaxy and $5\times 10^{4}$ for a cluster,
thus only this term remains in the square root to very high accuracy (better than $0.02\%$ for clusters
and better than $0.00005\%$ for galaxies). Therefore, the differential equation (\ref{jeth}) for
any practical reason is approximated by the simple equation
\begin{equation}
\frac{d\tilde{\psi}}{dz}=\frac{3^{\frac{1}{b}}b}{\xi\tilde{\gamma}^{\frac{1}{b}}}
(G_{\!N}H_{0}^{2})^{\frac{1}{b}}\frac{r_{\Sigma}a_{0}}{\sqrt{G_{\!N}}}\,
\frac{\tilde{\psi}^{1+\frac{1}{b}}}{(1\!+\!z)^{2}}\,.
\label{yheb}
\end{equation}
Integration of (\ref{yheb}) gives
\begin{equation}
\tilde{\psi}=\Big[\frac{3^{\frac{1}{b}}}{\xi\tilde{\gamma}^{\frac{1}{b}}}
(G_{\!N}H_{0}^{2})^{\frac{1}{b}}\frac{r_{\Sigma}a_{0}}{\sqrt{G_{\!N}}}\,\frac{1}{1\!+\!z}+c\Big]^{-b}\,,
\label{hyrm}
\end{equation}
where $c$ is integration constant. From the value of $\tilde{\psi}_{0}$ we find $c$, and finally we
have for the evolution of dark energy
\begin{equation}
\tilde{\psi}=\Big[\Omega_{DE,0}^{-\frac{1}{b}}-\frac{3^{\frac{1}{b}}}{\xi\tilde{\gamma}^{\frac{1}{b}}}
(G_{\!N}H_{0}^{2})^{\frac{1}{b}}\frac{r_{\Sigma}a_{0}}{\sqrt{G_{\!N}}}\,\frac{z}{1\!+\!z}
\Big]^{-b}\,.
\label{hyrt}
\end{equation}
The positiveness of $\psi$ implies that
\begin{equation}
z^{-1}>\Big(\frac{3\Omega_{DE,0}}{\xi^{b}\tilde{\gamma}}\Big)^{\!\frac{1}{b}}
(G_{\!N}H_{0}^{2})^{\frac{1}{b}}\frac{r_{\Sigma}a_{0}}{\sqrt{G_{\!N}}}-1\,.
\label{loey}
\end{equation}
As a result of the inequality (\ref{loey}) we get that
\begin{equation}
\xi^{b}\tilde{\gamma}>\frac{3\Omega_{DE,0}}{(1\!+\!z_{max}^{-1})^{b}}G_{\!N}H_{0}^{2}
\Big(\frac{r_{\Sigma}a_{0}}{\sqrt{G_{\!N}}}\Big)^{\!b}\,,
\label{lbyp}
\end{equation}
where $z_{max}=\mathcal{O}(1)$ is a redshift such that in the interval $(0,z_{max})$ the model
should definitely make sense.

For concreteness we write explicitly the Hubble evolution $H(z)$
\begin{equation}
\frac{H^{2}}{H_{0}^{2}}=\Omega_{m0}(1\!+\!z)^{3}
+\Big[\Omega_{DE,0}^{-\frac{1}{b}}-\frac{3^{\frac{1}{b}}}{\xi\tilde{\gamma}^{\frac{1}{b}}}
(G_{\!N}H_{0}^{2})^{\frac{1}{b}}\frac{r_{\Sigma}a_{0}}{\sqrt{G_{\!N}}}\,\frac{z}{1\!+\!z}\Big]^{-b}
+\Omega_{\kappa 0}(1\!+\!z)^{2}\,,
\label{kieb}
\end{equation}
which is extremely accurate for all relevant recent redshifts that our model is applicable.
The only unknown quantities in equation (\ref{kieb}) are $\xi^{b}\tilde{\gamma}$ and $b$.
Equation (\ref{kieb}) is a new Hubble evolution which can be tested against observations at the
background level. We note that the dark energy term in (\ref{kieb}) is quite different than the one
of equation (\ref{jrfd}). It is obvious from (\ref{kieb}) that $\Omega_{m0}+\Omega_{DE,0}
+\Omega_{\kappa 0}=1$. When $\xi^{b}\tilde{\gamma}$ is much larger than the right hand side of
(\ref{lbyp}), the dark energy of equation (\ref{kieb}) is approximately a cosmological constant.
The most interesting case is certainly when
$\xi^{b}\tilde{\gamma}\sim G_{\!N}H_{0}^{2}(\frac{r_{\Sigma}}{\sqrt{G_{\!N}}})^{b}$, and then,
all the terms in (\ref{kieb}) - with the exception of the spatial curvature $\Omega_{\kappa}$ - are
equally important and give a non-trivial dark energy evolution. In this case, various combinations of
$\tilde{\gamma},b$ are allowed such that (\ref{lbyp}) is satisfied, introducing in general new scales.
Moreover, due to the presence of the integration constant $c$ which arranges $\Omega_{DE,0}$, the
terse relation (\ref{keiw}) has now been replaced by the loose inequality (\ref{lbyp}), and
therefore the precise value of the quantity $\xi^{b}\tilde{\gamma}$ can be used in order to
accommodate some other observation, e.g. the acceleration. For a general $b$, although the dark energy
may sufficiently be attributed to the varying cosmological constant, the coincidence problem is
not particularly alleviated. However, what we find particularly interesting for the explanation of the
coincidence problem, as was discussed above, is the situation with $\tilde{\gamma}\sim 1$, because then,
no new mass scale is introduced for the explanation of the dark energy, other than the astrophysical
scales. This is our favorite case, where the relation (\ref{mutn}) is valid with the index $b$ having
values close to 2.1, as explained. Such values of $b$ are also theoretically interesting since they are
close to the IR fixed-point value $b=2$ of AS theory and it remains to AS to find if such values are
predicted at the astrophysical scales. For example, for a galaxy with $b=2.13$, inequality (\ref{lbyp})
provides that $\xi^{b}\tilde{\gamma}>2.2(1\!+\!z_{max}^{-1})^{-b}$, while for a cluster with $b=2.08$
it is $\xi^{b}\tilde{\gamma}>1.8(1\!+\!z_{max}^{-1})^{-b}$, relations which can easily be satisfied
with suitable $\tilde{\gamma}\sim 1$.

Combining equations
(\ref{kiet}), (\ref{pewr}) we obtain
\begin{equation}
G_{\!N}H_{0}^{2}\Big(\frac{D_{S,0}}{\sqrt{G_{\!N}}}\Big)^{b}=\frac{\xi^{b}\tilde{\gamma}}
{3\Omega_{DE,0}}\,.
\label{ngtd}
\end{equation}
For $\tilde{\gamma}$ such that inequality (\ref{lbyp}) is saturated, which means that the dark
energy in (\ref{kieb}) is non-trivial (different than a cosmological constant), we conclude from
(\ref{ngtd}) that the value of $D$ at the today Schucking surface is $D_{S,0}\sim r_{\Sigma}$.
This is true, either for our favorite $b$ or more generally. Note that $D_{S,0}$ is of the order of
$r_{\Sigma}$ and not approximately equal to $r_{\Sigma}$, as might be guessed initially. It is
obvious that the relation (\ref{keiw}) between the parameters has now disappeared in (\ref{ngtd}) due
to the freedom introduced by the presence of $D_{S,0}$. The precise value of $D_{S,0}$ is given by
(\ref{ngtd}) and depends on $\Omega_{DE,0}$. The quantity $D_{S}$ acts as an extra independent
cosmological field in the system (\ref{ksrg}), (\ref{kewr}), whose initial condition $D_{S,0}$ is set
today in agreement with the amount of measured dark energy. This initial condition measures the proper
distance of the current matching surface of the Swiss cheese model and the extra interesting thing,
which also sheds light to the coincidence problem, is that it is of the order of the radius of the
astrophysical structure. If it was not, then it would be just the integration constant of an extra
field that should be selected appropriately to create the measured dark energy, and therefore, would
introduce another scale. We can also find from (\ref{kiet}), (\ref{hyrt}) the expression of $D_{S}(z)$
to high accuracy
\begin{equation}
\frac{D_{S}}{r_{\Sigma}a_{0}}=\frac{\xi\tilde{\gamma}^{\frac{1}{b}}}{3^{\frac{1}{b}}
\Omega_{DE,0}^{\frac{1}{b}}}\,\frac{1}{(G_{\!N}H_{0}^{2})^{\frac{1}{b}}}\,\frac{\sqrt{G_{\!N}}}
{r_{\Sigma}a_{0}}-\frac{z}{1\!+\!z}\,.
\label{jerm}
\end{equation}
Equation (\ref{jerm}) reduced to (\ref{ngtd}) for $z=0$. For a non-trivial dark energy evolution in
(\ref{kieb}), the function $D_{S}$ remains for all relevant $z$ of the order of $r_{\Sigma}$.

In the interior of the object, the proper distance $D$ is a function of the position $R$, i.e. it
is $D(R)$, and the equation governing this evolution is (\ref{hteg}). Since $D_{S,0}$ is known,
the initial condition of equation (\ref{hteg}) is set at $R=R_{S,0}=r_{\Sigma}$ as
$D(r_{\Sigma})=D_{S,0}$. In terms of the normalized variables $\hat{D}=D/r_{\Sigma}$ and
$\hat{R}=R/r_{\Sigma}$, we have
\begin{equation}
\frac{d\hat{D}}{d\hat{R}}=\Bigg{\{}1-\frac{\xi^{b}\tilde{\gamma}}{3}\Big(\frac{\sqrt{G_{\!N}}}
{r_{\Sigma}}\Big)^{\!b-2}\frac{1}{\hat{D}_{S,0}^{b}}\,\Big[\frac{a_{0}^{3}\Omega_{m0}}{\Omega_{DE,0}}
\,\frac{1}{\hat{R}}+
\Big(\frac{\hat{D}_{S,0}}{\hat{D}}\Big)^{\!b}\hat{R}^{2}\Big]\Bigg{\}}^{\!-\frac{1}{2}}
\label{nyrb}
\end{equation}
with the initial condition set at $\hat{R}=1$ as $\hat{D}=\hat{D}_{S,0}=(\frac{\xi^{b}\tilde{\gamma}}
{3\Omega_{DE,0}})^{\frac{1}{b}}(G_{\!N}H_{0}^{2})^{-\frac{1}{b}}
\frac{\sqrt{G_{\!N}}}{r_{\Sigma}}$. We note that the fact that $D_{S}$ changes in time does not
mean that the interior solution is time-dependent. The interior solution is static and simply
$\hat{D}_{S,0}$ is used as initial condition for the integration of (\ref{nyrb}), since this is known
from the today amount of dark energy. Since for our favorite values of $\tilde{\gamma},b$ which validate
equation (\ref{mutn}), the factor $(\frac{\sqrt{G_{\!N}}}{r_{\Sigma}})^{b-2}$ is various orders of
magnitude smaller than one and $\hat{D}_{S,0}\sim 1$, the initial value of $\frac{d\hat{D}}{d\hat{R}}$
at the Schucking surface is one to high accuracy. Therefore, close to $r_{\Sigma}$, the function
$D(R)-R$ remains constant. We will see numerically that this is approximately true more
generally and find the integration constant.

Concerning the acceleration we find
\begin{equation}
\frac{\ddot{a}}{a}=-\frac{1}{2}\Omega_{m0}H_{0}^{2}(1\!+\!z)^{3}
+\psi-\frac{3^{\frac{1}{b}}b\psi^{1+\frac{1}{b}}}
{2\xi\gamma^{\frac{1}{b}}(1\!+\!z)\sqrt{\frac{1}{r_{\Sigma}^{2}a_{0}^{2}}
-\Omega_{m0}H_{0}^{2}(1\!+\!z)-\frac{\psi}{(1+z)^{2}}}}\,.
\label{kwtf}
\end{equation}
In terms of $\tilde{\psi}$ we have
\begin{equation}
\frac{\ddot{a}}{H_{0}^{2}a}=-\frac{1}{2}\Omega_{m0}(1\!+\!z)^{3}
+\tilde{\psi}-\frac{3^{\frac{1}{b}}b(G_{\!N}H_{0}^{2})^{\frac{1}{b}-\frac{1}{2}}
\tilde{\psi}^{1+\frac{1}{b}}}
{2\xi\tilde{\gamma}^{\frac{1}{b}}(1\!+\!z)\sqrt{\frac{1}{r_{\Sigma}^{2}a_{0}^{2}H_{0}^{2}}
-\Omega_{m0}(1\!+\!z)-\frac{\tilde{\psi}}{(1+z)^{2}}}}\,,
\label{lier}
\end{equation}
which is very well approximated, as before, by
\begin{equation}
\frac{\ddot{a}}{H_{0}^{2}a}=-\frac{1}{2}\Omega_{m0}(1\!+\!z)^{3}
+\tilde{\psi}-\frac{3^{\frac{1}{b}}b}
{2\xi\tilde{\gamma}^{\frac{1}{b}}} (G_{\!N}H_{0}^{2})^{\frac{1}{b}}\,\frac{r_{\Sigma}a_{0}}
{\sqrt{G_{\!N}}}\,\frac{\tilde{\psi}^{1+\frac{1}{b}}}{1\!+\!z}\,,
\label{ikne}
\end{equation}
or explicitly in terms of $z$ as
\begin{equation}
\frac{\ddot{a}}{H_{0}^{2}a}=-\frac{1}{2}\Omega_{m0}(1\!+\!z)^{3}+\Big[\Omega_{DE,0}^{-\frac{1}{b}}
-\frac{3^{\frac{1}{b}}}{2\xi\tilde{\gamma}^{\frac{1}{b}}}(G_{\!N}H_{0}^{2})^{\frac{1}{b}}
\frac{r_{\Sigma}a_{0}}{\sqrt{G_{\!N}}}\frac{b\!+\!2z}{1\!+\!z}\Big]
\Big[\Omega_{DE,0}^{-\frac{1}{b}}
-\frac{3^{\frac{1}{b}}}{\xi\tilde{\gamma}^{\frac{1}{b}}}(G_{\!N}H_{0}^{2})^{\frac{1}{b}}
\frac{r_{\Sigma}a_{0}}{\sqrt{G_{\!N}}}\frac{z}{1\!+\!z}\Big]^{-1-b}\,.
\label{ther}
\end{equation}
The transition redshift $z_{t}$ from deceleration to acceleration is found from (\ref{ther}) setting
$\ddot{a}=0$. The today deceleration parameter takes the following very accurate expression if we use
equation
(\ref{ther})
\begin{equation}
q_{0}=\frac{1}{2}\Omega_{m0}-\Omega_{DE,0}+\frac{3^{\frac{1}{b}}b}
{2\xi\tilde{\gamma}^{\frac{1}{b}}}\Omega_{DE,0}^{1+\frac{1}{b}}
(G_{\!N}H_{0}^{2})^{\frac{1}{b}}\frac{r_{\Sigma}a_{0}}{\sqrt{G_{\!N}}}\,,
\label{tbem}
\end{equation}
and therefore, $q_{0}$ takes values larger than the $\Lambda$CDM value $-0.55$.
From (\ref{kwtf}) or (\ref{lier}), the condition $\ddot{a}|_{0}>0$ is written as
\begin{equation}
\xi^{b}\gamma>\frac{3b^{b}\,\Omega_{DE,0}^{1+b}}
{(2\Omega_{DE,0}\!-\!\Omega_{m0})^{b}}H_{0}^{2-b}\Big(\frac{1}
{r_{\Sigma}^{2}a_{0}^{2}H_{0}^{2}}-1+\Omega_{\kappa 0}\Big)^{\!\!-\frac{b}{2}}
\label{keur}
\end{equation}
and is very well approximated by
\begin{equation}
\xi^{b}\tilde{\gamma}>\frac{3b^{b}\,\Omega_{DE,0}^{1+b}}
{(2\Omega_{DE,0}\!-\!\Omega_{m0})^{b}}\,G_{\!N}H_{0}^{2}\Big(\frac{r_{\Sigma}a_{0}}
{\sqrt{G_{\!N}}}\Big)^{\!b}\,,
\label{iwrf}
\end{equation}
which also arises from (\ref{tbem}). The inequality (\ref{iwrf}) can be seen that is sufficient
in order to have the significant condition $\dot{\Omega}_{m}|_{0}<0$. For our favorite
values of $\tilde{\gamma},b$ which validate equation (\ref{mutn}), the right hand side of
(\ref{iwrf}) is of order one and $\xi^{b}\tilde{\gamma}$ should be chosen accordingly, still being of
order unity. For example, for galaxies with $b=2.13$ we take from (\ref{iwrf}) that
$\xi^{b}\tilde{\gamma}>4.2$, while for clusters with $b=2.08$ it should be $\xi^{b}\tilde{\gamma}>3.2$.
These conditions are stronger that those implied by (\ref{lbyp}) for any $z_{max}$, so (\ref{lbyp})
can be forgotten. For these values of $b$ we provide now some indicative numerical results for
$q_{0}$ from equation (\ref{tbem}) and $z_{t}$.
For galaxies with $b=2.13$ we have: for $\xi=9$, $\tilde{\gamma}=5$ it is $z_{t}=0.68$ and $q_{0}=-0.49$;
for $\xi=3$, $\tilde{\gamma}=2$ it is $z_{t}=0.69$ and $q_{0}=-0.29$;
for $\xi=1$, $\tilde{\gamma}=6$ it is $z_{t}=0.51$ and $q_{0}=-0.09$.
For clusters with $b=2.08$ we have: for $\xi=9$, $\tilde{\gamma}=5$ it is $z_{t}=0.68$ and
$q_{0}=-0.50$; for $\xi=3$, $\tilde{\gamma}=2$ it is $z_{t}=0.69$ and $q_{0}=-0.32$;
for $\xi=1$, $\tilde{\gamma}=6$ it is $z_{t}=0.60$ and $q_{0}=-0.14$.
In all these cases, the inequality (\ref{loey}) does not provide any restriction on $z$.
If $b$ increases more than $2\%$ relatively to the above indicative values, then the quantity
$\xi^{b}\tilde{\gamma}$ moves to higher orders of magnitude to assure acceleration, while
if $b$ gets values smaller than the above indicative, the inequality (\ref{iwrf}) is easily
satisfied since its right hand side becomes suppressed.

Finally, the dark energy pressure and its equation-of-state parameter are
\begin{eqnarray}
&&8\pi G_{\!N}p_{DE}=-3\psi+\frac{3^{\frac{1}{b}}b\psi^{1+\frac{1}{b}}}
{\xi\gamma^{\frac{1}{b}}(1\!+\!z)\sqrt{\frac{1}{r_{\Sigma}^{2}a_{0}^{2}}
-\Omega_{m0}H_{0}^{2}(1\!+\!z)-\frac{\psi}{(1+z)^{2}}}}\\
\label{geyh}
&&w_{DE}=-1+\frac{3^{\frac{1}{b}-1}b\psi^{\frac{1}{b}}}
{\xi\gamma^{\frac{1}{b}}(1\!+\!z)\sqrt{\frac{1}{r_{\Sigma}^{2}a_{0}^{2}}
-\Omega_{m0}H_{0}^{2}(1\!+\!z)-\frac{\psi}{(1+z)^{2}}}}\,.
\label{heif}
\end{eqnarray}
In terms of $\tilde{\psi}$ we have
\begin{eqnarray}
&&\frac{8\pi G_{\!N}}{H_{0}^{2}}p_{DE}=
-3\tilde{\psi}+\frac{3^{\frac{1}{b}}b(G_{\!N}H_{0}^{2})^{\frac{1}{b}-\frac{1}{2}}
\tilde{\psi}^{1+\frac{1}{b}}}
{\xi\tilde{\gamma}^{\frac{1}{b}}(1\!+\!z)\sqrt{\frac{1}{r_{\Sigma}^{2}a_{0}^{2}H_{0}^{2}}
-\Omega_{m0}(1\!+\!z)-\frac{\tilde{\psi}}{(1+z)^{2}}}}\label{bydp}\\
&&w_{DE}=-1+\frac{3^{\frac{1}{b}-1}b(G_{\!N}H_{0}^{2})^{\frac{1}{b}-\frac{1}{2}}
\tilde{\psi}^{\frac{1}{b}}}
{\xi\tilde{\gamma}^{\frac{1}{b}}(1\!+\!z)\sqrt{\frac{1}{r_{\Sigma}^{2}a_{0}^{2}H_{0}^{2}}
-\Omega_{m0}(1\!+\!z)-\frac{\tilde{\psi}}{(1+z)^{2}}}}\,,
\label{tbmo}
\end{eqnarray}
which are very well approximated, as before, by
\begin{eqnarray}
&&\frac{8\pi G_{\!N}}{H_{0}^{2}}p_{DE}=
-3\tilde{\psi}+\frac{3^{\frac{1}{b}}b}
{\xi\tilde{\gamma}^{\frac{1}{b}}} (G_{\!N}H_{0}^{2})^{\frac{1}{b}}\,\frac{r_{\Sigma}a_{0}}
{\sqrt{G_{\!N}}}\,\frac{\tilde{\psi}^{1+\frac{1}{b}}}{1\!+\!z}\label{klov}\\
&&w_{DE}=-1+\frac{3^{\frac{1}{b}-1}b}
{\xi\tilde{\gamma}^{\frac{1}{b}}} (G_{\!N}H_{0}^{2})^{\frac{1}{b}}\,\frac{r_{\Sigma}a_{0}}
{\sqrt{G_{\!N}}}\,\frac{\tilde{\psi}^{\frac{1}{b}}}{1\!+\!z}\,,
\label{beot}
\end{eqnarray}
or explicitly in terms of $z$ as
\begin{eqnarray}
&&\frac{8\pi G_{\!N}}{H_{0}^{2}}p_{DE}=
\Big[\frac{3^{\frac{1}{b}}}{\xi\tilde{\gamma}^{\frac{1}{b}}}(G_{\!N}H_{0}^{2})^{\frac{1}{b}}
\frac{r_{\Sigma}a_{0}}{\sqrt{G_{\!N}}}\frac{b\!+\!3z}{1\!+\!z}-3\Omega_{DE,0}^{-\frac{1}{b}}\Big]
\Big[\Omega_{DE,0}^{-\frac{1}{b}}
-\frac{3^{\frac{1}{b}}}{\xi\tilde{\gamma}^{\frac{1}{b}}}(G_{\!N}H_{0}^{2})^{\frac{1}{b}}
\frac{r_{\Sigma}a_{0}}{\sqrt{G_{\!N}}}\frac{z}{1\!+\!z}\Big]^{-1-b}\label{pmgu}\\
&&w_{DE}=\Big[\frac{3^{\frac{1}{b}-1}}{\xi\tilde{\gamma}^{\frac{1}{b}}}(G_{\!N}H_{0}^{2})^{\frac{1}{b}}
\frac{r_{\Sigma}a_{0}}{\sqrt{G_{\!N}}}\frac{b\!+\!3z}{1\!+\!z}-\Omega_{DE,0}^{-\frac{1}{b}}\Big]
\Big[\Omega_{DE,0}^{-\frac{1}{b}}
-\frac{3^{\frac{1}{b}}}{\xi\tilde{\gamma}^{\frac{1}{b}}}(G_{\!N}H_{0}^{2})^{\frac{1}{b}}
\frac{r_{\Sigma}a_{0}}{\sqrt{G_{\!N}}}\frac{z}{1\!+\!z}\Big]^{-1}\,.
\label{gbwp}
\end{eqnarray}
The present-day dark energy equation-of-state parameter takes the following very accurate expression
if we use equation (\ref{gbwp})
\begin{equation}
w_{DE,0}= -1+\frac{3^{\frac{1}{b}-1}b}
{\xi\tilde{\gamma}^{\frac{1}{b}}}\Omega_{DE,0}^{\frac{1}{b}}
(G_{\!N}H_{0}^{2})^{\frac{1}{b}}\frac{r_{\Sigma}a_{0}}{\sqrt{G_{\!N}}}\,,
\label{ghkp}
\end{equation}
and therefore, $w_{DE,0}$ takes values larger than the $\Lambda$CDM value $-1$ (phantom values
smaller than -1 can only be obtained if $b<0$). According to (\ref{iwrf}), it can take any value
$w_{DE,0}<-\frac{1}{3}(1+\frac{\Omega_{m0}}{\Omega_{DE,0}})\approx -0.48$.
In our typical example, using $b=2.13$ at a galaxy structure, for $\xi=9$, $\tilde{\gamma}=5$ it is
$w_{DE,0}=-0.95$, for $\xi=3$, $\tilde{\gamma}=2$ it is $w_{DE,0}= -0.75$, while for $\xi=1$,
$\tilde{\gamma}=6$ it is $w_{DE,0}= -0.56$. Similarly, using $b=2.08$ at a cluster structure, for
$\xi=9$, $\tilde{\gamma}=5$ it is $w_{DE,0}=-0.95$, for $\xi=3$, $\tilde{\gamma}=2$ it is
$w_{DE,0}= -0.78$, while for $\xi=1$, $\tilde{\gamma}=6$ it is $w_{DE,0}= -0.61$.

\begin{figure}[ht]
\begin{center}
\includegraphics[width=0.85\textwidth]{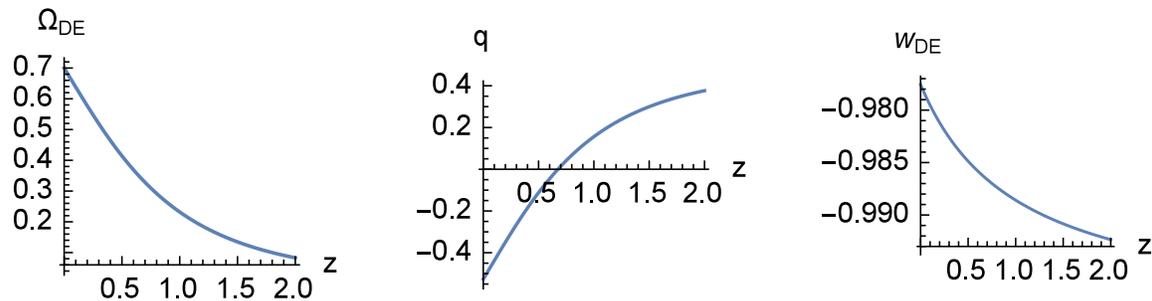}
\caption{{\it{The late-times cosmological evolution for a spatially flat
universe, for the parameter choice $b=2.06$, $\xi=5$, $\tilde{\gamma}=5$,
$M=10^{15}M_{\odot}$ and $\Omega_{m0}=0.3$.
In the left graph we depict the evolution of the
dark energy density parameter as a function of the redshift $z$. In the middle graph we
present the evolution of the deceleration parameter.
Finally, the dark energy equation-of-state
parameter is depicted in the right graph.}}} \label{basicplot3}
\end{center}
\end{figure}

To capture the results of this case, we present in Fig. \ref{basicplot3} the cosmological evolution
as a function of $z$ for a spatially flat universe, with the quantum cosmological constant originated
at the cluster level with $M=10^{15}M_{\odot}$, $\Omega_{m0}=0.3$, and for the parameter choice
$b=2.06$, $\xi=5$, $\tilde{\gamma}=5$, which provide $z_{t}=0.67$, $q_{0}=-0.52$, $w_{DE,0}=-0.978$.
Of course, these are just indicative values of the parameters, which however look successful at
least at first glance (in a forthcoming paper \cite{fitting}, we will perform a numerical
analysis at the background level with data from SNIa, $H(z)$ measurements, BAO, e.tc., where we
can say in advance that the fittings show excellent accuracy).
We depict in the left graph the evolution of the dark energy density
parameter $\Omega_{DE}$ from equation (\ref{munf}), where a typical decreasing behaviour for larger
redshifts is shown. In the middle graph the evolution of the deceleration parameter $q$ is shown,
where the passage from deceleration to acceleration at late times can be seen.
The dark energy equation-of-state parameter $w_{DE}$ in the right graph shows a non-constant
evolution with a today value $w_{DE,0}$ close to -1. We also note that if we solve numerically
the differential equation (\ref{wskt}), instead of using the analytical expression (\ref{hyrt}), the
results are exactly the same due to the high degree of accuracy of our analytical expressions.
Notice that the two reasons mentioned in the end of subsection VII-B-1, about the time variability
of $b$ and the further decrease of $w_{DE}$, are also valid here. Namely, a more realistic
study of the RG flow inside the structure, together with a model of structure formation on
one side, and the study of more realistic inhomogeneous/anisotropic cosmological models through
averaging processes on the other side, would reduce $w_{DE}$, possibly even to phantom values.

Because we are taking our scenario seriously, we would like to finish with an estimate of the
potentials and the forces inside and at the border of an astrophysical object.
As for the border, the situation is clear and precise. Since the mass of the structure
can be considered as being gathered at the origin, the Schwarzschild term is exact and
provides the Newtonian force at the border. If this force and its potential are dominant compared to
the new cosmological constant term at the border (which will indeed be the case), this will already
be an indication that inside the object the interior dynamics will also not be severely disturbed.
More precisely, in the interior of the object, either galaxy or cluster, there is a profile
of the matter distribution (luminous, gas, dark matter) which will give some deviations from the
central $1/R$ potential. We will first study what is the situation in Milky Way, which is a well
studied galaxy. A very good fit of dark matter halo profile (which is the dominant matter component)
to available data for Milky Way is performed \cite{Nesti:2013uwa} using the Universal Rotation
Curve (Burkert) profile with matter density $\varrho(R)=\varrho_{c}\big(1\!+\!\frac{R}{R_{c}}\big)^{-1}
\big[1\!+\!\big(\frac{R}{R_{c}}\big)^{2}\big]^{-1}$, where the density scale
$\varrho_{c}\sim 4\times 10^{7}M_{\odot}/\text{kpc}^{3}$ and the radius scale $R_{c}\sim 10\text{kpc}$.
The use of a matter profile is necessary in order to get precise values of the Newtonian forces and
velocities inside the galaxy ($R<R_{b}$). It can be easily seen that in the inner regions of the galaxy
($R\lesssim 0.4 R_{b}$), the real amount of matter is much larger than the matter predicted by a
constant energy density profile, so this constant profile gives a poor underestimation of the Newtonian
force. On the other hand, an interesting result is that away from the very center
of the galaxy ($R\gtrsim 0.1 R_{b}$) the real Newtonian force is of the same order as the Newtonian
force due to the idealized picture with all the galaxy mass gathered at the origin. This result of
Milky Way indicates that, although we will not enter the complicated discussion to study the precise
Newtonian force inside other galaxies or clusters, this force will be estimated by the central
$1/R^{2}$ force. This way, we will be in position to compare the varying cosmological constant force
relatively to the Newtonian force, and see how long the new force does not give obvious
inconsistencies with internal dynamics. Of course, a more detailed study and
comparison to existing or upcoming data is necessary.

With the explanations of the previous paragraph, we assume that the gravitational field given by the
modified Schwarzschild metric (\ref{ASBH}) in the interior of the object (far from its very center,
e.g. for $R\gtrsim 0.1 R_{b}$) provides a sufficient estimate to anticipate what are the interior
Newtonian and cosmological constant forces. We use the indicative parameters $b=2.08$, $\xi=1$,
$\tilde{\gamma}=6$ to numerically integrate equation (\ref{nyrb}). The solution of this equation
gives the function $D(R)$, which turns out to be an increasing function of $R$, as it should be.
The first observation is that the function $D(R)$ in all the relevant distances, throughout the
interior of the astrophysical object up to the today Schucking radius $R_{S,0}=r_{\Sigma}$, has values
of the order of $r_{\Sigma}$. Therefore, not only $D_{S}(z)$ remains of order $r_{\Sigma}$, but also
the whole $D(R)$ is so, and provides a variable with values natural to the dimension of the object,
without acquiring unnatural very large or very small values. In general, at distances up to a few
decades of $r_{\Sigma}$, a reliable approximation for the solution, as it is seen numerically, is
$\hat{D}\approx \hat{R}+0.79$, thus $D$ has approximately a constant difference from $R$. Of course,
this expression of $\hat{D}$ can be used to obtain some intuition, but the detailed structure of
$\hat{D}$ could also be significant elsewhere.
The next important thing is to study the ratio of the potential term due to the
varying cosmological constant to the Newtonian potential, as they arise from equation (\ref{nyrb}).
For small clusters, this ratio is less than 1\textperthousand \, either at the border or inside.
For the largest possible clusters, the ratio becomes $15\%$ at the border and less inside. Therefore,
for the latest clusters, we have a non-negligible contribution to the pure Newtonian potential with
possible observable signatures, still without an obvious inconsistency. Since the Newtonian potential
is negligible relatively to unity for $R\gtrsim 0.1 R_{b}$ up to the border, thus both potentials
are very weak and $F$ is approximately one to high accuracy. For all clusters,
with any diameter, the two potentials become of the same order at the Schucking radius and this was
the reason above for the successful explanation of dark energy. At even larger distances
($R\gtrsim 3 r_{\Sigma}$) the cosmological constant becomes the dominant term
(although in the Swiss-cheese model, at such distances the cosmological patch is present instead of
the static one). Finally, the ratio of the cosmological constant force to the Newtonian force
is $\frac{\Omega_{DE,0}}{a_{0}^{3}\Omega_{m0}}\big(b\frac{\hat{R}}{\hat{D}\sqrt{F}}\!-\!2\big)
\big(\frac{\hat{D}_{S,0}}{\hat{D}}\big)^{\!b}\hat{R}^{3}$.
It turns out that this ratio is negative, which means that the new force is repulsive, as expected.
For small clusters, this ratio is again less than 1\textperthousand \, either at the border or inside.
For the largest possible clusters, the ratio becomes $20\%$ at the border and less inside. Again,
for such clusters, a non-negligible contribution to the Newtonian force arises, which should be
studied more thoroughly in comparison with real data. Similar results with all the above occur for
clusters with different values of the parameters $\xi,\tilde{\gamma}$, where most usually the extra
force and potential are further suppressed. As for galaxies, it can be seen that the force or the
potential of the extra term is always restricted to a contribution of a few percent, independently of
the parameters.

Some remarks are in order about what may happen with the formation of large scale structures of the
universe in the context of our model. It is well known that modifications of gravity that are
almost $\Lambda$CDM at the background level may have very different evolution of structures.
In our present understanding of the theory, we cannot analyse the whole structure evolution because
the scenario starts somehow suddenly with the appearance of the structures. The present work provides
a study only at small redshifts. If in the future we manage to understand better the behavior of the
RG flow at different energies and scales, we will able to quantify more accurately the antigravity
effects during the formation and the evolution of structures. This knowledge will require
from AS theory the correct RG flow of gravity with matter at the infrared energy scales and also the
exact relation of the amount of energy/mass that is associated with the value of the varying
cosmological constant. However, we foresee that General Relativity will be preserved somehow
accurately during structure formation because, in the initial clouds that collapse, the antigravity
quantum effects are not expected to be very important and they become significant recently where
structures are denser and smaller. But this, of course, must be studied with an exact RG flow.

\section{Discussion and Conclusions}

We have proposed that the dark energy and the recent cosmic acceleration can be the result of the
existence of local antigravity sources associated with astrophysical matter configurations distributed
throughout the universe. This is a tempting proposal in relation to the coincidence problem since
in that case the dark energy naturally emanates from the recent formation of structure.

The cosmic evolution can arise through some interrelation between the local and the cosmic patches.
In the present work we have assumed a Swiss cheese model to derive the cosmological equations, where
the interior spherically symmetric metric matches smoothly to an FRW exterior across a spherical
boundary. This Schucking surface has a fixed coordinate radius in the cosmic frame, but expands with
time in the local frame.

Various gravitational theories can be implemented in the above context and see if the
corresponding intermediate distance infrared phenomena can provide the necessary cosmic acceleration.
This is not always an easy task since the appropriate spherically symmetric solutions should be
used along with the correct matching conditions. Our main concern in this work, in order to test
our proposal, is to consider quantum modified spherically symmetric metrics, and more precisely,
quantum improved Schwarzschild-de Sitter metrics, which are used for modeling the metric of galaxies
or clusters of galaxies.

Asymptotically safe (AS) gravity provides specific forms for the quantum
corrections of the cosmological and Newton's constants depending on the energy scale. In the far
infrared (IR) regime of AS evolution, which certainly corresponds to the cosmological scales, there
are encouraging indications for the existence of a fixed point of a specific form. In the
intermediate infrared scales of our astrophysical objects, it is therefore quite reasonable that
some small deviations from this IR law occur, and on this behaviour our most successful model was built
containing the appropriate antigravity effect.
This model uses dimensionless order one parameters of AS, the Newton's constant and the astrophysical
length scale (which enters through the Schucking radius of matching) and provides a recent dark energy
comparable to dark matter. At the same time, sufficient cosmic acceleration emerges at small redshifts,
while the freedom of the order one parameters has to be constrained by observational data in the future.
To the best of our knowledge, this is the first solution of the dark energy problem without using
fine-tuning or introducing add-hoc energy scales.
Although this cosmology, given by equation (\ref{kieb}), has a quite different functional form than
the $\Lambda$CDM cosmology, the modified Schwarzschild-de Sitter interior metric allows us to
interpret the encountered quantity (\ref{jurd}) as giving approximately the correct order of
magnitude of the standard cosmological constant $\Lambda$, which can thus be considered as a composite
quantity.

As a more technical point concerning the above cosmology, there appears, from the relation of the
AS energy scale with a length scale, a coupled geometrical field with its own equation of motion,
which is identified as the proper distance. The integration constant of this field, which is the
proper distance at the today Schucking radius, arranges the precise amount of dark energy, and since
its value is of the order of the length of the astrophysical object, yet no new scale is introduced
from this stage. Finally, we have presented a crude estimation of the antigravity effects in the
interior of the structure and it appears that they stay at sufficiently small values, so that not
to create obvious conflicts with the local dynamics of the object. This is an interesting issue
that deserves a more thorough investigation.

As a future work, it is worth investigating the same scenario with inhomogeneous/anisotropic Swiss
cheese models. The present work uses the simplest Swiss cheese model as a first simple approach.
It is necessary to begin with it in order to isolate the magnitude of the effects of antigravity
sources. Inhomogeneous Swiss cheese type models are certainly more realistic, and we expect that
they will enhance the produced effective amount of acceleration. Moreover, the evolution of structure
will add a more refined picture of the passage from the deceleration to the acceleration regime.

\begin{acknowledgments}

We wish to thank E.V. Linder for useful discussions. V.Z. acknowledges the hospitality of
Nazarbayev university and S.P.G.

\end{acknowledgments}

\begin{appendix}

\section{}
\label{Appendix A}

\subsection{Asymptotically safe gravity : Theory}

The elusive theory of quantum gravity is associated not only with mathematical challenges but also
with many conceptual problems, including a measurements scheme and several epistemological issues.
This is something expected since quantum gravity will most probably provide the Theory of Everything,
a model that most certainly will include revolutionary new mathematical and physical concepts.
Nevertheless, some part of the scientific community last decades has been focused, as a first,
on attempts to propose solutions to the well-known result that the quantization of the
Einstein-Hilbert action leads to a quantum field theory which is perturbatively non-renormalizable
\cite{Weinberg:2009bg}.

In general, a mathematical modeling of a system is greatly simplified if one allows for more
parameters, more dimensions or more symmetries. Remarkably, there is one serious attempt of
quantum gravity that works in four dimensions using only the symmetries of conventional
quantum field theory and of General Relativity. An effective quantum field theory of General
Relativity can give answers about the calculations of amplitudes at energy scales below the
Planck scale. This is a result of the fact that higher-derivative terms are suppressed by powers
of the Planck mass. However, for energies close or larger than Planckian scales, the effective
theory requires a fixing of an infinite number of free coupling constants from experimental input.
This equivalently means that at every loop more experiments must be performed and this finally leads
to loss of predictability.

Asymptotic Safety (AS) exists in the space of theories that includes the corresponding effective
field theory. The AS program recovers ``predictivity'' by imposing the demand/principle that the
physically accepted quantum theory is located within the ultraviolet (UV) critical hypersurface
of a Renormalization Group (RG) fixed point that is called Non-Gaussian fixed point (NGFP).
The existence of the latter point guarantees that the UV description of the theory furnishes all
dimensionless coupling constants to be finite. Now, determining the trajectory uniquely (which
means to pick up a specific universe RG flow) requires a number of experimental input
parameters equal to the dimensionality of the hypersurface. This has been proved, under some
simplified approximations, that it is indeed possible and there is the NGFP where a trajectory
begins and generates General Relativity at low energy
\cite{Reuter:1996cp,Niedermaier:2006wt,Codello:2008vh,Litim:2011cp,
Percacci:2011fr,Reuter:2012id,Reuter:2012xf}.
Approximations of the gravitational RG flow can be carried out with the help of the functional
renormalization group equation (FRGE) \cite{Reuter:1996cp}
\begin{equation}
\label{FRGE}
\partial_k \Gamma_k[g,\bar{g}]=\frac{1}{2}{\rm Tr}\left[
\big(\Gamma_k^{(2)}+\mathcal{R}_k \big)^{-1} \partial_k \mathcal{R}_k \right]\,,
\end{equation}
regarding the effective average gravity action $\Gamma_k$, where $\Gamma_k^{(2)}$, $\bar{g}_{\mu\nu}$
and $\mathcal{R}_k$ are defined in the context of the background field formalism. This methodology
splits the metric $g_{\mu\nu}$ into a fixed background $\bar{g}_{\mu\nu}$ and fluctuations
$h_{\mu\nu}$. The quantity $\Gamma_k^{(2)}$ is the second order functional derivative of $\Gamma_k$
with respect to the fluctuation field $h_{\mu\nu}$ and $\mathcal{R}_k$ gives a scale-dependent mass
term for fluctuations with momenta $p^2 \ll k^2$, where the RG scale $k$ is constructed from the
background metric. This RG equation implements Wilson's idea which suggests integrating out momenta
$p^2 \ll k^2$, i.e. small fluctuations. In this way, $\Gamma_k$ provides an effective description of
the system for scales $k^2$. Remarkably, this is a background independent method
\cite{Benedetti:2010nr}.

The simplest estimation of the RG flow, concerning gravity field, arises after projecting the
FRGE onto the gravity action approximated by the following $\Gamma_k$
\begin{equation}
\Gamma_k=\frac{1}{16\pi G_k} \int d^4x\,\sqrt{|g|}\,\big(\!-R+2\Lambda_k \big)\,,
\end{equation}
where gauge fixing and ghost terms are of course included. This approximation includes two
energy-dependent couplings, the Newton's constant $G_k$ and the cosmological constant $\Lambda_k$.
For convenience we define their dimensionless counterparts
\begin{equation}
g_k \equiv k^2\,G_k\,\,\,\,,\,\,\,\,\lambda_k \equiv k^{-2}\Lambda_k\,,
\label{dimless}
\end{equation}
which should respect the beta functions.

In the absence of knowledge of the real functional scale dependence of $g_{k}, \lambda_{k}$, it is
not clear what is the correct trajectory in the space of $g_{k},\lambda_{k}$ that was followed by
the universe. In other words, we do not really know yet the detailed path along which the classical
General Relativity regime at the present-day epoch with a constant $G_{\!N}$ and negligible $\Lambda$
can be obtained. In the transplanckian regime the NGFP is present \cite{Reuter:2009kq} and the
behaviour of the couplings near this point is given by constant values,
$g_{k}=g_{\ast},\lambda_{k}=\lambda_{\ast}$, so in
the deep ultraviolet ($k\rightarrow \infty$), $G$ approaches zero and $\Lambda$ diverges.
There is another fixed point, the Gaussian fixed point (GFP) \cite{Reuter:2001ag}, which is saddle and
is located at $g=\lambda=0$. In the linear regime of the GFP, where the dimensionless couplings are
pretty small, the analysis predicts that $G$ is approximately constant, while
$\Lambda$ displays a running proportional to $k^{4}$. To the other edge of the far infrared limit
($k \rightarrow 0$) \cite{Bonanno:2001hi}, the behaviour of the RG flow
trajectories with positive $G,\Lambda$ is not so well understood since the approximation breaks
down (divergence of beta functions) when $\lambda_{k}$ approaches $1/2$ at a non-zero $k$, where
an unbounded growth of $G$ appears
together with a vanishingly small $\Lambda$ (interestingly enough, this happens near $k=H_{0}$).
The exact value of the current $\Lambda$ is unknown due to this break down.

\subsection{Asymptotically safe gravity : Cosmology}

The framework of AS in principle describes a modified gravitational force at all length scales,
something that makes the cosmological model building feasible \cite{Bonanno:2001xi},
\cite{Bonanno:2001hi}, \cite{Reuter:2005kb}-\cite{Bonanno:2008xp} (see also review
\cite{Reuter:2012xf}). Consequently, phenomenological studies that use cosmological data can
constraint the various free parameters appearing in AS, including the values of the cosmological
constant and Newton's constant. The various research studies that appear in the literature incorporate
the AS property of energy-dependent couplings in two ways.

In the first approach, the scale laws of the couplings are taken from the rigor RG computation of AS
close to the NGFP or GFP, or at some infrared range. Then, either these laws are incorporated in
General Relativity solutions or they are included in properly modified Einstein equations that respect
Bianchi identities. In this approach, there is the advantage of using rigorous and trusted
results from RG studies of AS. However, this approach is supposed to concentrate on the study of a
relatively restricted energy scale range, i.e. for big bang regime or for infrared regime, and
typically is not used for the description of the whole cosmological evolution.

In the second approach, RG improved techniques are used either in the equations of motion or in
the machinery of the  effective average action. The models are not implemented at the same level of
rigor as the full RG flow studies forming the core of AS. However, they allow for the construction
of interesting cosmological scenarios with extended cosmological evolution.

It is common in the AS literature to set $G$ and $\Lambda$ as functions of the energy $k$ in the
existing solutions of Einstein equations in order to improve their behaviour. The simple input
of $G(k)$ and $\Lambda(k)$ into the classical vacuum equations results to violation of the Bianchi
identities, while this same input into a classical solution creates a metric which is not solution
of a well-defined theory. In \cite{bianchi}, the formalism of obtaining RG improved solutions that
respect Bianchi identities was presented at the action level. In \cite{Kofinas:2015sna}, an
alternative and mathematically more solvable approach was developed at the level of equations of
motion, consistent with Bianchi identities, where the appropriate covariant kinetic terms that
support an arbitrary source field $\Lambda(k)$ was included without any symmetry assumption.

Many AS cosmological studies have analyzed the early cosmological evolution or the dark energy
problem and it was even possible to propose solutions to the cosmic entropy issue
\cite{Bonanno:2008xp}. Of particular interest are studies where ``RG improved'' cosmologies admit
exponential or power-law inflationary solutions \cite{Bonanno:2007wg}. The initial vacuum state of
cosmos is characterized by an energy-dependent cosmological constant, and subsequently, Einstein
equations, modified according to AS, include a non-zero matter energy-momentum tensor with an
energy-dependent Newton's constant (matter is expected to appear due to energy transfer from vacuum
to matter fields). Both these $\Lambda$ and $G$ respect the energy dependence that is predicted in
the context of AS at the NGFP. In \cite{Kofinas:2016lcz}, extending the formalism presented in
\cite{Kofinas:2015sna} beyond the vacuum case to also include matter, quantum gravity inspired
modified Einstein equations were realized, capable to describe both absence of matter cases and
configurations with matter contributions. There are also studies discussing the singularity problem
\cite{Bonanno:2017gji} or the assumption that the universe had a quantum vacuum birth
\cite{Kofinas:2015sna}.

An important question is the association of the RG scale parameter $k$ to the cosmological
time or proper length, in order for the model to be reasonable. First works have chosen the RG
scale to be inversely proportional to cosmological time \cite{Bonanno:2001xi}, while later, the more
popular connection with the Hubble scale was developed. In some other works, the RG scale is linked
with the plasma temperature or with the fourth root of the energy density \cite{Guberina:2002wt},
the cosmological event/particle horizons \cite{Bauer:2005rpa}, or curvature invariants like Ricci
scalar \cite{Frolov:2011ys}-\cite{Copeland:2013vva}.

\section{}
\label{Appendix B}

Here we perform an analytical study of section \ref{kiee} concerning the
Gaussian fixed point. Because of the extreme fine-tuning in $\alpha$, numerical study of the
equations of section \ref{kiee} is not possible. For $\kappa=0$ we define the quantities
\begin{equation}
\tilde{\chi}=\frac{\chi-(1\!-\!\Omega_{m0})H_{0}^{2}}{(1\!-\!\Omega_{m0})H_{0}^{2}-\frac{\alpha}{3}}
\,\,\,\,\,\,\,\,\,\,,\,\,\,\,\,\,\,\,\,\,\tilde{\alpha}=\frac{\alpha}{H_{0}^{2}}>0\,.
\label{lsur}
\end{equation}
Note that $\tilde{\chi}+1>0$.
From (\ref{poew}), the dimensionless quantity $\frac{\tilde{\alpha}}{3}$ is fine-tuned very close to
$1-\Omega_{m0}$, i.e. $0<1-\Omega_{m0}-\frac{\tilde{\alpha}}{3}\lesssim 10^{-22}$.
From (\ref{bdet}) it is $\tilde{\chi}_{0}=0$, while equation (\ref{sery}) becomes
\begin{equation}
\frac{d\tilde{\chi}}{dz}=\frac{4\zeta(\tilde{\chi}\!+\!1)^{\frac{5}{4}}}{(1\!+\!z)^{2}\sqrt{1\!-\!
\Big[\frac{\tilde{\alpha}}{3}\!+\!\Omega_{m0}(1\!+\!z)^{3}\!+\!(1\!-\!\Omega_{m0}\!-\!
\frac{\tilde{\alpha}}{3})(\tilde{\chi}\!+\!1)\Big]\frac{r_{\Sigma}^{2}a_{0}^{2}
H_{0}^{2}}{(1+z)^{2}}}}\,,
\label{wrfs}
\end{equation}
where
\begin{equation}
\zeta\equiv\frac{3^{\frac{1}{4}}}{\xi\nu^{\frac{1}{4}}}
\Big(\frac{r_{\Sigma}^{2}a_{0}^{2}H_{0}^{2}}{H_{0}\sqrt{G_{\!N}}}\Big)^{\!\frac{1}{2}}
\Big(1\!-\!\Omega_{m0}\!-\!\frac{\tilde{\alpha}}{3}\Big)^{\!\frac{1}{4}}\,.
\label{jwrw}
\end{equation}
Due to (\ref{poew}) it is $\zeta\lesssim 10^{23}$.

Equation (\ref{heth}) takes the form
\begin{equation}
\Omega_{m}=\frac{\Omega_{m0}(1\!+\!z)^{3}}
{\frac{\tilde{\alpha}}{3}\!+\!\Omega_{m0}(1\!+\!z)^{3}\!+\!(1\!-\!\Omega_{m0}\!-\!
\frac{\tilde{\alpha}}{3})(\tilde{\chi}\!+\!1)}\,,
\label{wkrd}
\end{equation}
where the first $\frac{\tilde{\alpha}}{3}$ in the denominator can be practically replaced by
$1-\Omega_{m0}$. Then, since $\tilde{\chi}_{0}=0$, equation (\ref{wkrd}) is consistent today. For
a recent range of redsifts $z$ it is $(1-\Omega_{m0}-\frac{\tilde{\alpha}}{3})(\tilde{\chi}+1)\ll 10^{5}$.
This condition actually defines this recent range of $z$. This is a very weak condition
which is also physically reasonable. Indeed, in the opposite case, $\Omega_{m}$ from (\ref{wkrd})
would become extremely small for recent $z$, which is unacceptable, since the universe would be
practically empty of matter. Therefore, this condition is expected to be valid
for all relevant recent $z$. Thus, it arises that
$(1-\Omega_{m0}-\frac{\tilde{\alpha}}{3})r_{\Sigma}^{2}a_{0}^{2}H_{0}^{2}(\tilde{\chi}
+1)\ll 1$ and (\ref{wrfs}) is well approximated by
\begin{equation}
\frac{d\tilde{\chi}}{dz}=\frac{4\zeta(\tilde{\chi}\!+\!1)^{\frac{5}{4}}}{(1\!+\!z)^{2}}\,,
\label{wkww}
\end{equation}
with general solution
\begin{equation}
\tilde{\chi}=\Big(\tilde{c}+\frac{\zeta}{1\!+\!z}\Big)^{\!-4}-1\,,
\label{sofe}
\end{equation}
where $\tilde{c}$ is integration constant. From the integration of (\ref{wkww}) it arises that
it should be $\tilde{c}+\frac{\zeta}{1+z}>0$.
Since $\tilde{\chi}_{0}=0$, the solution takes the form
\begin{equation}
\tilde{\chi}=\Big[\frac{1+z}{1-(\zeta\!-\!1)z}\Big]^{4}-1\,,
\label{sler}
\end{equation}
under the condition $(\zeta-1)z<1$.

For $\zeta\leq 1$ the condition $(\zeta-1)z<1$ is satisfied for any $z$.
From (\ref{sler}) it arises that $\tilde{\chi}+1=\mathcal{O}(1)$, thus
$(1-\Omega_{m0}-\frac{\tilde{\alpha}}{3})(\tilde{\chi}+1)\lesssim 10^{-22}$. Therefore, the
previous inequality $(1-\Omega_{m0}-\frac{\tilde{\alpha}}{3})(\tilde{\chi}+1)\ll 10^{5}$
is indeed satisfied, and moreover, (\ref{wkrd}) takes the form
\begin{equation}
\Omega_{m}=\frac{\Omega_{m0}(1\!+\!z)^{3}}
{1\!-\!\Omega_{m0}\!+\!\Omega_{m0}(1\!+\!z)^{3}}\,,
\label{kwwe}
\end{equation}
which is the $\Lambda$CDM behaviour. Therefore, in this case the behaviour of $\Omega_{m}(z)$
cannot be discerned from the $\Lambda$CDM behaviour.

For $\zeta>1$, the condition $(\zeta-1)z<1$ is satisfied for $z<z_{\zeta}$,
where $z_{\zeta}=(\zeta-1)^{-1}$. Therefore, for $\zeta>1$ the model is valid only for $z<z_{\zeta}$.
It is obvious that as $\zeta$ increases, $z_{\zeta}$ decreases and $z$ is only meaningful for a
short range around $z=0$. Thus, physically the only reasonable values of $\zeta$ are those which are
of order one. Furthermore, for any $\zeta>1$, the redshift $z$ should not be extremely close to
$z_{\zeta}$, otherwise $\tilde{\chi}\!+\!1$ would become very large, and as stated above,
$\Omega_{m}$ would become extremely suppressed, which is not acceptable.
To be more precise, let us define the quantity
\begin{equation}
\varepsilon=\Big(\!1\!-\!\Omega_{m0}\!-\!\frac{\tilde{\alpha}}{3}\Big)^{\!\frac{1}{4}}\,,
\label{ksat}
\end{equation}
where $\varepsilon\lesssim 10^{-5.5}$.
Thus, it should be $z\lesssim (1\!-\!\varepsilon) z_{\zeta}$, which is the regime of
applicability of the model, and then $(1-\Omega_{m0}-\frac{\tilde{\alpha}}{3})
(\tilde{\chi}+1)\lesssim 1$. This means that very close to the higher value of $z$, i.e. very close to
$(1\!-\!\varepsilon) z_{\zeta}$, there is a deviation from $\Lambda$CDM, while shortly later,
as $z$ reduces, the $\Omega_{m}$ behaviour is not discerned from that of $\Lambda$CDM.
When we say shortly later, we mean that for $z\lesssim (1\!-\!5\varepsilon)z_{\zeta}$
$\Lambda$CDM is established. In the initial era
$(1\!-\!5\varepsilon)z_{\zeta}\lesssim z \lesssim (1\!-\!\varepsilon)z_{\zeta}$ the full equation
(\ref{wkrd}) is valid.

The evolution of the Hubble parameter is found from (\ref{jwfr}) to be
\begin{equation}
\frac{H^{2}}{H_{0}^{2}}=\frac{\tilde{\alpha}}{3}+\Omega_{m0}(1\!+\!z)^{3}
+\Big(1\!-\!\Omega_{m0}\!-\!\frac{\tilde{\alpha}}{3}\Big)(\tilde{\chi}+1)\,.
\label{ltwr}
\end{equation}
The first term $\frac{\tilde{\alpha}}{3}$ on the r.h.s. of (\ref{ltwr}) can be practically
replaced by $1-\Omega_{m0}$.
As above, for $\zeta\leq 1$ the $\Lambda$CDM expression arises
\begin{equation}
\frac{H^{2}}{H_{0}^{2}}=1-\Omega_{m0}+\Omega_{m0}(1\!+\!z)^{3}\,,
\label{lwrw}
\end{equation}
while for $\zeta>1$ the full expression (\ref{ltwr}) is kept, which however reduces to
(\ref{lwrw}) when $z\lesssim (1\!-\!5\varepsilon)z_{\zeta}$.

In order to study the acceleration properties of the model, equation (\ref{ower}) is written as
\begin{equation}
\frac{\ddot{a}}{H_{0}^{2}a}\!=\!\frac{\tilde{\alpha}}{3}\!-\!
\frac{\Omega_{m0}}{2}(1\!+\!z)^{3}\!+\!
\Big(\!1\!-\!\Omega_{m0}\!-\!\frac{\tilde{\alpha}}{3}\Big)(\tilde{\chi}\!+\!1)
\Bigg[1\!-\!\frac{2\zeta(\tilde{\chi}\!+\!1)^{\frac{1}{4}}}{(\!1\!+\!z)\sqrt{1\!-\!
\Big[\frac{\tilde{\alpha}}{3}\!+\!\Omega_{m0}(1\!+\!z)^{3}\!+\!(1\!-\!\Omega_{m0}\!-\!
\frac{\tilde{\alpha}}{3})(\tilde{\chi}\!+\!1)\Big]\frac{r_{\Sigma}^{2}a_{0}^{2}
H_{0}^{2}}{(1+z)^{2}}}}\Bigg]\,,
\label{kwrf}
\end{equation}
while equation (\ref{kjet}) becomes
\begin{equation}
w_{DE}=-1+\frac{4\zeta(1\!-\!\Omega_{m0}\!-\!
\frac{\tilde{\alpha}}{3})(\tilde{\chi}\!+\!1)^{\frac{5}{4}}}{3
(1\!+\!z)\big[\frac{\tilde{\alpha}}{3}\!+\!(1\!-\!\Omega_{m0}\!-\!
\frac{\tilde{\alpha}}{3})(\tilde{\chi}\!+\!1)\big]\sqrt{1\!-\!
\Big[\frac{\tilde{\alpha}}{3}\!+\!\Omega_{m0}(1\!+\!z)^{3}\!+\!(1\!-\!\Omega_{m0}\!-\!
\frac{\tilde{\alpha}}{3})(\tilde{\chi}\!+\!1)\Big]\frac{r_{\Sigma}^{2}a_{0}^{2}
H_{0}^{2}}{(1+z)^{2}}}}\,.
\label{wesv}
\end{equation}
According to the previous results, the long square root in (\ref{kwrf}) can be set to unity, while
the term $\frac{\tilde{\alpha}}{3}$ in the beginning of the r.h.s. can be replaced by $1-\Omega_{m0}$.
Similar simplifications occur also in (\ref{wesv}).
If we define for convenience the quantity
\begin{equation}
\mu=\frac{3^{\frac{1}{4}}}{\xi\nu^{\frac{1}{4}}}
\Big(\frac{r_{\Sigma}^{2}a_{0}^{2}H_{0}^{2}}{H_{0}\sqrt{G_{\!N}}}\Big)^{\!\frac{1}{2}}\,,
\label{kset}
\end{equation}
then $\zeta=\mu\varepsilon$. For galaxies it is $\mu\sim 10^{27}$, while for clusters
$\mu\sim 10^{28}$.

For $\zeta\leq 1\Leftrightarrow \varepsilon < 10^{-28}$ equation (\ref{kwrf}) obtains the
$\Lambda$CDM behaviour
\begin{equation}
\frac{\ddot{a}}{H_{0}^{2}a}\!=\!1\!-\!\Omega_{m0}\!-\!
\frac{\Omega_{m0}}{2}(1\!+\!z)^{3}\,.
\label{wkee}
\end{equation}
Additionally, equation (\ref{wesv}) gives $w_{DE}=-1$. Therefore, for $\zeta\leq 1$ the
acceleration properties of the model coincide with those of $\Lambda$CDM.

For $\zeta> 1\Leftrightarrow \varepsilon> 10^{-28}$ equations (\ref{kwrf}), (\ref{wesv}) become
\begin{equation}
\frac{\ddot{a}}{H_{0}^{2}a}\!=\!1\!-\!\Omega_{m0}\!-\!
\frac{\Omega_{m0}}{2}(1\!+\!z)^{3}\!+\!
\varepsilon^{4}(\tilde{\chi}\!+\!1)
\Bigg[1\!-\!\frac{2\mu\varepsilon(\tilde{\chi}\!+\!1)^{\frac{1}{4}}}{\!1\!+\!z}\Bigg]\,,
\label{hyeb}
\end{equation}
\begin{equation}
w_{DE}=-1+\frac{4\mu\varepsilon^{5}(\tilde{\chi}\!+\!1)^{\frac{5}{4}}}{3
(1\!+\!z)\big[1\!-\!\Omega_{m0}\!+\!\varepsilon^{4}(\tilde{\chi}\!+\!1)\big]}\,.
\label{kyeb}
\end{equation}
In these equations there are some characteristic intervals of $z$ with the following hierarchy:
$(1-\mu^{\frac{1}{5}}\varepsilon)z_{\zeta}<
(1-0.5\mu^{\frac{1}{5}}\varepsilon)z_{\zeta}<(1-5\varepsilon)z_{\zeta}<(1-\varepsilon)z_{\zeta}$.
In the initial regime
$(1\!-\!0.5\mu^{\frac{1}{5}}\varepsilon)z_{\zeta}\lesssim z\lesssim(1\!-\!\varepsilon)z_{\zeta}$,
equations (\ref{hyeb}), (\ref{kyeb}) can well be approximated by only their very last terms, which
means that in this regime a deceleration is present. Especially for
$(1\!-\!5\varepsilon)z_{\zeta}\lesssim z\lesssim (1\!-\!\varepsilon)z_{\zeta}$ this deceleration
is large and is basically controlled by the parameter $\mu$. Progressively, as $z$ reduces towards
the value $(1\!-\!\mu^{\frac{1}{5}}\varepsilon)z_{\zeta}$, these last terms
become smaller, towards some values of order one, and thus, these terms are comparable to the
conventional $\Lambda$CDM terms of (\ref{hyeb}), (\ref{kyeb}). Therefore, in the redshift interval around
$z=(1\!-\!\mu^{\frac{1}{5}}\varepsilon)z_{\zeta}$, $w_{DE}$ gets a positive, order one correction of the
$\Lambda$CDM value $-1$. As we can see, for the astrophysical and cosmological values encountered in
our model, the term with the unit 1 in the bracket of (\ref{hyeb}) can always be omitted
and also the quantity $\varepsilon^{4}(\tilde{\chi}\!+\!1)$ in the denominator of
(\ref{kyeb}) is only significant for $z\sim (1-\varepsilon)z_{\zeta}$. Finally, for $z=0$
we get $\frac{\ddot{a}_{0}}{H_{0}^{2}a_{0}}=1-\frac{3}{2}\Omega_{m0}-2\mu\varepsilon^{5}$,
$w_{DE,0}=-1+\frac{4\mu\varepsilon^{5}}{3(1-\Omega_{m0})}$.
Depending on the numerical value of the quantity $\mu\varepsilon^{5}$, the values
$\ddot{a}_{0}$, $w_{DE,0}$ coincide or not with the $\Lambda$CDM ones.
Of course, in order to have acceleration today it should be
$\frac{4\mu\varepsilon^{5}}{3(1-\Omega_{m0})}<1$.
Therefore, from the beginning of the AS effect, the functions $\ddot{a}(z)$, $w_{DE}(z)$ evolve
in a non-$\Lambda$CDM way up to $z\sim (1\!-\!\mu^{\frac{1}{5}}\varepsilon)z_{\zeta}$ or
up to $z=0$, while on the contrary, as seen above, the functions $H(z)$, $\Omega(z)$
have already passed into the $\Lambda$CDM behaviour for $z\lesssim(1\!-\!5\varepsilon)z_{\zeta}$.
This peculiar phenomenon is due to the presence of the higher time derivatives contained in the
acceleration, which can lead to significant contribution from terms which are negligible in
$H$, $\Omega_{m}$. For the most interesting case with $\zeta\sim 1$, the today values of
$\ddot{a}_{0}$, $w_{DE,0}$ are the same with the $\Lambda$CDM ones, which means that after passing
the era with $z\sim (1\!-\!\mu^{\frac{1}{5}}\varepsilon)z_{\zeta}$ the model reduces to $\Lambda$CDM.
Therefore, the behaviour in the evolution of the model around the passage from deceleration to
acceleration differs from the $\Lambda$CDM one and could in principle be discerned using precise
observational data. However, this is not the case since it is
$\mu^{\frac{1}{5}}\varepsilon=\zeta\mu^{-\frac{4}{5}}\ll 1$
for the numerical values of the astrophysical and cosmological quantities we are interested in.
Only in the case that the quantity $\nu$ is (\ref{kset}) is substantially enlarged, which
is not predicted by the theory, could $\mu$ be reduced essentially.
Therefore, the previous eras of deviation from $\Lambda$CDM cannot be observed and the
model is practically identical to $\Lambda$CDM in all the range of its validity.
For $1\ll \zeta\ll 10^{22}$, it is still $\mu^{\frac{1}{5}}\varepsilon=\zeta\mu^{-\frac{4}{5}}\ll 1$,
which means that the model is indistinguishable from $\Lambda$CDM, beyond the fact that $z_{\zeta}$
is already unphysically small. Finally, for $\zeta\sim 10^{22}$, which means
$\varepsilon\sim 10^{-5}$, it is $\mu^{\frac{1}{5}}\varepsilon=\zeta\mu^{-\frac{4}{5}}\sim 1$.
Therefore, in this case the today value of $w_{DE}$ is different from $-1$ and the model
is always different from $\Lambda$CDM. However, the model in this case is meaningless since
$z_{\zeta}$ in extraordinarily small. As a result we can summarize saying that in the physically
meaningful case with $\zeta\sim 1$, the acceleration properties of the model cannot be discerned
from $\Lambda$CDM.

\end{appendix}


\begin{thebibliography}{99}

\bibitem{Weinberg:1988cp}
  S.~Weinberg,
  Rev.\ Mod.\ Phys.\  {\bf 61}, 1 (1989)
  doi:10.1103/RevModPhys.61.1.

\bibitem{Sahni:1999gb}
  V.~Sahni and A.~A.~Starobinsky,
  Int.\ J.\ Mod.\ Phys.\ D {\bf 9}, 373 (2000)
  doi:10.1142/S0218271800000542
  [astro-ph/9904398];
  S.~M.~Carroll,
  Living Rev.\ Rel.\  {\bf 4}, 1 (2001)
  doi:10.12942/lrr-2001-1
  [astro-ph/0004075];
  T.~Padmanabhan,
  Phys.\ Rept.\  {\bf 380}, 235 (2003)
  doi:10.1016/S0370-1573(03)00120-0
  [hep-th/0212290];
  S.~Nojiri and S.~D.~Odintsov,
  eConf C {\bf 0602061}, 06 (2006)
  [Int.\ J.\ Geom.\ Meth.\ Mod.\ Phys.\  {\bf 4}, 115 (2007)]
  doi:10.1142/S0219887807001928
  [hep-th/0601213];
  M.~Li, X.~D.~Li, S.~Wang and Y.~Wang,
  Commun.\ Theor.\ Phys.\  {\bf 56}, 525 (2011)
  doi:10.1088/0253-6102/56/3/24
  [arXiv:1103.5870 [astro-ph.CO]];
  J.~Martin,
  Comptes Rendus Physique {\bf 13}, 566 (2012)
  doi:10.1016/j.crhy.2012.04.008
  [arXiv:1205.3365 [astro-ph.CO]].

\bibitem{Peebles:2002gy}
  P.~J.~E.~Peebles and B.~Ratra,
  Rev.\ Mod.\ Phys.\  {\bf 75}, 559 (2003)
  doi:10.1103/RevModPhys.75.559
  [astro-ph/0207347];
  E.~J.~Copeland, M.~Sami and S.~Tsujikawa,
  Int.\ J.\ Mod.\ Phys.\ D {\bf 15}, 1753 (2006)
  doi:10.1142/S021827180600942X
  [hep-th/0603057].

\bibitem{Sola:2013gha}
  J.~Sola,
  J.\ Phys.\ Conf.\ Ser.\  {\bf 453}, 012015 (2013)
  doi:10.1088/1742-6596/453/1/012015
  [arXiv:1306.1527 [gr-qc]].

\bibitem{Riess:1998cb}
  A.~G.~Riess {\it et al.} [Supernova Search Team],
  Astron.\ J.\  {\bf 116}, 1009 (1998)
  doi:10.1086/300499
  [astro-ph/9805201].

\bibitem{Perlmutter:1998np}
  S.~Perlmutter {\it et al.} [Supernova Cosmology Project Collaboration],
  Astrophys.\ J.\  {\bf 517}, 565 (1999)
  doi:10.1086/307221
  [astro-ph/9812133].

\bibitem{Knop:2003iy}
  R.~A.~Knop {\it et al.} [Supernova Cosmology Project Collaboration],
  Astrophys.\ J.\  {\bf 598}, 102 (2003)
  doi:10.1086/378560
  [astro-ph/0309368];
  A.~G.~Riess {\it et al.} [Supernova Search Team],
  Astrophys.\ J.\  {\bf 607}, 665 (2004)
  doi:10.1086/383612
  [astro-ph/0402512].

\bibitem{Spergel:2006hy}
  D.~N.~Spergel {\it et al.} [WMAP Collaboration],
  Astrophys.\ J.\ Suppl.\  {\bf 170}, 377 (2007)
  doi:10.1086/513700
  [astro-ph/0603449].

\bibitem{Komatsu:2010fb}
  E.~Komatsu {\it et al.} [WMAP Collaboration],
  Astrophys.\ J.\ Suppl.\  {\bf 192}, 18 (2011)
  doi:10.1088/0067-0049/192/2/18
  [arXiv:1001.4538 [astro-ph.CO]].

\bibitem{Ade:2013zuv}
  P.~A.~R.~Ade {\it et al.} [Planck Collaboration],
  Astron.\ Astrophys.\  {\bf 571}, A16 (2014)
  doi:10.1051/0004-6361/201321591
  [arXiv:1303.5076 [astro-ph.CO]].

\bibitem{Huterer:1998qv}
  D.~Huterer and M.~S.~Turner,
  Phys.\ Rev.\ D {\bf 60}, 081301 (1999)
  doi:10.1103/PhysRevD.60.081301
  [astro-ph/9808133].

\bibitem{ASreviews}
M.~Niedermaier and M.~Reuter,
 Living Rev.\ Rel.\  {\bf 9}, 5 (2006);
 R.~Percacci,
 In {\textit{Oriti, D. (ed.): Approaches to quantum gravity}} 111-128
 [arXiv:0709.3851 [hep-th]];
 O.~Lauscher and M.~Reuter,
 In {\textit{Fauser, B. (ed.) et al.: Quantum gravity}} 293-313 [hep-th/0511260];
 M.~Reuter and F.~Saueressig,
 New J.\ Phys.\  {\bf 14}, 055022 (2012)
 [arXiv:1202.2274 [hep-th]];
 A.~Bonanno,
 PoS CLAQG {\bf 08}, 008 (2011) [arXiv:0911.2727 [hep-th]];
 M.~Niedermaier,
 Class.\ Quant.\ Grav.\  {\bf 24}, R171 (2007) [gr-qc/0610018].

\bibitem{Einstein:1946ev}
  A.~Einstein and E.~G.~Strauss,
  Annals Math.\  {\bf 47}, 731 (1946)
  doi:10.2307/1969231.

\bibitem{Dymnikova:2001fb}
  I.~Dymnikova,
  Class.\ Quant.\ Grav.\  {\bf 19}, 725 (2002)
  doi:10.1088/0264-9381/19/4/306
  [gr-qc/0112052].

\bibitem{Hayward:2005gi}
  S.~A.~Hayward,
  Phys.\ Rev.\ Lett.\  {\bf 96}, 031103 (2006)
  doi:10.1103/PhysRevLett.96.031103
  [gr-qc/0506126];
  T.~De Lorenzo, C.~Pacilio, C.~Rovelli and S.~Speziale,
  Gen.\ Rel.\ Grav.\  {\bf 47}, no. 4, 41 (2015)
  doi:10.1007/s10714-015-1882-8
  [arXiv:1412.6015 [gr-qc]].

\bibitem{Kiritsis:2009rx}
  E.~B.~Kiritsis and G.~Kofinas,
  JHEP {\bf 1001}, 122 (2010)
  doi:10.1007/JHEP01(2010)122
  [arXiv:0910.5487 [hep-th]];
  J.~Z.~Tang and B.~Chen,
  Phys.\ Rev.\ D {\bf 81}, 043515 (2010)
  doi:10.1103/PhysRevD.81.043515
  [arXiv:0909.4127 [hep-th]].

\bibitem{Babichev:2013cya}
  E.~Babichev and C.~Charmousis,
  JHEP {\bf 1408}, 106 (2014)
  doi:10.1007/JHEP08(2014)106
  [arXiv:1312.3204 [gr-qc]];
  C.~Bambi, D.~Malafarina and L.~Modesto,
  Phys.\ Rev.\ D {\bf 88}, 044009 (2013)
  doi:10.1103/PhysRevD.88.044009
  [arXiv:1305.4790 [gr-qc]].

\bibitem{Rodrigues:2015hba}
  D.~C.~Rodrigues, B.~Chauvineau and O.~F.~Piattella,
  JCAP {\bf 1509}, no. 09, 009 (2015)
  doi:10.1088/1475-7516/2015/09/009
  [arXiv:1504.05119 [gr-qc]].

\bibitem{inho}
  G.~F.~R.~Ellis and W.~Stoeger,
  Class.\ Quant.\ Grav.\  {\bf 4}, 1697 (1987)
  doi:10.1088/0264-9381/4/6/025;
S.~R.~Green and R.~M.~Wald,
Phys.\ Rev.\ D {\bf 83}, 084020 (2011)
doi:10.1103/PhysRevD.83.084020
[arXiv:1011.4920 [gr-qc]];
T.~Buchert, M.~J.~France and F.~Steiner,
Class.\ Quant.\ Grav.\  {\bf 34}, no. 9, 094002 (2017),
doi:10.1088/1361-6382/aa5ce2
[arXiv:1701.03347 [astro-ph.CO]];
T.~Buchert, A.~A.~Coley, H.~Kleinert, B.~F.~Roukema and D.~L.~Wiltshire,
Int.\ J.\ Mod.\ Phys.\ D {\bf 25}, no. 03, 1630007 (2016),
doi:10.1142/S021827181630007X, 10.1142/9789813226609-0034
[arXiv:1512.03313 [astro-ph.CO]].

\bibitem{inhoswiss}
S.~M.~Koksbang,
Phys.\ Rev.\ D {\bf 95}, no. 6, 063532 (2017)
doi:10.1103/PhysRevD.95.063532
[arXiv:1703.03572 [astro-ph.CO]];
K.~Bolejko and M.~N.~Celerier,
Phys.\ Rev.\ D {\bf 82}, 103510 (2010)
doi:10.1103/PhysRevD.82.103510
[arXiv:1005.2584 [astro-ph.CO]];
P.~Mishra, M.~N.~Celerier and T.~P.~Singh,
Phys.\ Rev.\ D {\bf 86}, 083520 (2012)
doi:10.1103/PhysRevD.86.083520
[arXiv:1206.6026 [astro-ph.CO]];
T.~Biswas and A.~Notari,
JCAP {\bf 0806}, 021 (2008)
doi:10.1088/1475-7516/2008/06/021
[astro-ph/0702555];
V.~Marra, E.~W.~Kolb and S.~Matarrese,
Phys.\ Rev.\ D {\bf 77}, 023003 (2008)
doi:10.1103/PhysRevD.77.023003
[arXiv:0710.5505 [astro-ph]].

\bibitem{structure}
  S.~Rasanen,
  EAS Publ.\ Ser.\  {\bf 36}, 63 (2009)
  doi:10.1051/eas/0936008
  [arXiv:0811.2364 [astro-ph]];
  S.~Rasanen,
  arXiv:1012.0784 [astro-ph.CO];
R.~A.~Sussman,
Class.\ Quant.\ Grav.\  {\bf 28}, 235002 (2011)
doi:10.1088/0264-9381/28/23/235002
[arXiv:1102.2663 [gr-qc]];
  M.~Lavinto, S.~Rasanen and S.~J.~Szybka,
  JCAP {\bf 1312}, 051 (2013)
  doi:10.1088/1475-7516/2013/12/051
  [arXiv:1308.6731 [astro-ph.CO]].

\bibitem{Kofinas:2011pq}
G.~Kofinas and V.~Zarikas,
Eur.\ Phys.\ J.\ C {\bf 73}, no. 4, 2379 (2013)
doi:10.1140/epjc/s10052-013-2379-9
[arXiv:1107.2602 [hep-th]].

\bibitem{Israel:1966rt}
  W.~Israel,
  Nuovo Cim.\ B {\bf 44S10}, 1 (1966)
  [Nuovo Cim.\ B {\bf 44}, 1 (1966)],
  Erratum: [Nuovo Cim.\ B {\bf 48}, 463 (1967)]
  doi:10.1007/BF02710419, 10.1007/BF02712210;
G. Darmois, {\em M\'{e}morial des Sciences Math\'{e}matiques\/},
Fascicule XXV (Gauthier-Villars, Paris, 1927), Chap. V.

\bibitem{Wetterich:1994bg}
  C.~Wetterich,
  Astron.\ Astrophys.\  {\bf 301}, 321 (1995)
  [hep-th/9408025];
  L.~Amendola,
  Phys.\ Rev.\ D {\bf 60}, 043501 (1999)
  doi:10.1103/PhysRevD.60.043501
  [astro-ph/9904120];
  L.~P.~Chimento, A.~S.~Jakubi, D.~Pavon and W.~Zimdahl,
  Phys.\ Rev.\ D {\bf 67}, 083513 (2003)
  doi:10.1103/PhysRevD.67.083513
  [astro-ph/0303145].

\bibitem{Kofinas:2005hc}
  G.~Kofinas, G.~Panotopoulos and T.~N.~Tomaras,
  JHEP {\bf 0601}, 107 (2006)
  doi:10.1088/1126-6708/2006/01/107
  [hep-th/0510207].

\bibitem{Ade:2015xua}
  P.~A.~R.~Ade {\it et al.} [Planck Collaboration],
  Astron.\ Astrophys.\  {\bf 594}, A13 (2016)
  doi:10.1051/0004-6361/201525830
  [arXiv:1502.01589 [astro-ph.CO]].

\bibitem{Torres:2014gta}
  R.~Torres,
  Phys.\ Lett.\ B {\bf 733}, 21 (2014)
  doi:10.1016/j.physletb.2014.04.010
  [arXiv:1404.7655 [gr-qc]].

\bibitem{Koch:2014cqa}
B.~Koch and F.~Saueressig,
Int.\ J.\ Mod.\ Phys.\ A {\bf 29}, no. 8, 1430011 (2014)
doi:10.1142/S0217751X14300117
[arXiv:1401.4452 [hep-th]].

\bibitem{Bonanno:2001xi}
A.~Bonanno and M.~Reuter,
Phys.\ Rev.\ D {\bf 65}, 043508 (2002)
doi:10.1103/PhysRevD.65.043508
[hep-th/0106133].

\bibitem{Bhattacharya:2013tq}
  S.~Bhattacharya and A.~Lahiri,
  Eur.\ Phys.\ J.\ C {\bf 73}, 2673 (2013)
  doi:10.1140/epjc/s10052-013-2673-6
  [arXiv:1301.4532 [gr-qc]].

\bibitem{Reuter:2001ag}
M.~Reuter and F.~Saueressig,
Phys.\ Rev.\ D {\bf 65}, 065016 (2002)
doi:10.1103/PhysRevD.65.065016
[hep-th/0110054].

\bibitem{Bonanno:2001hi}
A.~Bonanno and M.~Reuter,
Phys.\ Lett.\ B {\bf 527}, 9 (2002)
doi:10.1016/S0370-2693(01)01522-2
[astro-ph/0106468];
A.~Bonanno and M.~Reuter,
Int.\ J.\ Mod.\ Phys.\ D {\bf 13}, 107 (2004)
doi:10.1142/S0218271804003809
[astro-ph/0210472];
E.~Bentivegna, A.~Bonanno and M.~Reuter,
JCAP {\bf 0401}, 001 (2004)
doi:10.1088/1475-7516/2004/01/001
[astro-ph/0303150];
I.~Donkin and J.~M.~Pawlowski,
arXiv:1203.4207 [hep-th].

\bibitem{Reuter:2009kq}
M.~Reuter and H.~Weyer,
 Gen.\ Rel.\ Grav.\  {\bf 41}, 983 (2009) [arXiv:0903.2971 [hep-th]];
 M.~Reuter and H.~Weyer,
Phys.\ Rev.\ D {\bf 79}, 105005 (2009) [arXiv:0801.3287 [hep-th]];
 P.~F.~Machado and R.~Percacci,
 Phys.\ Rev.\ D {\bf 80}, 024020 (2009) [arXiv:0904.2510 [hep-th]];
 E.~Manrique and M.~Reuter,
 PoS CLAQG {\bf 08}, 001 (2011) [arXiv:0905.4220 [hep-th]].

\bibitem{Reuter:2004nx}
M.~Reuter and H.~Weyer,
JCAP {\bf 0412}, 001 (2004)
doi:10.1088/1475-7516/2004/12/001
[hep-th/0410119];
M.~Reuter and H.~Weyer,
Phys.\ Rev.\ D {\bf 70}, 124028 (2004)
doi:10.1103/PhysRevD.70.124028
[hep-th/0410117].

\bibitem{Esposito:2007xz}
G.~Esposito, C.~Rubano and P.~Scudellaro,
Class.\ Quant.\ Grav.\  {\bf 24}, 6255 (2007)
doi:10.1088/0264-9381/24/24/008
[arXiv:0709.1403 [gr-qc]].

\bibitem{fitting}
F. Anagnostopoulos, S. Basilakos, G. Kofinas, V. Zarikas, to appear.

\bibitem{Nesti:2013uwa}
F.~Nesti and P.~Salucci,
JCAP {\bf 1307}, 016 (2013)
doi:10.1088/1475-7516/2013/07/016
[arXiv:1304.5127 [astro-ph.GA]].

\bibitem{Weinberg:2009bg}
  S.~Weinberg,
  PoS CD {\bf 09}, 001 (2009)
  [arXiv:0908.1964 [hep-th]].

\bibitem{Reuter:1996cp}
  M.~Reuter,
  Phys.\ Rev.\ D {\bf 57}, 971 (1998)
  doi:10.1103/PhysRevD.57.971
  [hep-th/9605030].

\bibitem{Niedermaier:2006wt}
M.~Niedermaier, M.~Reuter,
Living Rev.\ Rel.\ 9 (2006) 5.

\bibitem{Codello:2008vh}
  A.~Codello, R.~Percacci and C.~Rahmede,
  Annals Phys.\  {\bf 324}, 414 (2009)
  doi:10.1016/j.aop.2008.08.008
  [arXiv:0805.2909 [hep-th]].

\bibitem{Litim:2011cp}
  D.~F.~Litim,
  Phil.\ Trans.\ Roy.\ Soc.\ Lond.\ A {\bf 369}, 2759 (2011)
  doi:10.1098/rsta.2011.0103
  [arXiv:1102.4624 [hep-th]].

\bibitem{Percacci:2011fr}
  R.~Percacci,
  arXiv:1110.6389 [hep-th].

\bibitem{Reuter:2012id}
  M.~Reuter and F.~Saueressig,
  New J.\ Phys.\  {\bf 14}, 055022 (2012)
  doi:10.1088/1367-2630/14/5/055022
  [arXiv:1202.2274 [hep-th]].

\bibitem{Reuter:2012xf}
  M.~Reuter and F.~Saueressig,
  Lect.\ Notes Phys.\  {\bf 863}, 185 (2013)
  doi:10.1007/978-3-642-33036-08
  [arXiv:1205.5431 [hep-th]].

\bibitem{Benedetti:2010nr}
  D.~Benedetti, K.~Groh, P.~F.~Machado and F.~Saueressig,
  JHEP {\bf 1106}, 079 (2011)
  doi:10.1007/JHEP06(2011)079
  [arXiv:1012.3081 [hep-th]].

\bibitem{Reuter:2005kb}
  M.~Reuter and F.~Saueressig,
  JCAP {\bf 0509}, 012 (2005)
  doi:10.1088/1475-7516/2005/09/012
  [hep-th/0507167].

\bibitem{Weinberg:2009wa}
  S.~Weinberg,
  Phys.\ Rev.\ D {\bf 81}, 083535 (2010)
  doi:10.1103/PhysRevD.81.083535
  [arXiv:0911.3165 [hep-th]].

\bibitem{Koch:2010nn}
  B.~Koch and I.~Ramirez,
  Class.\ Quant.\ Grav.\  {\bf 28}, 055008 (2011)
  doi:10.1088/0264-9381/28/5/055008
  [arXiv:1010.2799 [gr-qc]].

\bibitem{Casadio:2010fw}
  R.~Casadio, S.~D.~H.~Hsu and B.~Mirza,
  Phys.\ Lett.\ B {\bf 695}, 317 (2011)
  doi:10.1016/j.physletb.2010.10.060
  [arXiv:1008.2768 [gr-qc]].

\bibitem{Bonanno:2010bt}
  A.~Bonanno, A.~Contillo and R.~Percacci,
  Class.\ Quant.\ Grav.\  {\bf 28}, 145026 (2011)
  doi:10.1088/0264-9381/28/14/145026
  [arXiv:1006.0192 [gr-qc]].

\bibitem{Hindmarsh:2011hx}
  M.~Hindmarsh, D.~Litim and C.~Rahmede,
  JCAP {\bf 1107}, 019 (2011)
  doi:10.1088/1475-7516/2011/07/019
  [arXiv:1101.5401 [gr-qc]].

\bibitem{Bonanno:2011yx}
  A.~Bonanno and S.~Carloni,
  New J.\ Phys.\  {\bf 14}, 025008 (2012)
  doi:10.1088/1367-2630/14/2/025008
  [arXiv:1112.4613 [gr-qc]].

\bibitem{Ahn:2011qt}
  C.~Ahn, C.~Kim and E.~V.~Linder,
  Phys.\ Lett.\ B {\bf 704}, 10 (2011)
  doi:10.1016/j.physletb.2011.08.075
  [arXiv:1106.1435 [astro-ph.CO]].

\bibitem{Cai:2011kd}
  Y.~F.~Cai and D.~A.~Easson,
  Phys.\ Rev.\ D {\bf 84}, 103502 (2011)
  doi:10.1103/PhysRevD.84.103502
  [arXiv:1107.5815 [hep-th]].

\bibitem{Fang:2012ca}
  C.~Fang and Q.~G.~Huang,
  Eur.\ Phys.\ J.\ C {\bf 73}, no. 4, 2401 (2013)
  doi:10.1140/epjc/s10052-013-2401-2
  [arXiv:1210.7596 [hep-th]].

\bibitem{Bonanno:2013dja}
  A.~Bonanno and M.~Reuter,
  Phys.\ Rev.\ D {\bf 87}, no. 8, 084019 (2013)
  doi:10.1103/PhysRevD.87.084019
  [arXiv:1302.2928 [hep-th]].

\bibitem{Kaya:2013bga}
  A.~Kaya,
  Phys.\ Rev.\ D {\bf 87}, 123501 (2013)
  doi:10.1103/PhysRevD.87.123501
  [arXiv:1303.5459 [hep-th]].

\bibitem{Becker:2014jua}
  D.~Becker and M.~Reuter,
  JHEP {\bf 1412}, 025 (2014)
  doi:10.1007/JHEP12(2014)025
  [arXiv:1407.5848 [hep-th]].

\bibitem{Xianyu:2014eba}
  Z.~Z.~Xianyu and H.~J.~He,
  JCAP {\bf 1410}, 083 (2014)
  doi:10.1088/1475-7516/2014/10/083
  [arXiv:1407.6993 [astro-ph.CO]].

\bibitem{Nielsen:2015una}
  N.~G.~Nielsen, F.~Sannino and O.~Svendsen,
  Phys.\ Rev.\ D {\bf 91}, 103521 (2015)
  doi:10.1103/PhysRevD.91.103521
  [arXiv:1503.00702 [hep-ph]].

\bibitem{Bonanno:2015fga}
  A.~Bonanno and A.~Platania,
  Phys.\ Lett.\ B {\bf 750}, 638 (2015)
  doi:10.1016/j.physletb.2015.10.005
  [arXiv:1507.03375 [gr-qc]].

\bibitem{Frolov:2011ys}
A.~V.~Frolov and J.~Q.~Guo,
[arXiv:1101.4995 [astro-ph.CO]].

\bibitem{Hindmarsh:2012rc}
M.~Hindmarsh and I.~D.~Saltas,
Phys.\ Rev.\ D {\bf 86}, 064029 (2012), [arXiv:1203.3957 [gr-qc]].

\bibitem{Copeland:2013vva}
E.~J.~Copeland, C.~Rahmede and I.~D.~Saltas,
Phys.\ Rev.\ D {\bf 91}, no. 10, 103530 (2015), [arXiv:1311.0881 [gr-qc]].

\bibitem{c3}
S.-H.~H.~Tye and J.~Xu,
Phys.\ Rev.\ D {\bf 82}, 127302 (2010)
[arXiv:1008.4787 [hep-th]];
B.~F.~L.~Ward,
PoS ICHEP {\bf 2010}, 477 (2010)
[arXiv:1012.2680 [gr-qc]];
R.~J.~Yang,
Eur.\ Phys.\ J.\ C {\bf 72}, 1948 (2012)
[arXiv:1108.0227 [gr-qc]].

\bibitem{Bonanno:2008xp}
A.~Bonanno and M.~Reuter,
J.\ Phys.\ Conf.\ Ser.\  {\bf 140}, 012008 (2008), [arXiv:0803.2546 [astro-ph]].

\bibitem{bianchi}
M.~Reuter and H.~Weyer,
Phys.\ Rev.\ D {\bf 69}, 104022 (2004), [hep-th/0311196].

\bibitem{Kofinas:2015sna}
  G.~Kofinas and V.~Zarikas,
  JCAP {\bf 1510}, no. 10, 069 (2015)
  doi:10.1088/1475-7516/2015/10/069
  [arXiv:1506.02965 [hep-th]].

\bibitem{Bonanno:2007wg}
  A.~Bonanno and M.~Reuter,
  JCAP {\bf 0708}, 024 (2007)
  doi:10.1088/1475-7516/2007/08/024
  [arXiv:0706.0174 [hep-th]].

\bibitem{Kofinas:2016lcz}
  G.~Kofinas and V.~Zarikas,
  Phys.\ Rev.\ D {\bf 94}, no. 10, 103514 (2016)
  doi:10.1103/PhysRevD.94.103514
  [arXiv:1605.02241 [gr-qc]].

\bibitem{Bonanno:2017gji}
  A.~Bonanno, S.~J.~Gabriele Gionti and A.~Platania,
  Class.\ Quant.\ Grav.\  {\bf 35}, no. 6, 065004 (2018)
  doi:10.1088/1361-6382/aaa535
  [arXiv:1710.06317 [gr-qc]].

\bibitem{Guberina:2002wt}
  B.~Guberina, R.~Horvat and H.~Stefancic,
  Phys.\ Rev.\ D {\bf 67}, 083001 (2003)
  doi:10.1103/PhysRevD.67.083001
  [hep-ph/0211184].

\bibitem{Bauer:2005rpa}
  F.~Bauer,
  Class.\ Quant.\ Grav.\  {\bf 22}, 3533 (2005)
  doi:10.1088/0264-9381/22/17/012
  [gr-qc/0501078].


\end{thebibliography}
\end{document}